\documentclass[aps,superscriptaddress,nofootinbib,eqsecnum,prd,notitlepage,showkeys]{revtex4-1} 

\pdfoutput=1

\usepackage{amsfonts}
\usepackage{amsmath}
\usepackage{amssymb}
\usepackage{graphicx,color}
\usepackage{float}
\usepackage{hyperref}
\usepackage{subfigure}
\usepackage{dcolumn}
\usepackage{soul}
\usepackage{ulem}
\usepackage{verbatim}
\usepackage{graphicx}
\usepackage{color}
\usepackage{xparse}

\begin{document}

\title{Recent developments in warm inflation}

\author{Vahid  Kamali}
\email{vkamali@basu.ac.ir}
\affiliation{Department of Physics, Bu-Ali Sina (Aviccena) University,
  Hamedan 65178, 016016, Iran}
\affiliation{School of Physics,
Institute for Research in Fundamental Sciences (IPM),
19538-33511, Tehran, Iran}
\affiliation{Physics Department, McGill University, Montreal, Quebec, H3A 2T8, Canada}

\author{Meysam Motaharfar}
\email{mmotah4@lsu.edu}
\affiliation{Department of Physics and Astronomy, Louisiana State University,
  Baton Rouge, LA 70803, USA}

\author{Rudnei O. Ramos} \email{rudnei@uerj.br}
\affiliation{Departamento de Fisica Teorica, Universidade do Estado do
  Rio de Janeiro, 20550-013 Rio de Janeiro, RJ, Brazil }
\affiliation{Physics Department, McGill University, Montreal, Quebec, H3A 2T8, Canada}

\begin{abstract}

Warm inflation, its different particle physics model implementations
and  the implications of dissipative particle production for its
cosmology are  reviewed. First, we briefly present the background
dynamics of warm inflation and contrast it with the cold inflation
picture. An exposition of the space of
parameters  for different well-motivated potentials, which are ruled
out, or severely constrained in the cold inflation scenario, but not
necessarily in warm inflation, is provided. 
Next, the quantum field theory aspects in realizing explicit
microscopic models for warm inflation are
given. This includes the derivation of dissipation coefficients relevant
in warm inflation for different particle field theory models.  The
dynamics of cosmological perturbations in warm inflation are then described.
The general expression for the curvature scalar power spectrum is shown.
We then discuss in details the relevant regimes of warm inflation,  the weak and strong
dissipative regimes. We also discuss the results predicted in these regimes
of warm inflation and how they are confronted with the observational data. 
We explain 
how the dissipative dynamics in warm inflation can address several long-standing 
issues related to (post-) inflationary cosmology. This includes recent 
discussions concerning the so-called swampland criteria and how warm inflation 
can belong to the landscape of string theory. 

\end{abstract}

\maketitle

\section{Introduction} 
\label{intro}

Among the different proposals that attempted to implement consistent
inflationary dynamics within an explicit quantum field theory
realization, the warm inflation (WI)
paradigm~\cite{Berera:1995wh,Berera:1995ie,Berera:1998px} is one of
the most attractive. Warm inflation explores the fact that the
inflationary dynamics is inherently a multifield problem, since the
vacuum energy that drives inflation eventually must be converted to
radiation, which generally is comprised of a variety of particle
species. Thus, WI model realizations explore those associated
dissipative processes to realize radiation production concurrently
with the inflationary expansion\footnote{{}For an earlier review on the
  microphysics of warm inflation see  Ref.~\cite{Berera:2008ar},
  while for its phenomenology, see
  Ref.~\cite{BasteroGil:2009ec}.}.  This is the opposite of the more
usual scenario of cold (supercooled) inflation
(CI)~\cite{Guth:1980zm,Sato:1981ds,Albrecht:1982wi,Linde:1981mu,Linde:1983gd},
where a separated period of radiation production after the end of
inflation (graceful exit) is required.  

{}From a model building perspective, the recent developments have
aimed at overcoming some of the important issues found in earlier
particle physics realizations of warm inflation. In order to be able
to sustain  a nearly-thermal bath during WI, a sufficiently strong
dissipation is typically required, such that some of the energy
density in the inflaton can be converted to radiation.  {}For this to
happen, earlier particle physics realizations of WI required  large
field multiplicities~\cite{BasteroGil:2009ec,Bartrum:2013fia}. These
large field multiplicities can be difficult to be generated in simple
models while keeping perturbativity and unitarity in these
models~\cite{Bastero-Gil:2012akf} (see, however, for natural
realizations in the context of brane
models~\cite{Bastero-Gil:2011zxb}, or in  extra-dimensional models
with a Kaluza-Klein tower~\cite{Matsuda:2012kc}).  One other
difficulty in WI model building is to properly control both quantum
and thermal corrections to the inflaton such as not to spoil the
flatness of its potential, which could otherwise prevent inflation to
happen. Earlier models for WI have made use, for instance, of both
supersymmetry and heavy intermediate fields coupled to the inflaton
for this purpose~\cite{Berera:2002sp,Berera:2008ar}. More recent
implementations of WI have focused, instead, in using symmetry
properties such as to be able to efficiently control the corrections
to the inflaton
potential~\cite{Bastero-Gil:2016qru,Bastero-Gil:2019gao,Berghaus:2019whh}.
{}Finally, from an effective field theory point of view, WI
constructions that can be able to achieve strong dissipative regimes have
been shown to display quite appealing features.  {}For example,
already in some of the first studies in
WI~\cite{Berera:1999ws,Berera:2004vm} it has been claimed that WI in
the strong dissipative regime can also prevent super-Planckian field
excursions for the inflaton,  thus making WI potentially attractive in
terms of an effective field theory consistent with an UV-complete
realization in terms of quantum
gravity~\cite{Das:2018hqy,Motaharfar:2018zyb,Das:2018rpg,Das:2019hto,Kamali:2019xnt,Brandenberger:2020oav,Das:2020xmh,Berera:2019zdd,Berera:2020dvn,Kamali:2019hgv,Kamali:2019wdh,Berera:2020iyn,Kamali:2018ylz,Kamali:2021gkl}.
{}Finally, the dissipative effects in WI can lower the energy scale of
inflation and, as a result of this, the tensor-to-scalar ratio can be
decreased with respect to what it would be in CI for the same type of
primordial inflaton potential. This makes several
primordial inflaton potential models that would otherwise be and had been discarded
in CI, to be in line with the CMB observations in the context of
WI~\cite{Benetti:2016jhf}.  The above are just a few examples of
recent developments in WI and which have been attracting increasingly
interest in this intriguing alternative picture of inflation. In this
paper we review some of these major developments achieved in the area
in the recent years. 

This paper is divided as follows.  In Section~\ref{section2},  we
start by briefly reviewing the WI background dynamics and contrasting
it to the CI picture. We discuss how a supplementary friction term in
the Klein-Gordon equation is able to bring about a richer dynamics for
WI. The smooth connection of the end of WI with the radiation
dominated regime is discussed.  We show that there are several
different possibilities for graceful exit depending on the form of the
inflaton potential, the dissipation coefficient and whether being in
the weak or strong dissipative regimes.  In Section~\ref{section3},
we describe the necessary tools for calculating dissipation
coefficients in the context of non-equilibrium quantum field theory
and which are applied to WI. Some of the most recent microscopic
realizations of WI are discussed and the respective derivation of the
dissipation coefficients for these models is outlined.  In
Section~\ref{section4}, we discuss the cosmological perturbation
theory for WI. Several important issues are discussed and the general
derivation of the scalar of curvature power spectrum in WI is
given. The bispectrum and non-Gaussianities in WI are also discussed.
In Section~\ref{section5}, we discuss the observational constraints
and other applications of the WI dissipative dynamics. It is shown how
the dissipative particle production addresses/alleviates some of the
long-outstanding problems in cosmology, e.g., related to the 
inflationary and post-inflationary phases and which CI cannot directly answer.  In
Section~\ref{section5}, we also discuss the connections which WI 
recently made with the so-called swampland criteria. We start by briefly
reviewing the motivation behind the swampland
conjectures~\cite{Obied:2018sgi,Ooguri:2018wrx,Kinney:2018nny,Bedroya:2019snp,Agrawal:2018own,Bedroya:2019tba}. 
We discuss why the dynamics of WI allows it to satisfy the swampland
conjectures. Given the constraints imposed by the swampland conjectures, we
find under which conditions WI is able to simultaneously satisfy the
swampland conjectures and the implications and the implications of this for building
inflationary models in string theory in the context of WI.  An overview of different WI
implementations and applications, including in the context of
noncanonical models, is also given.  {}Finally, in
Section~\ref{section6}, we give our concluding remarks.


\section{Background Dynamics of WI}
\label{section2}

A WI regime is typically realized when the inflaton field is able to
dissipate its energy into other light degrees of freedom with a rate
that is faster than the Hubble expansion. Thus, the produced particles
have enough time to thermalize and become radiation. During this time,
where the inflaton is decaying into the radiation particles and that can
subsequently thermalize, one can
then model their contributions as simply a radiation fluid with
$\rho_{r} = \pi^2 g_{*} T^4/30$, with $\rho_{r}$, $T$ and $g_{*}$
being the radiation energy density, the temperature and the effective
number of relativistic degrees of freedom of the produced
particles. Hence, the total energy density of the universe in the WI
scenario contains both the inflaton field and a primordial radiation
energy density, i.e., $\rho = \rho_{\phi} + \rho_{r}$, where
$\rho_{\phi}$ is the inflaton field energy density.  Energy
conservation then demands that the energy lost by the inflaton field
must be gained by the radiation fluid. Therefore, the evolution
equations can be obtained from the conservation of the energy-momentum
tensor $T^{\mu\nu}$ \cite{Bastero-Gil:2011rva},
\begin{align}\label{conservation}
\nabla_{\mu} T^{\mu\nu} =0.
\end{align}

We work in the spatially flat {}Friedmann-Lema\^{\i}tre-Robertson-Walker
(FLRW) metric,  $ds^2 = - dt^2 + a(t)^2 \delta_{ij} dx^i dx^j$, where
$a(t)$ is the scale factor. Hence, Eq.~(\ref{conservation}) leads to a 
set of continuity equations for each component of the cosmological fluid,
\begin{align}
\dot \rho_{\alpha} + 3H(\rho_{\alpha} + p_{\alpha}) = Q_{\alpha},
\label{fluid}
\end{align}
where a dot here means  a derivative with respect to the cosmic time
$t$, with $\rho_{\alpha}$ and $p_{\alpha}$ being the energy density
and pressure for each fluid component $\alpha$, respectively, and
$H \equiv \dot a/a = \sqrt{\rho/3}/M_{\rm Pl}$ is the Hubble expansion rate. 
Here, $M_{\rm Pl} = (8\pi G)^{-\frac{1}{2}} \simeq 2.44\times 10^{18}$ GeV 
is the reduced Planck mass and $G$ is Newton's gravitational constant.
Moreover, $Q_{\alpha}$ in Eq.~(\ref{fluid}) is a the source term, which describes
the energy conversion between the species $\alpha$ accounted in the
theory. The conservation of energy assures that $\sum_{\alpha}
Q_{\alpha}=0$. Therefore, the conversion of the inflaton energy
density into radiation energy density in the WI scenario is, hence,
described by the following set of equations~\cite{BasteroGil:2009ec}
\begin{align}
\dot \rho_{\phi} + 3 H (\rho_{\phi} + p_{\phi}) &= - \Upsilon
(\rho_{\phi} + p_{\phi}) \label{a3}, \\ \dot \rho_{r} + 3 H (\rho_{r}
+ p_{r}) &=  \Upsilon (\rho_{\phi} + p_{\phi}) \label{a4},
\end{align}
where $\Upsilon$ is the dissipation coefficient, which can generally
be a function of the inflaton field $\phi$ and temperature $T$ and
whose functional form depends on how WI is being described in terms of
the microscopic physics~\cite{Bastero-Gil:2012akf,Bartrum:2013fia,
  Bastero-Gil:2016qru, Berghaus:2019whh, Bastero-Gil:2019gao}.
Considering the energy density and pressure for a standard canonical
inflaton field, i.e., $\rho_{\phi} = \dot\phi^2/2  +  V(\phi)$ and
$p_{\phi} =  \dot\phi^2/2 - V(\phi)$, with $p_{r} =  \rho_{r}/3$,
Eqs.~(\ref{a3}) and (\ref{a4}) reduce to
\begin{align}
\ddot\phi + (3H+ \Upsilon)\dot\phi + V_{\phi} &=
0 \label{inflaton-equation},\\  \dot \rho_{r} + 4H\rho_{r} = \Upsilon
\dot\phi^2 \label{radiation-equation}, 
\end{align}
where $V_{\phi}$ is the derivative of the inflaton potential with
respect to $\phi$. Although inflation happens when the energy density
is dominated by the inflaton field potential $V$, i.e., $\rho_{r},
\  \dot\phi^2/2\ll V$, such that the radiation energy density is
sub-dominant, even so the produced radiation energy density can still
satisfy $\rho^{1/4}_{r}> H$. Assuming thermalization, this condition
then translates into $T>H$, which is usually considered as a condition
for WI to happen. This condition is easy to understand. Since the
typical mass for the inflaton field during inflation is $m_{\phi}
\simeq H$, hence, when $T>H$, thermal fluctuations of the inflaton
field will become important. Looking at Eq.~(\ref{inflaton-equation}),
one can immediately see that dissipative particle production effects
manifest as an extra friction term in the equation of motion for the
inflaton. Therefore, radiation will not be necessarily redshifted
during inflation, because it can be continuously fed by the inflaton
through dissipation. As a consequence, this can result in a
sustainable quasi-stationary thermal bath during the inflationary
dynamics.  Such radiation production also results in entropy
production. The entropy density $s$ is related to the radiation energy
density by $T s = 4\rho_{r}/3$, i.e., it is related to the
temperature as $s = {2\pi^2 g_{*}}T^3/45$, where we have considered a
thermalized radiation bath as it is typically the case in the WI
scenario. Then, Eq.~(\ref{radiation-equation}) can be rewritten in
terms of the entropy density as follows~\cite{Moss:2008yb}
\begin{align}
T(\dot s + 3 H s) = \Upsilon \dot\phi^2.
\end{align}

With the inflaton's potential dominating, an inflationary phase sets
in. Thus, to solve Eqs.~(\ref{inflaton-equation})  and
(\ref{radiation-equation}), one can use the so-called slow-roll
approximation, which consists in dropping the leading derivative term
in each equation, i.e., $\ddot \phi \ll 3H\dot\phi, V_{\phi}$ and
$\dot\rho_{r}\ll 4 H\rho_{r}, \Upsilon \dot\phi^2$. Hence, the
slow-roll equations read as follows,
\begin{align}
3 H(1+Q) \dot\phi + V_{\phi} & \simeq 0 \label{c1}, \\ \rho_{r} &
\simeq \frac{3}{4}Q\dot\phi^2 \label{a5},
\end{align}
and $H\simeq \sqrt{V/3}/M_{\rm Pl}$. In Eqs.~(\ref{c1}) and
(\ref{a5}), we have introduced $Q$,
\begin{equation}
Q = \frac{\Upsilon}{3H},
\label{Qratio}
\end{equation}
which defines the dissipation ratio in WI and it measures the strength
of the dissipative particle production effects in comparison to the
spacetime expansion. There are two different regimes in WI that can be
defined depending on the value of the dissipation ratio $Q$. The {\it
  weak dissipative regime} is when $Q < 1$. In this regime, dissipation is not
expected to modify significantly the background dynamics. Hence, in this case
the background dynamics is similar as in CI. However, as we will see
in the Section~\ref{section4}, thermal fluctuations of the radiation
energy density can still strongly affect the field fluctuations, and
also the primordial spectrum of perturbations as long as $T>H$. The
other regime of WI is the {\it strong dissipative regime}, $Q >1$. In
this case, dissipation strongly modifies both the background dynamics
and the primordial fluctuations. Note that $Q$ is not necessarily
constant. In fact, it can increase or decrease depending on the form
of dissipation coefficient and inflaton potential. Therefore, there is
also a possibility that a model can start in the weak dissipative
regime, but later on to transit into the strong dissipative regime, or
vice versa. Let us discuss in more details these different possibilities
that can appear in WI and how dissipation ultimately affects the
dynamics.

{}From the slow-roll equations, one can express the Hubble
slow-roll parameter $\epsilon_{H}$ in terms of the so-called potential
slow-roll parameter $\epsilon_{V}$ as follows,
\begin{align}
\epsilon_{H} \equiv - \frac{\dot H}{H^2} \simeq \frac{1}{2} M_{\rm
  Pl}^{-2}(1+Q) \frac{\dot\phi^2}{H^2} \simeq
\frac{\epsilon_{V}}{(1+Q)} \label{c2},
\end{align}
where $\epsilon_{V} = {M^2_{\rm Pl}} (V_{\phi}/V)^{2}/2$. To reach the
second equality in Eq.~(\ref{c2}), we have used the second {}Friedmann
equation, i.e., $-2 M^2_{\rm Pl}\dot H = \rho + P$, together with
Eq.~(\ref{a5}) and then used Eq.~(\ref{c1}) to reach the last
equality. One can realize from Eq.~(\ref{c2}) that inflation will end
when $\epsilon_{H} \simeq 1$ or, equivalently, when $\epsilon_{V}
\simeq 1+Q$. Moreover, looking at the last equality in Eq.~(\ref{c2}),
one can immediately see that the equality between the Hubble slow-roll
and the potential slow-roll parameters as observed in the CI,
$\epsilon_{H} \simeq \epsilon_{V}$, does not hold in the WI
scenario. In the CI scenario, the inflationary phase occurs when the
Hubble slow-roll parameter is smaller than unity, i.e., $\epsilon_{H}
\ll 1$, which means that the potential slow-roll parameter should also
be smaller than unity. However, in the WI scenario, the inflationary
phase occurs even when the potential slow-roll parameter is bigger
than one (or even much bigger than one), provided that the dissipation
ratio is large enough. This in particular alleviates the need for very
flat potentials, as far as the background dynamics is
concerned. Moreover, defining the number of e-folding as $N = \ln a$,
i.e., $d N = H dt = (H/\dot \phi) d\phi$, which measures the expansion
of the universe, one can see from the last two expressions in
Eq.~(\ref{c2}) that $d\phi/dN = M_{\rm Pl}\sqrt{2\epsilon_{V}}/(1+Q)$,
which means that the inflaton field excursion can be much smaller in the WI
scenario than in comparison to the CI scenario for the same variation of
$\epsilon_{H}$. In fact, the inflaton field excursion in WI can be sub-Planckian even
for very steep potentials. We will discuss later in section~\ref{section5}
how such novel background dynamics will allow WI to reside in the
landscape of the string theory.

When using the slow-roll approximation in Eqs.~(\ref{a3}) and
(\ref{a4}), one should carefully consider its consistency. 
This consistency check can be performed, for
instance, using a linear stability analysis to determine under which
conditions the system remains close to the slow-roll solution for many
Hubble times~\cite{Moss:2008yb}. Through this procedure, one finds
that the sufficient conditions for the slow-roll approximation to hold
are\footnote{See also Refs.~\cite{Bastero-Gil:2012vuu,delCampo:2010by} for the stability analysis in the presence of
  radiation viscous effects and also
  Refs.~\cite{Zhang:2013waa,Zhang:2014dja,Motaharfar:2017dxh,Cid:2015ota,Peng:2016yvb}
  for generalizations into non-canonical kinetic terms.}
\begin{align}
\epsilon_{V}, \eta_{V}, \beta_{V} \ll 1+Q,  \ \ \ \  \ \ \ 0<b\ll
\frac{Q}{1+Q}, \ \ \ \ \ \ \ \ \ |c|<4 \label{c4}, 
\end{align}
where we have defined the additional quantities $\eta_{V} = M^2_{\rm
  Pl} V_{\phi \phi}/V$, $\beta_{V} = M^2_{\rm Pl} V_{\phi}
\Upsilon_{\phi}/V\Upsilon$, $b = T V_{\phi T} /V_{\phi}$ and $c = T
\Upsilon_{T}/\Upsilon$. The condition on the parameter $b$ states that
the thermal correction to the inflaton potential should be small,
while the condition on the parameter $c$ reflects the fact that
radiation has to be produced at a rate larger than the red-shift due
to the expansion of the universe. Therefore, slow-roll WI occurs when
all conditions in Eq.~(\ref{c4}), together with $T>H$, are
satisfied. However, one may wonder that the condition for WI to happen
is in conflict with the conditions for the slow-roll
approximation. Using Eq.~(\ref{a5}) and given that $T>H$ during
inflation, one finds that 
 \begin{align}
 \frac{\dot\phi^2/2}{V(\phi)}> \frac{\pi^2 g_{*}}{135} Q^{-1}
 \frac{H^2}{M^2_{\rm Pl}}.
 \end{align}
Since $H\ll M_{\rm Pl}$ in most inflationary models, one can find
 that the WI scenario is consistent with the slow-roll approximation
 even for the weak dissipation regime, $Q<1$. Moreover, one shall
 further show that the radiation energy density will never exceed the
 potential energy in the slow-roll regime, thus guaranteeing a period
 of accelerated expansion. To this end, one can calculate the
 radiation energy density to potential energy density ratio as
 follows, 
  \begin{align}\label{rhoV}
 \frac{\rho_{r}}{V} \simeq \frac{1}{2}\frac{\epsilon_{V}}{1+Q}
 \frac{Q}{1+Q}.
 \end{align}
During inflation $\epsilon_{H}\ll 1$, meaning that $\rho_{r}\ll V$,
  even for a large dissipation ratio, while at the end of inflation
  $\epsilon_{H} \simeq 1$, i.e., $\epsilon_{V} \simeq 1+Q$, implying
  that $\rho_{r} \simeq V$, if the strong dissipation regime can be
  achieved.  Therefore, radiation will not be diluted and can even
  become dominant at the end of inflation. As a consequence, the
  universe can smoothly enter into the radiation dominated epoch
  without the need of a separate reheating phase as required in the CI
  scenario. Therefore, there is a possibility that even potentials
  without a minimum can also be embedded into a WI scenario without
  any difficulty. In other words, those inflationary potentials
  without minimum, which have attracted considerable attention due to
  recently proposed swampland conjectures inspired from string theory,
  usually result in an ever-lasting inflationary phase in the CI
  scenario and they require another mechanism for termination of
  inflation. However,  inflation will end due to dissipative particle
  production in the WI scenario even if the inflaton potential has no
  minimum. Hence, larger classes of inflationary potentials can be
  embedded into the WI scenario due to its richer dynamics in comparison
  to the CI case. 

Having discussed under which conditions a slow-roll WI dynamics can be
consistently achieved, one next consistency check is to investigate
under which conditions the inflationary phase can end in this
context. Looking at Eq.~(\ref{c2}), one can see that there are several
possibilities for graceful exit in the WI scenario. The end of WI
depends on the form of the potential, on the dissipation coefficient
and whether the regime of weak or strong dissipation has been
achieved. In other words, the inflationary phase can continue as long
as $\epsilon_{V}<1+Q$ and it will end when $\epsilon_{V} \simeq
1+Q$. Although in the CI scenario inflation ends when $\epsilon_{V}$
increases during inflation, in the WI scenario, there is a possibility
that even potentials with constant and decreasing $\epsilon_{V}$ have
graceful exit, since $Q$ is also a dynamical parameter. In fact,
depending on the evolution of $\epsilon_{V}$ and $Q$ there are
generally three possibilities for graceful exit in WI
scenario. {}First, if $\epsilon_{V}$ increases, inflation ends when
$Q$ decreases, remains constant or not increases faster than
$\epsilon_{V}$. In fact, potentials such as the monomial potentials,
hilltop potentials, natural inflation potential and the Starobinsky potential, which
have a graceful exit in CI scenario, also have graceful exit in WI
depending on the form of dissipation coefficient. Second, if
$\epsilon_{V}$ is constant, in the exponential type of potentials,
inflation ends only when $Q$ is a decreasing function. Third, if
$\epsilon_{V}\gg1$ and it is a decreasing function, inflation ends
when $Q$ decreases much faster than $\epsilon_{V}$ and cross
$\epsilon_{V}$ before it become less than unity. Although the last
possibility exists in the WI scenario, it is very challenging from a
model building point of view and these models usually do not end as in
the case of the CI scenario (see Ref.~\cite{Das:2020lut} for more
details). All of the aforementioned possibilities can be summarized in
terms of the Hubble slow-roll parameter in such a way that inflation
ends when $\epsilon_{H}$ increases with the number of
e-folding. Therefore, taking the derivative with respect to the number
of e-fold from Eq.~(\ref{c2}), one obtains the following
inequality~\cite{Das:2020lut}
\begin{align}\label{b1}
\frac{d \ln \epsilon_{V}}{d\ln N} > \frac{Q}{1+Q} \frac{d\ln Q}{dN}.
\end{align}
As long as the inequality (\ref{b1}) is satisfied, inflation will
end. To understand under which conditions WI goes through graceful
exit, one needs to find the evolution of $\epsilon_{V}$ and $Q$ during
inflation. To this end, we need to fix both the dissipation
coefficient and the potential function. Several different forms of
dissipation coefficients were derived from first principles in quantum
field theory during the development of WI scenario. In the early
development of WI, an inverse temperature dependent dissipation
coefficient, i.e., $\Upsilon \sim \phi^2/T$ was derived. But soon
after that it was realized that such model suffers from large thermal
corrections affecting the inflaton potential~\cite{Gleiser:1993ea,
  Berera:1998gx}. To overcome such difficulty, a two-stage mechanism
was proposed (see, e.g., Ref.~\cite{Berera:2008ar}) in which
the thermal bath can be produced and sustained without introducing large
thermal corrections to the inflaton potential. In this case, the dissipation
coefficient has a cubic temperature dependence, i.e., $\Upsilon \sim
T^3/\phi^2$. Later~\cite{Bastero-Gil:2016qru}, a model with
a dissipation coefficient with linear temperature dependence, i.e.,
$\Upsilon \sim T$ was also constructed and also a variant
model~\cite{Bastero-Gil:2019gao} with an inverse temperature
dependence in the high temperature regime to obtain $\Upsilon \sim T^{-1}
$. More recently~\cite{Berghaus:2019whh}, a dissipation coefficient
with a simple cubic temperature dependence without field dependence,
i.e., $\Upsilon \sim T^3$ was also realized.  The models originating
these forms of dissipation coefficients and others will be discussed
in more details in Sec.~\ref{section3}.  Almost all of the aforementioned
dissipation coefficients can generally be parameterized as follows
\begin{align}
\Upsilon(\phi, T) = C_{\Upsilon} T^{c} \phi^p M^{1-p-c},\label{g1}
\end{align}
where $C_{\Upsilon}$ is a dimensionless constant that carries the
details of the microscopic model used to derive the dissipation
coefficient, like the different coupling constants of the model (see,
e.g., Sec.~\ref{section3}), $M$ is some mass scale in the model and
depends on its construction, while $c$ and $p$ are numerical powers,
which can be either positive or negative powers (note that the
dimensionality of the dissipation coefficient in Eq.~(\ref{g1}) is
[$\Upsilon$] = [energy]). Given this general form for the dissipation
coefficient, one can find the dynamical evolution of the relevant
parameters of the system, $\epsilon_{V}$, $Q$, $T/H$ and $T$ in terms
of the potential slow-roll parameters as follows 
 \begin{align}
 \frac{d \ln \epsilon_{V}}{d N} &= \frac{4 \epsilon_{V} - 2
   \eta_{V}}{1+Q},\label{b2}
\\ 
\frac{d \ln Q}{d N} & = C_{Q}^{-1}
 \left[(2c+ 4) \epsilon_{V} - 2c\eta_{V} - 4 p
   \kappa_{V}\right], \label{b3}
\\ 
\frac{d\ln (T/H)}{d N} &= C_{Q}^{-1}
 \left[\frac{7-c + (5+c)Q}{1+Q} \epsilon _{V} - 2\eta_{V} -
   \frac{1-Q}{1+Q} p \kappa_{V}\right], \label{b4}
\\ 
\frac{d \ln T}{d
   N} &=   C_{Q}^{-1} \left(\frac{3+Q}{1+ Q} \epsilon_{V} - 2 \eta_{V}
 - \frac{1-Q}{1 +Q} p \kappa_{V}\right), \label{b5}
 \end{align}
where $\kappa_{V} = M^2_{Pl} {V_{\phi}}/{(\phi V)}$ and $C_{Q} = 4-c + (4+c) Q $  
is a positive quantity, since Q is always positive and $-4<c<4$ from stability 
conditions~\cite{Moss:2008yb,Bastero-Gil:2012vuu,delCampo:2010by}. 
 
\begin{center}
\begin{figure*}[!htb]
\subfigure[ \ 3d parameter space
  $(c,p,n)$]{\includegraphics[width=5.2cm]{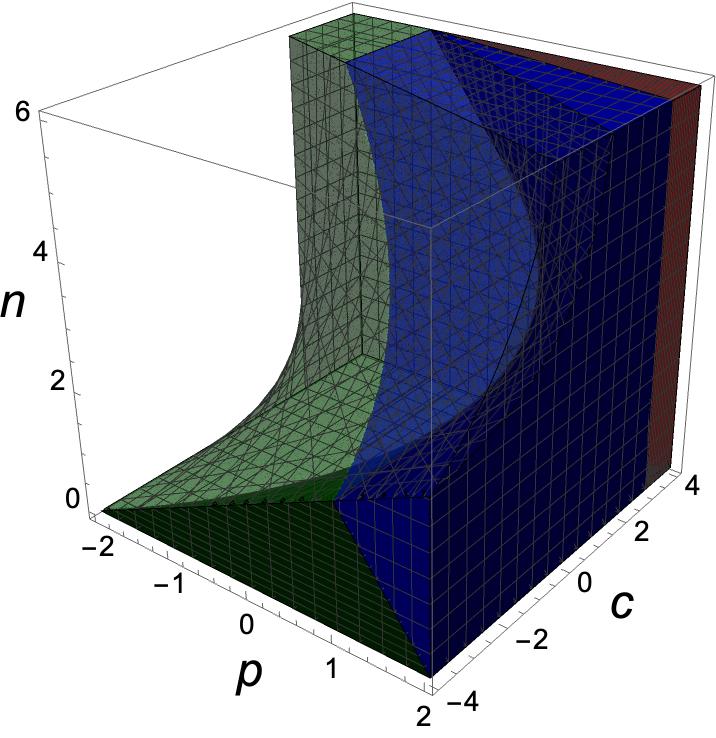}}
\ \ \ \ \ \ \  \subfigure[ \ Plane
  $p=0$]{\includegraphics[width=5.2cm]{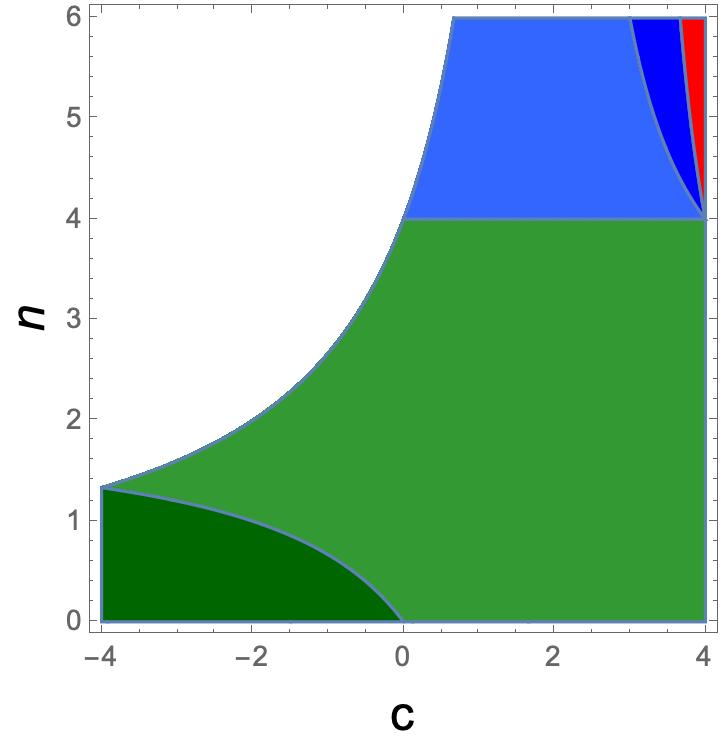}}
\ \ \ \ \ \ \  \subfigure[  \ Plane
  $p=2$]{\includegraphics[width=5.2cm]{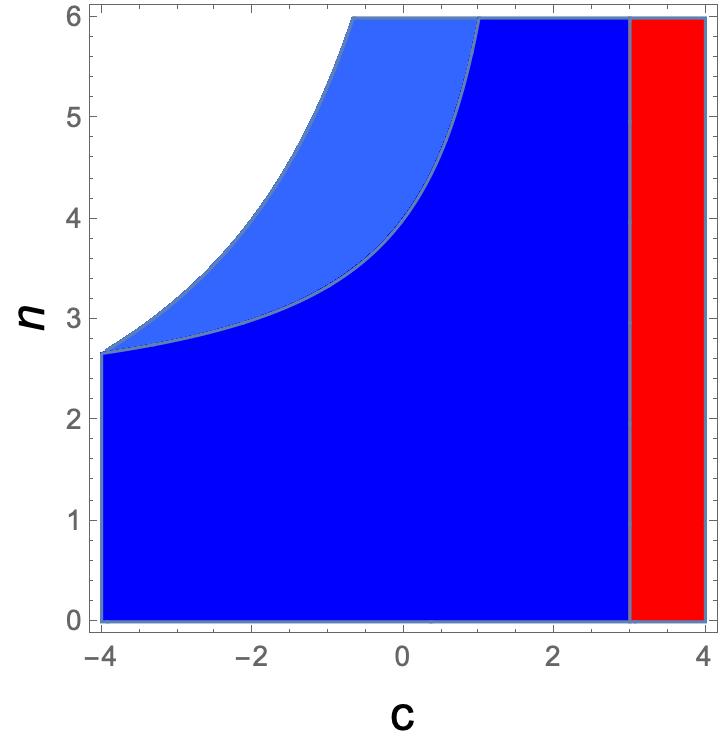}}  \\ \subfigure[  \ Plane
  $c=-1$]{\includegraphics[width=5.2cm]{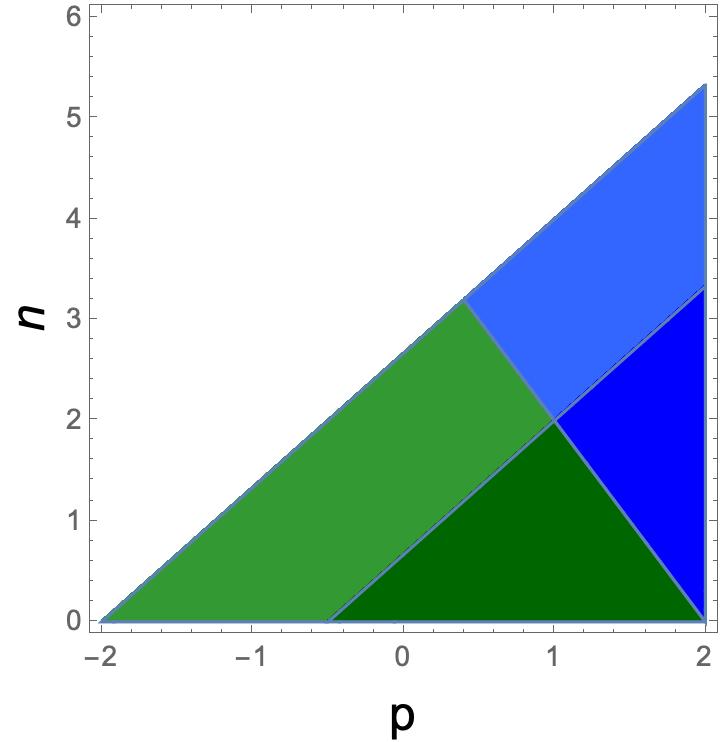}}
\ \ \ \ \ \ \  \subfigure[  \ Plane
  $c=1$]{\includegraphics[width=5.2cm]{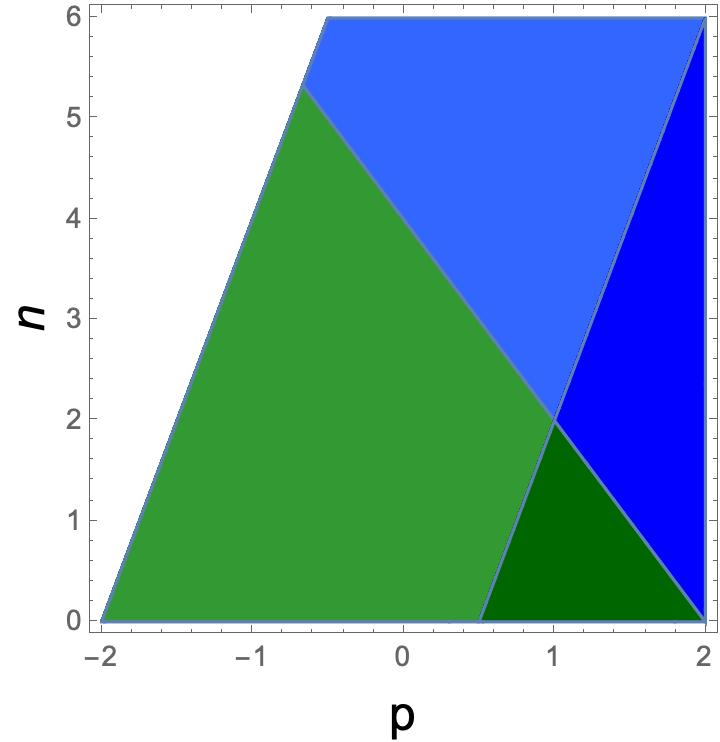}}\ \ \ \ \ \ \  \subfigure[
  \ Plane $c=3$]{\includegraphics[width=5.2cm]{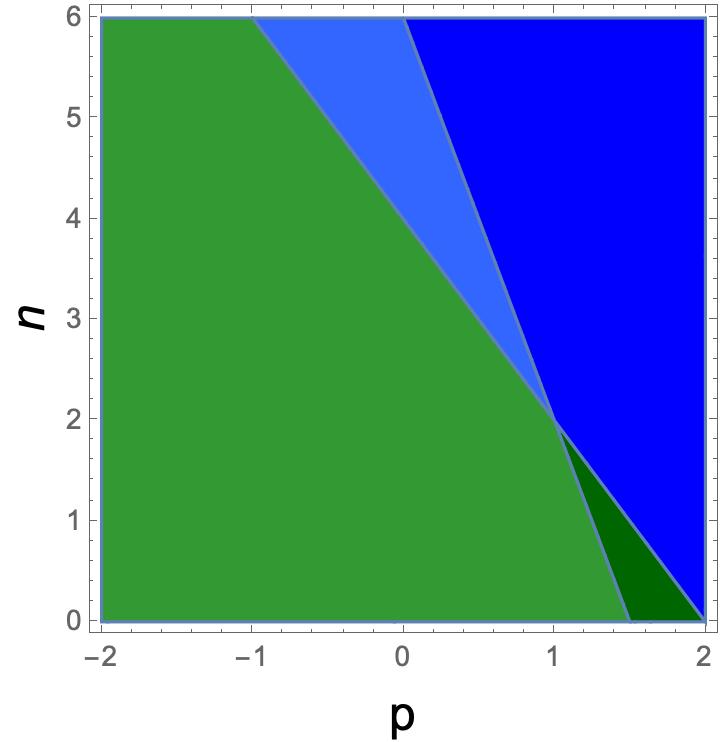}}
\caption{The parameter space of WI in the weak dissipative regime
  ($Q< 1$) and for the monomial class of inflaton potentials. The
  shaded areas indicate WI has a graceful exit, while empty (white)
  space indicates where there is no graceful exit. The shaded areas
  are classified based on the behavior of $Q$, $T/H$ and $T $ during
  inflation when all three increase (light green), $Q$ decreases and
  both $T/H$ and  $T$ increase (dark green), $T$ decreases and both
  $Q$ and $T/H$ increases (light blue), $T/H$ increases and both $Q $
  and $T$ decreases (dark blue), and all three decrease (dark
  red). {}Figure based on Ref.~\cite{Das:2020lut}.}\label{f1}
\end{figure*}
\end{center}

    \begin{center}
\begin{figure*}[!htb]
\subfigure[ \ 3d parameter space
  $(c,p,n)$]{\includegraphics[width=5.2cm]{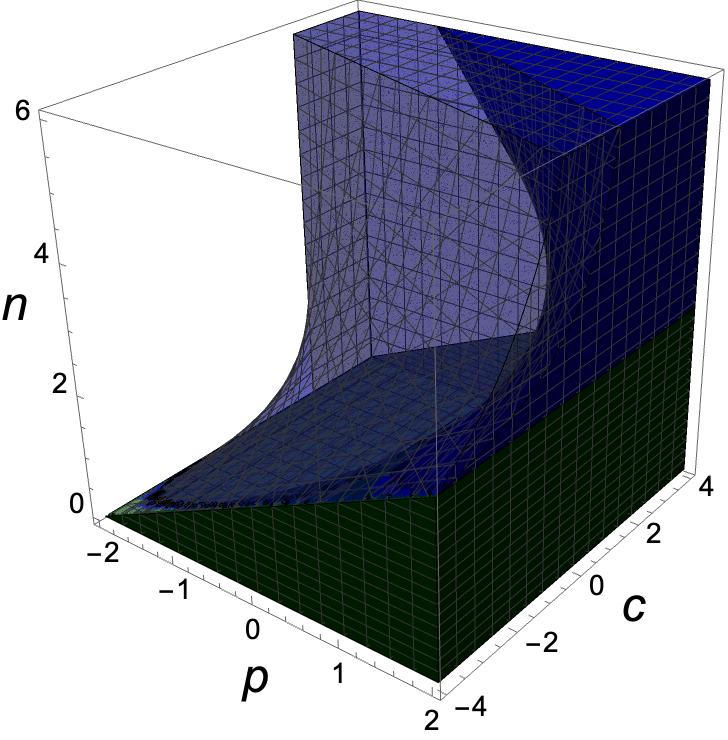}}
\ \ \ \ \ \ \  \subfigure[ \ Plane
  $p=0$]{\includegraphics[width=5.2cm]{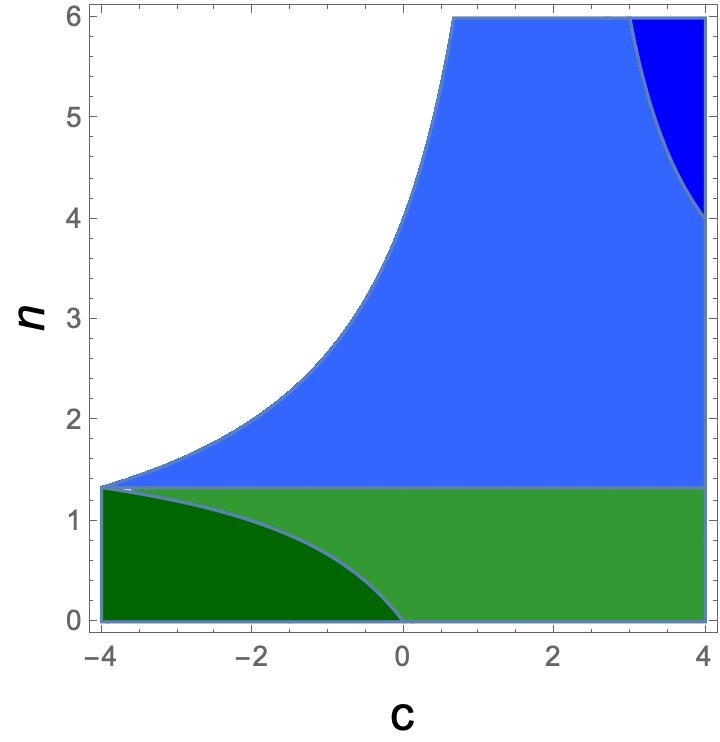}}
\ \ \ \ \ \ \  \subfigure[ \ Plane
  $p=2$]{\includegraphics[width=5.2cm]{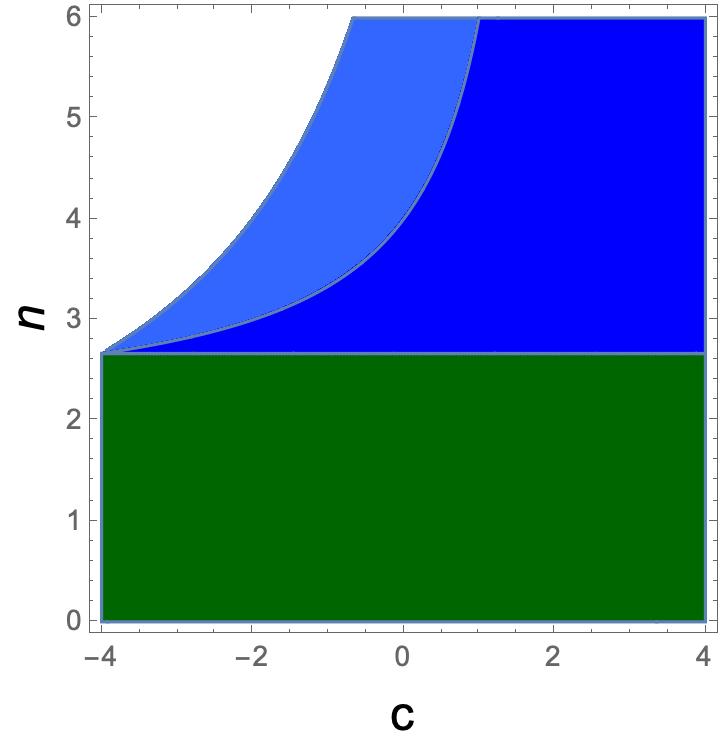}} \\ \subfigure[ \ Plane
  $c=-1$]{\includegraphics[width=5.2cm]{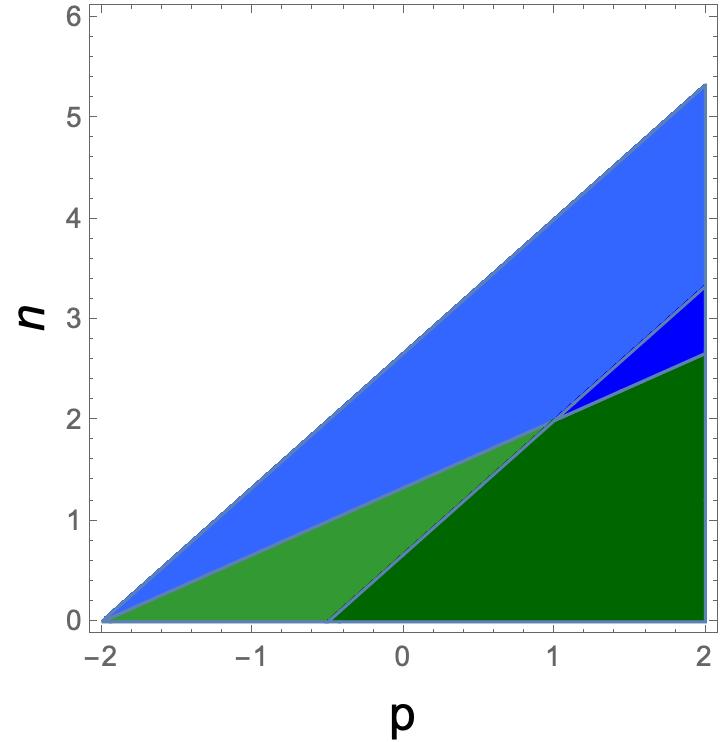}}
\ \ \ \ \ \ \  \subfigure[ \ Plane
  $c=1$]{\includegraphics[width=5.2cm]{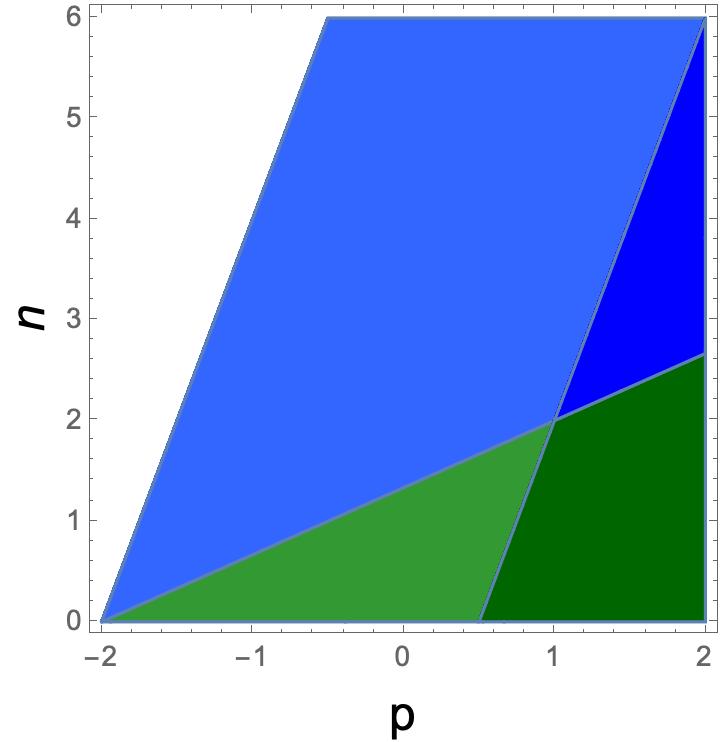}}
\ \ \ \ \ \ \  \subfigure[ \ Plane
  $c=3$]{\includegraphics[width=5.2cm]{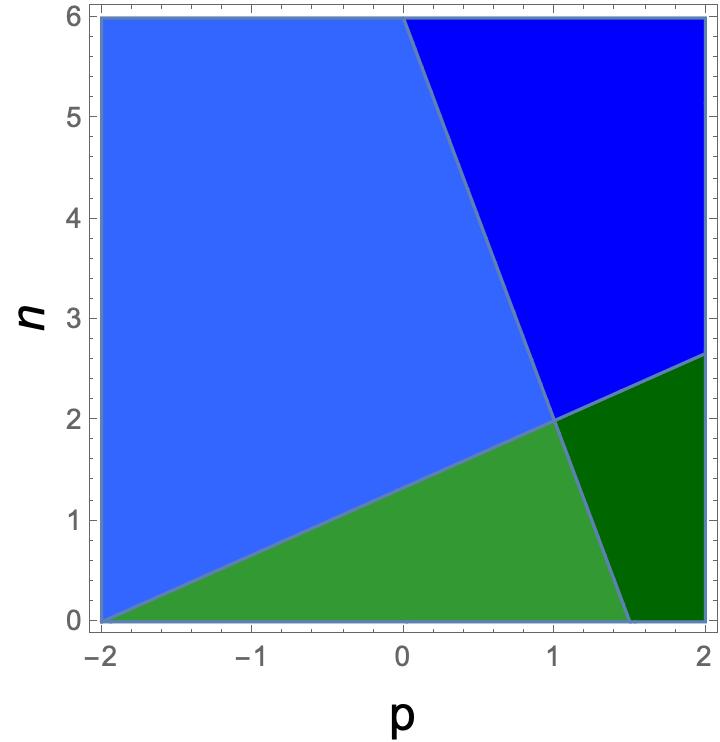}}
\caption{ The parameter space of WI in the strong dissipative regime
  ($Q > 1$) and for the monomial class of inflaton potentials. The
  shaded areas is classified based on the behavior of $Q$, $T/H$ and
  $T$ during inflation when all three increase (light green), $Q $
  decreases and both $T/H $and $T$ increase (dark green), $T $
  decreases and both $Q$ and $T/H$ increase (light blue), and $T/H$
  increases and both $Q$ and $T$ decrease (dark blue).  {}Figure based
  on Ref.~\cite{Das:2020lut}.  }\label{f2}
\end{figure*}
\end{center}
 \begin{figure}[!htb]
\centerline{\includegraphics[scale= 0.6]{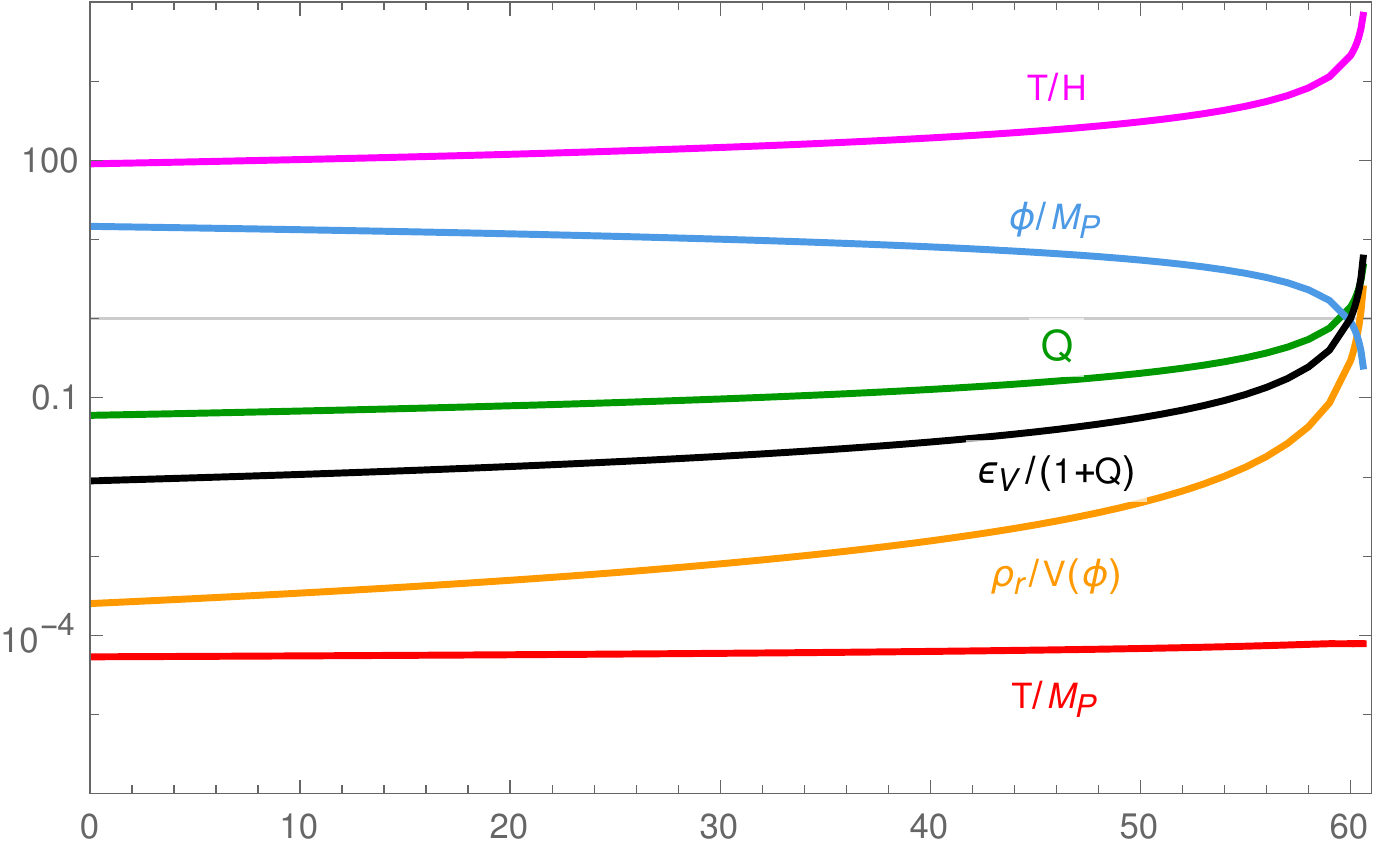}}
\caption{Evolution of the different quantities in the WI scenario for the case of
$V(\phi) =
  \frac{1}{2} m^2 \phi^2$, $\Upsilon = C_{T} T$,  with $m = 10^{-7} M_{\rm
    Pl}$ and $C_{T}=0.002$.}\label{r1}
\end{figure}

Given that,  in {}Fig.~\ref{f1} we show the space of parameters leading to different
scenarios in WI when considering monomial potentials,   
 \begin{align}
 V(\phi) = \frac{V_{0}}{n} \left(\frac{\phi}{M_{\rm Pl}}\right)^{n},
 \end{align}
and for different parametric forms for the dissipation coefficient
 in the weak dissipative regime.
 Using Eqs.~(\ref{b2}) and (\ref{b3}) and the inequality (\ref{b1}),
 we find for which values in the space of parameters $(n, c, p)$, WI
 has graceful exit. One should note that the conditions for WI to have
 graceful exit is independent of being in the weak or in the strong
 dissipation regime as it is clear from Eq.~(\ref{b3}). Moreover, we
 also used Eqs.~(\ref{b4}) and (\ref{b5}) to see how $T/H$ and $T$
 evolve in the region that WI has graceful exit. 

In {}Fig.~\ref{f2},
 we also illustrate the space of parameters for WI in the strong
 dissipation regime to allow to see how the region for evolution for
 $T/H$ and $T$ change. As we have discussed previously, $Q$ is a
 dynamical parameter, hence WI may start in the weak dissipative
 regime and be able to end in the strong dissipative regime, or vice
 versa. Therefore, $T/H$ and $T$ should not necessarily monotonically
 increase or decrease during inflation. 

Comparing {}Figs.~\ref{f1} and
 \ref{f2}, one may easily observe that there are possibilities for
 which $T/H$ and $T$ first increase and then decrease, or vice versa
 depending on being in the weak or strong dissipation regime and $Q$
 increases/decreases. 

In {}Fig.~\ref{r1}, for illustration purposes,
 the dynamical parameters of WI for a  quadratic potential and linear
 temperature dependent dissipation coefficient are plotted versus the
 number of e-folding. One can obviously see that the dissipation ratio
 $Q$, the temperature $T$ and the ratio $T/H$ all increases during the
 inflationary phase, as it was expected from
 {}Fig.~\ref{f1}. Moreover, the radiation to potential energy density
 increases and become order unity shortly after the end of the
 inflation. That is because the condition $\epsilon_{V} \simeq 1+Q$ is
 a weaker condition than $\epsilon_{H} = 1$ to specify the end of the
 inflationary phase in WI scenario. In fact, the first condition
 points out that inflation ends when $\rho_{r} \simeq V/2$ while in
 reality one expects that inflation ends when the radiation energy
 density equals or suppresses the potential energy density as one can
 see in {}Fig.~\ref{r1}. In this sense, $\epsilon_{V} \simeq 1+Q$
 predicts the end of inflation slightly earlier.
In general, the weaker condition only underestimates the end of
 inflation by less than one e-folding. Thus, the condition $\epsilon_{V} = 1+Q$ 
is still good enough 
as a way of estimating the instant where WI ends for all practical purposes.  
{}Furthermore, the inflaton field starts from
 super-Planckian values and ends sub-Planckian, with an overall
 super-Planckian field excursion, which is a typical feature of large
 field inflationary potentials and, in this case, for WI in the weak dissipative
regime, as considered in the example shown in {}Fig.~\ref{r1}. Sub-Planckian
field excursions in WI is possible in the strong dissipative regime of WI,
as we will discuss later.

\begin{center}
\begin{figure*}[!htb]
\subfigure[ \ 3d parameter space $(c,p,n)$  in weak dissipative
  regime]{\includegraphics[width=5.2cm]{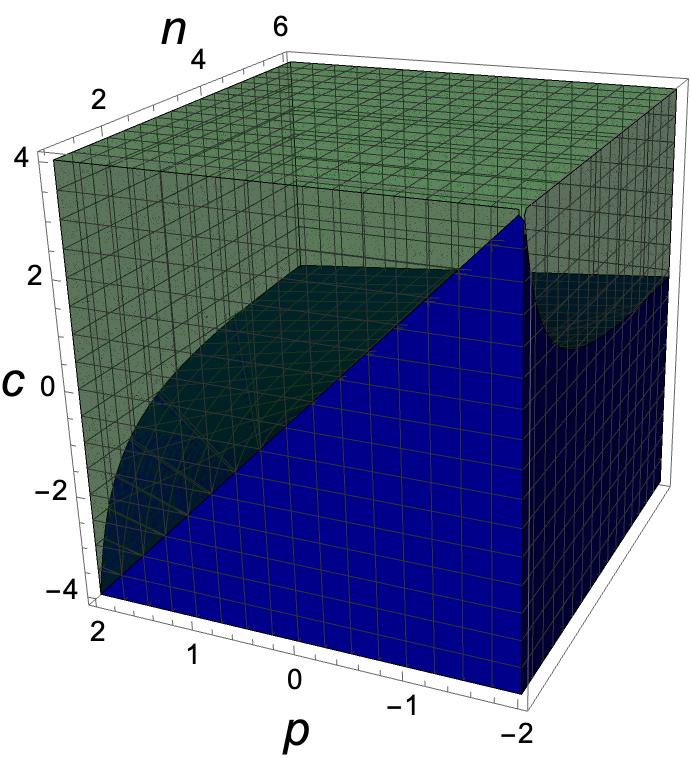}}
\ \ \ \ \ \ \ \ \ \ \ \ \ \ \ \ \ \ \ \ \ \ \  \subfigure[ \ 3d
  parameter space $(c,p,n)$ in strong dissipative regime
]{\includegraphics[width=5.2cm]{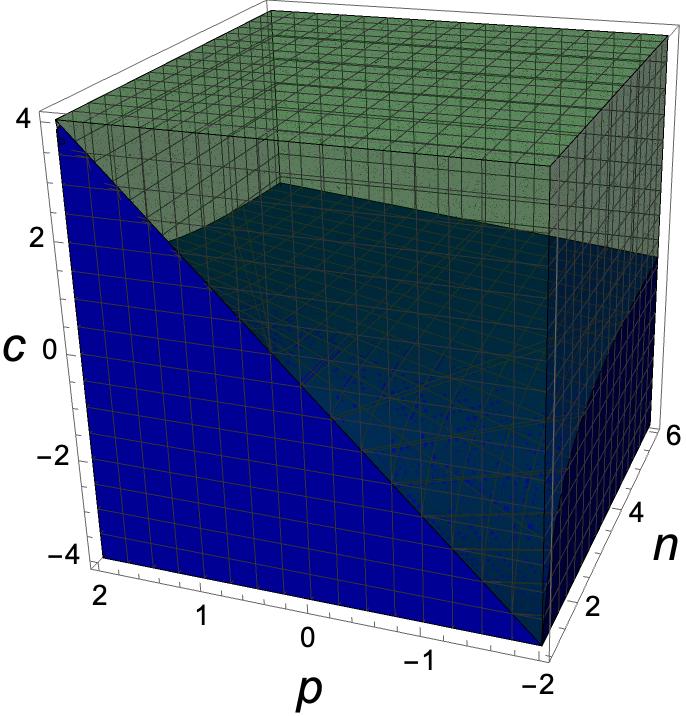}}
\caption{The parameter space in WI for the hiltop-like class of
  inflaton potentials. The shaded areas indicate WI has a graceful
  exit, while empty (white) space indicates where there is no graceful
  exit. The shaded areas is classified based on the behavior of $Q$,
  $T/H$ and $T$ during inflation when all three increase (dark green),
  and $Q$ decrease and both $T/H$ and $T$ increase (dark blue).
  {}Figure based on Ref.~\cite{Das:2020lut}.}
\label{f3}
    \label{fig:1}
\end{figure*}
\end{center}

As an additional example, in {}Fig.~\ref{f3}, we plotted the space of
parameters for another class of potential,  hiltop-like potentials,
given by 
\begin{align}
 V(\phi)= V_{0}[1- (\phi/\phi_{0})^{2n}]^2,
 \end{align}
with $n \ge 1$ and with the inflation taking place around the top
(plateau) of the potential, $|\phi| \ll \phi_{0}$. We also consider
that $\phi_{0}$ is sufficiently large such that inflation ends before
the inflection point of the potential. Thus, we are considering that
inflation takes place exactly in the concave part of the
potential. Notice that here all the parameter space allows for
graceful exit, and $T$ and $T/H$ always increase for all space of
parameters. {}Finally, it deserves to be noticed that exponential potentials
like
\begin{align}
 V(\phi) = V_{0}\exp({-\alpha {\phi}/{M_{\rm Pl}}}),
 \end{align}
can lead to a power law inflation only for $\alpha<\sqrt{2}$ in CI
scenario, while it does not have a graceful exit for $\alpha\ge
\sqrt{2}$ \cite{Copeland:1997et}. However, looking at Eq.(\ref{c2}),
one can easily observe that the exponential potential not only can
result in an accelerated expansion even for $\alpha>\sqrt{2}$, but
also it has a graceful exit as long as the dissipation ratio is
decreasing during inflation, i.e., when $c>2$, regardless of the value
of $p$ for $\alpha \phi/M_{\rm Pl}>1$. This case and other forms of
primordial inflaton potentials have been studied in details in Ref.~\cite{Das:2020lut}.

\section{Deriving dissipation coefficients in WI}\label{Section III}
\label{section3}

The dynamics in WI is intrinsically a result of particle production
processes able to happen during inflation. The generic idea is that as
the inflaton  evolves in time, moving around its potential, it might
excite any other fields that are coupled to it. This in turn can
produce relativistic particles and maintain a quasi-stationary
radiation bath throughout the inflationary regime. At the end of the
accelerated inflationary regime, the universe can then smoothly
transit to the radiation dominated regime.

Crucial to the idea of WI is then the role played by the dissipation
coefficient $\Upsilon$, e.g., in the inflaton effective evolution
equation, Eq.~(\ref{inflaton-equation}).  As in any inflation model,
we expect the inflaton  to be necessarily coupled to other fields,
which similar to the case of CI as required for reheating, the energy
density in the inflaton will eventually be transferred to
radiation. The emergence of dissipative processes in this case is
reminiscent of the so-called Caldeira and Leggett type of
models~\cite{Caldeira:1982iu}.  There is a relevant degree of freedom,
in which we are interested in the dynamics and that here is the
background inflaton field.  The system (the inflaton background) is in
turn coupled to other degrees of freedom, i.e., any other fields
coupled to the inflaton, which are regarded as environmental degrees
of freedom. Below, we sketch the general idea for completeness, but
which can also be found in details in many other previous papers, in
particular in the
Refs.~\cite{Gleiser:1993ea,Berera:1998gx,Berera:2001gs,Berera:2004kc,Berera:2008ar}.
We can describe the inflaton field and environment fields through a
generic Lagrangian density of the form,
\begin{equation}
{\cal L}[\Phi,X,Y] = {\cal L}[\Phi] + {\cal L}[X] + {\cal L}[Y] +
{\cal L}_{\rm int}[\Phi,X] + {\cal L}_{\rm int}[X,Y],
\label{LPhiXY}
\end{equation}
where here $\Phi$ is the inflaton field, or more specifically its zero
mode, represented by the homogeneous background value, while $X$ is
describing any field or degrees of freedom  coupled directly to the
inflaton field (like fermion fields or other bosons that can be
scalars or vector fields), while $Y$ can be any other fields not
necessarily coupled to the inflaton, like additional bosons and/or
fermions,  but that can be coupled to $X$. The different interactions
are described by the terms  in Eq.~(\ref{LPhiXY}), ${\cal L}_{\rm int}[\Phi, X]$, for the interactions
of the inflaton with the $X$ fields,  and ${\cal L}_{\rm  int}[X,Y]$,
for the interactions between $X$ and $Y$, but not directly with the inflaton field.  

The evolution of the inflaton field is then determined from
Eq.~(\ref{LPhiXY})  by integrating over the environment fields $X$ and
$Y$. This can be done for example in the context of the in-in
Schwinger closed-time path functional formalism (for a textbook
account, see, for example, Ref.~\cite{Bellac:2011kqa})  or,
equivalently, through the influence functional
formalism~\cite{Calzetta:2008iqa}.  The formal expression for the
evolution for a background field value $\Phi(x)$ turns  out to be in
the form of a stochastic equation of motion of the
form~\cite{Gleiser:1993ea,Berera:1998gx,Berera:2008ar}
\begin{eqnarray}
\partial^2 \Phi + V_{\rm eff}'(\Phi) + \int d^4 x' {\cal D}^{(X)}
(x,x')  \dot{\Phi}_c(x')= \xi(x),
\label{langevin2}
\end{eqnarray}
where $V_{\rm eff}'(\Phi)= d V_{\rm eff}(\Phi)/d\Phi$, with $V_{\rm
  eff}(\Phi)$ being the effective potential for $\Phi$. The term with
${\cal D}^{(X)} (x,x')$, defined as
\begin{equation}
\Sigma^{(X)}_\rho(x,x') {\rm sgn}(t-t') = - \frac{\partial}{\partial
  t'} {\cal D}^{(X)} (x,x'),
\label{diss kernel}
\end{equation}
describes the dissipative effects due to the interaction of the
inflaton with the environment fields, while $\xi(x)$ represents 
stochastic noise fluctuations, which affect the dynamics of $\Phi_c$
and describe  Gaussian processes with the general properties of
having zero mean, $\langle \xi(x) \rangle =0$, and two-point
correlation function given by
\begin{equation}
\langle \xi(x) \xi(x') \rangle = \Sigma_F^{(X)}[\Phi_c] (x,x').
\label{noisetwopoint}
\end{equation}
In Eqs.~(\ref{diss kernel}) and (\ref{noisetwopoint}),
$\Sigma^{(X)}_\rho$ and  $\Sigma_F^{(X)}$ are the self-energy terms in
the Schwinger-Keldshy real time formalism of quantum field
theory~\cite{Calzetta:2008iqa}. {}Furthermore, both noise and
dissipation terms  are related to each other through a generalized
fluctuation dissipation relation~\cite{Berera:2008ar}, as expected for
a  Langevin-like evolution describing a stochastic process under both
dissipation and noise terms.

If the field $\Phi$ is slowly varying on the response timescale
$1/\Gamma$, with the $\Gamma$ derived from  the self-energies terms
coming from the $X$ and $Y$, then $\dot \Phi/\Phi \ll \Gamma$. Under
these conditions, which are typically referred to as the adiabatic
approximation, then a simple Taylor expansion of the non-local terms
in Eq.~(\ref{langevin2}) can be performed.  {}Furthermore, if $\Gamma
> H$, the produced radiation through the dissipative processes can be
thermalized sufficiently fast. Under the further adiabatic condition
that $\Gamma  \gg \dot{T}/T$, then it is expected that the produced
radiation can be maintained in a close to thermal equilibrium state at
a temperature $T$.  These conditions provide a clear separation of
timescales in the system, which allows to approximate
Eq.~(\ref{langevin2}) in the form of a local, Markovian equation of
motion~\cite{Berera:2007qm}, with  a local dissipation coefficient
$\Upsilon$ defined  by
\begin{equation}
\Upsilon= \int d^4 x' \, \Sigma_R[\Phi](x,x') \, (t'-t),
\label{Upsilon}
\end{equation}
where $\Sigma_{R}^{(X)} (x,x') = \Sigma^{(X)}_\rho(x,x') \theta(t-t')$
is a retarded self-energy term.  Overall, the Eq.~(\ref{langevin2}),
when working in a {}Friedmann-Lema\^{\i}tre-Robertson-Walker (FLRW)
background metric, then becomes of the form
\begin{equation}
\left[ \frac{\partial^2}{\partial t^2}  + (3H + \Upsilon)
  \frac{\partial}{\partial t} - \frac{1}{a^2}\nabla^2 \right] \Phi+
\frac{\partial V_{\rm eff}(\Phi)}{\partial \Phi} = \xi_T,
\label{eomWI}
\end{equation}
where $\xi_T$ describes thermal (Gaussian and white) noise
fluctuations in the local approximation as defined above and it is
connected to the dissipation coefficient through a Markovian
fluctuation-dissipation relation, 
\begin{equation}
\langle \xi_T ({\bf x},t) \xi_T({\bf x}',t') \rangle =  2 \Upsilon T
a^{-3} \delta({\bf x}- {\bf x}') \delta(t-t'),
\label{flucdiss}
\end{equation} 
where the average here is to be interpreted as been taken over the
statistical ensemble. Stochastic terms can also be ascribed for
quantum contributions and will be important when deriving the perturbations
equations. This will be discussed in the next section, Sec.~\ref{section4}.

The effective evolution equation for the inflaton,
Eq.~(\ref{inflaton-equation}), then follows from the localized form
for the effective equation of motion, Eq.~(\ref{eomWI}) by separating
the background field into its homogeneous term $\phi(t)$. On the other
hand, the non-homogeneous fluctuations over  $\phi(t)$, denoted as
$\delta\phi({\bf x},t)$, will describe the fluctuations of the
inflaton and to be considered in the density perturbation equations
(see, next section). The latter are explicitly dependent on the
stochastic noise term.  {}From the above definitions, once a
particular interaction Lagrangian density is given, the corresponding
dissipation term $\Upsilon$ can be computed explicitly. Various
examples of interactions and the resulting dissipation terms have been
derived and given, e.g., in Ref.~\cite{Bastero-Gil:2010dgy}.

Below, we will summarize some of the most important microscopic
particle physics constructions considered for WI.  As already
explained before, the particle physics implementation for WI seeks
models for which a significant dissipation can be generated during
inflation and at the same time the radiative and thermal corrections
for the inflaton potential are kept under control, such that the
flatness of the inflaton potential is not spoiled. The first particle
physics models constructions in WI achieved this using quantum field
theory models in supersymmetry (SUSY). This is the case of the first
two models described below.

\subsection{The distributed mass model}

In the distributed mass model (DMM)~\cite{Berera:1998px}, there are a
set of scalar $\chi_j$ and fermionic $\psi_j$ fields coupled to the
inflaton through a series of couplings of the form $g^2 (\phi-M_j)^2
\chi_j^2$ and $h (\phi - M_j) \bar \psi_j \psi_j$, respectively. The
idea is that as the inflaton is evolving, eventually the inflaton
satisfies $\phi \sim M_j$. At this point,  the masses of the fields
coupled to it can become very light.  In special, as these masses
become smaller than the temperature, those fields will get thermally
excited and the inflaton will be able to decay into those fields. A
sequence of mass distributions $M_j$ can then be constructed in such a
way that as the background inflaton field evolves, it is able to
dissipate energy into these fields throughout the inflationary regime.
This process can then be described by a dissipative term in the
inflaton evolution equation~\cite{Berera:1998gx}.  The idea is
reminiscent of string theory model constructions~\cite{Berera:1999wt}.
In this case, the inflaton is interpreted as an excited string zero
mode. This string mode can then be interacting with  massive string
levels. As the string levels can be highly degenerate, a distribution
of mass states can emerge as a type of fine structure splitting of
those levels. The DMM realization for WI was also revised more
recently in Ref.~\cite{Bastero-Gil:2018yen}.  Both radiative and
thermal corrections coming from the large set of bosonic and fermionic
fields coupled to the inflaton can be well controlled by constructing
the DMM in the context of SUSY.  Despite the inflaton background
field and thermal effects both break supersymmetry, there are still
large cancellations between fermions and bosons
contributions~\cite{Hall:2004zr}. The DMM can be implemented by a
superpotential in the context of SUSY, given by
\begin{equation}\label{superpotential0}
W=\sum_{j}\left[\frac{g}{2}(\Phi-M_{j})
  X_{j}^{2}+\frac{h}{2}X_{j}Y_{j}^{2} \right],
\end{equation}  
where $\Phi$, $X_{j}$ and $Y_{j}$ chiral superfields, with (complex)
scalar and fermion components $\varphi$ and $\psi_{\varphi}$ for
$\Phi$, $\chi_{j}$ and $\psi_{\chi_{j}}$ for $X_j$, and $\sigma_{j}$
and $\psi_{\sigma_{j}}$ for $Y_j$. The inflaton is associated with
the real component of $\varphi$.  The $g$ and $h$ are coupling
constants and the sum is taken over an arbitrary distribution of
supermultiplets $X_j$ and $Y_j$.  {}From Eq.~(\ref{superpotential0}),
we can derive the scalar $\mathcal{L}_{S}$ and fermionic
$\mathcal{L}_{F}$ Lagrangian density interaction terms, which are
defined, respectively, as~\cite{Hall:2004zr}
\begin{equation}\label{lagrangian-LS-definintion}
-\mathcal{L}_{S}=|\partial_{\Phi}W|^{2}+\sum_{j}|\partial_{X_{j}}W|^{2}+\sum_{j}|\partial_{Y_{j}}W|^{2},
\end{equation} 
and
\begin{equation}\label{lagrangian-LF-definition}
-\mathcal{L}_{F}=\frac{1}{2}\sum_{n,m}\frac{\partial^{2}W}{\partial\xi_{n}\partial\xi_{m}}\bar{\psi}_{n}P_{L}\psi_{m}+\frac{1}{2}\sum_{n,m}\frac{\partial^{2}W^{\dagger}}{\partial\xi^{\dagger}_{n}\partial\xi^{\dagger}_{m}}\bar{\psi}_{n}P_{R}\psi_{m},
\end{equation}
where $\xi_{n}$ is a superfield: $\Phi,X_{j}, Y_{j}$ and
$P_{L}=1-P_{R}=(1+\gamma_{5})/2$ are the chiral projection operators
acting on Majorana 4-spinors. The expressions
Eqs.~(\ref{lagrangian-LS-definintion}) and
(\ref{lagrangian-LF-definition}), then lead to the explicit Lagrangian
density relevant contributions that are given by~\cite{Bastero-Gil:2018yen}
\begin{eqnarray}
-\mathcal{L}_{S} &=&
g^{2}\sum_{j}\left(\phi-M_{j}\right)^{2}|\chi_{j}|^{2}+\frac{gh}{2}\sum_{j}\left(\phi-M_{j}\right)\left[\chi_{j}(\sigma_{j}^{\dagger})^{2}+\chi_{j}^{\dagger}\sigma_{j}^{2}\right]
\nonumber\\ && + h^{2}\sum_{j}|\chi_{j}|^{2}|\sigma_{j}|^{2}+
\frac{g^{2}}{4}\sum_{j}|\chi_{j}|^{4} +
\frac{h^{2}}{4}\sum_{j}|\sigma_{j}|^{4} ,
\label{lagrangians-fullS}
\end{eqnarray}
and
\begin{eqnarray}
-\mathcal{L}_{F} &=&
\frac{g}{2}\sum_{j}\left(\phi-M_{j}\right)\bar{\psi}_{\chi_{j}}P_{L}\psi_{\chi_{j}}
+
\frac{g}{2}\sum_{j}\left(\phi-M_{j}\right)\bar{\psi}_{\chi_{j}}P_{R}\psi_{\chi_{j}}
\nonumber\\ && +
\frac{h}{2}\sum_{j}\chi_{j}\bar{\psi}_{\sigma_{j}}P_{L}\psi_{\sigma_{j}}
+
\frac{h}{2}\sum_{j}\chi^{\dagger}_{j}\bar{\psi}_{\sigma_{j}}P_{R}\psi_{\sigma_{j}}
+ h\sum_{j}\sigma_{j}\bar{\psi}_{\sigma_{j}}P_{L}\psi_{\chi_{j}} +
h\sum_{j}\sigma^{\dagger}_{j}\bar{\psi}_{\sigma_{j}}P_{R}\psi_{\chi_{j}}.
\label{lagrangians-fullF} 
\end{eqnarray}

{}From the interactions terms in Eqs.~(\ref{lagrangians-fullS}) and
(\ref{lagrangians-fullF}), we can determine the dissipation
coefficient  $\Upsilon$, which will receive contributions both from
the scalar bosons~\cite{Berera:1998gx,Berera:1998px} (see also
Ref.~\cite{Bastero-Gil:2018yen} for details)
\begin{equation}\label{Upsilon_DMS}
  \Upsilon^{S}(\phi,T)= \sum_{j=1}^{N_{th}}\frac{32 g^{4}}{\pi
    h^{2}\left(
    16\frac{m_{\chi_{j}}^{2}}{\tilde{m}_{\chi_{j}}^{2}}+\frac{\tilde{m}_{\chi
        j}^{2}}{T^{2}}\right)
  }\ln\left(\frac{2T}{\tilde{m}_{\chi_{j}}}\right)\frac{\left(\phi-M_{j}\right)^{2}}{\tilde{m}_{\chi_{j}}},
\end{equation}
and from the fermions~\cite{Bastero-Gil:2016qru},
\begin{equation}\label{Upsilon_DMF}
  \Upsilon^{F}(T)=\sum_{j=1}^{N_{th}}C_{T}^{F}T \,, \quad C_{T}^{F}
  \simeq \frac{3g^{2}}{h^{2}[1-0.34\ln(h)]} \,,
\end{equation}
where in the above equations, the sum run over the number $N_{th}$ of
thermally excited field modes and the masses $m_{\chi_j}$ and
$\tilde{m}_{\chi_j}$ are defined, respectively, as
$m_{\chi_{j}}=g\left(\phi-M_{j}\right)$ and $\tilde{m}_{\chi_{j}}^2=
m_{\chi_{j}}^2 + g^2T^2/12+h^2T^2/8$. One notes that the dissipation
coefficient in the DMM depends on the mass distribution $M_j$.
Assuming~\cite{Bastero-Gil:2018yen} $M_j = \phi + j \Delta
M(\phi,T,m,g)$, where $\Delta M$ denotes the mass gap in the tower of
states, different functional forms for the dissipation coefficient in
the DMM can be generated through different choices of $ \Delta
M(\phi,T,m,g)$. This motivates the parametric choice Eq.~(\ref{g1})
adopted in many works in WI.

\subsection{The two-stage mechanism model}

Without a way of controlling the thermal corrections of the radiation
fields that are directly coupled to the inflaton, it is expected that
the finite temperature of the radiation bath will induce large thermal
corrections  to the inflaton mass, leading to $m_\phi \sim T > H$. If
this occurs, successful realizations of WI in the simplest models
are jeopardized. In this case,
the additional friction caused by the dissipation effects through
$\Upsilon$ cannot overcome the increase in the  inflaton’s mass. In the
DMM discussed above, even though there are couplings of the
inflaton directly to the radiation fields, it is able to evade this
problem by a judicious choice of the mass distribution function.  But
in other model realizations this is not a simple task to achieve, as
discussed originally in Refs.~\cite{Berera:1998gx,Yokoyama:1998ju}.  
However, we also
recall that fields that are directly coupled to the inflaton will tend
to acquire large masses during inflation due to the large background
field value $\phi$.  This then suggests that WI can be better
implemented in  scenarios where the inflaton does not couple directly
to the radiation  fields, but instead to heavy intermediate fields,
which can be either bosons or fermions, with masses such that $m_\chi,
\, m_\psi > T$. This naturally leads to thermal corrections to the
inflaton potential that are Boltzmann suppressed.  {}Furthermore, once
again, we can use SUSY to control potentially large radiative
corrections to the inflaton.  In turn, the heavy fields can be made
coupled to the radiation fields, which remain decoupled from the
inflaton sector.  Once again, as the inflaton dynamics change the
masses of the heavy fields, these can decay into the light radiation
fields. These processes provide a way through which the inflaton can dissipate its
energy into radiation. A significant dissipation can be generated
depending on the field multiplicities. This is the two-stage decay
mechanism model for WI.

The two-stage mechanism for WI can be implemented through a
supersymmetric model with chiral superfields $\Phi$, $X$ and $Y_i$,
$i=1,\ldots,N_Y$, and described by the superpotential
\begin{equation} \label{superpotential}
W={g\over2}\Phi X^2+{h_i\over2} XY_i^2+f(\Phi)~,
\end{equation}  
where a sum over the index $i$ is implicit.  The scalar component of
the superfield $\Phi$ describes the inflaton field, with an
expectation value $\phi/\sqrt{2}$, which we assume to be real, and the
generic holomorphic function $f(\Phi)$ describes the self-interactions
in the inflaton sector.  The superpotential Eq.~(\ref{superpotential})
leads to the Lagrangian density describing the interactions between
the inflaton field $\phi$ with the other scalars and fermions as given
by~\cite{Bastero-Gil:2012akf}
\begin{eqnarray} \label{scalar_lagrangian}
\mathcal{L}_{scalar}&=&V(\phi)+{1\over2}g^2\phi^2|\chi|^2+{g\over2}\sqrt{V(\phi)}\left(\chi^2+\chi^{\dagger
  2}\right)+{g^2\over4}|\chi|^4+\nonumber\\ &+&{h_i\over2}{g\phi\over
  \sqrt{2}}\left(\chi\sigma_i^{\dagger
  2}+\chi^\dagger\sigma_i^2\right)+{h_ih_j\over4}\sigma_i^2\sigma_j^{\dagger
  2}+h_i^2|\chi|^2|\sigma_i|^2~,
\end{eqnarray}  
and 
\begin{eqnarray} \label{fermion_lagrangian}
\mathcal{L}_{fermion}&=&{g\phi\over\sqrt2}\bar\psi_\chi
P_L\psi_\chi+h_i\chi\bar\psi_{\sigma_i}P_L\psi_{\sigma_i}+{h_i\over
  2}\sigma_i\bar\psi_{\sigma_i}P_L\psi_\chi+\mathrm{h.c.}~,
\end{eqnarray}  
where the scalar components of the superfields $X$ and $Y_i$ were
denoted by $\chi$ and $\sigma_i$, respectively,  the fermionic
components by $\psi_\chi$ and $\psi_{\sigma_i}$, respectively ,
$V(\phi)=|f'(\phi)|^2$ is the potential driving inflation and
$P_L=(1-\gamma_5)/2$ is the left-handed chiral projector. 

The leading dissipation coefficient obtained for this model and using
the quantum field theory expression Eq.~(\ref{Upsilon}), can be
explicitly derived and it is given by~\cite{Bastero-Gil:2010dgy}:
\begin{equation} \label{dissipation_general}
\Upsilon={4\over T}\left({g^2\over 2}\right)^2\varphi^2\int {d^4p
  \over (2\pi)^4}\rho_\chi^2n_B(1+n_B)~, 
\end{equation}
where $n_B(p_0)=[e^{p_0/T}-1]^{-1}$ is the Bose-Einstein distribution
and $\rho_\chi$ is the spectral function for the $\chi$ field,
\begin{equation} \label{spectral_function}
\rho_\chi(p_0,p)={4\omega_p\Gamma_\chi\over
  (p_0^2-\omega_p^2)^2+4\omega_p^2\Gamma_\chi^2}~, 
\end{equation}
with $\Gamma_\chi$ denoting the decay width for the heavy fields
$\chi$, which includes contributions from both the bosonic and
fermionic final states in the $Y_i$ multiplets,
$\omega_p=\sqrt{\tilde{m}_\chi^2+p^2}$ for modes of 3-momentum
$|\mathbf{p}|=p$ and energy $p_0$, while the thermal mass for the
$\chi$ fields is $\tilde{m}_{\chi}^{2}=m_\chi^2+ h^2N_Y T^2/8$ (where
here we are assuming all couplings $h_i=h$) and $m_\chi= g
\phi/\sqrt{2}$.  {}For larger values of the mass and effective
coupling, the main contribution to the dissipation coefficient comes
from {\it virtual} $\chi$ modes with low momentum and energy,
$p,p_0\ll m_\chi$, such that one can use the approximation
$(p_0^2-\omega_p^2)^2\simeq \tilde{m}_\chi^4$.  In the case where
these modes also have a narrow width and thermal mass corrections can
be neglected, $\Gamma_\chi\ll \tilde{m}_\chi\sim m_\chi$, then the
spectral function Eq.~(\ref{spectral_function}) can be simply
expressed as~\cite{Berera:2008ar,Bastero-Gil:2010dgy} $\rho_\chi\simeq
4\Gamma_\chi/m_\chi^3$. Under these circumstances, the dissipation
coefficient $\Upsilon$ describing the dissipation mediated by the
decay of virtual scalar modes $\chi\rightarrow {\rm light\, radiation
  \, fields}$, is given by~\cite{Bastero-Gil:2012akf}
\begin{equation} \label{dissipation_coefficient}
\Upsilon=C_\phi {T^3\over \phi^2}~,\qquad C_\phi\simeq
          {1\over4}\alpha_h N_X~,
\end{equation}
for $\alpha_h=h^2N_Y/4\pi\lesssim1$ and $N_{X,Y}$ are the chiral
multiplets. 

To obtain sufficient dissipation in these models, it is usually
required large values for the numbers of multiplets $N_{X,Y}$. While
these can be seen as a possible drawback for this model, there are
well motivated scenarios where this can be naturally achieved, like in
Ref.~\cite{Bastero-Gil:2011zxb} using brane constructions, or like in
Ref.~\cite{Matsuda:2012kc}, where large field multiplicities can be
allowed due to a Kaluza-Klein tower in extra-dimensional scenarios.

Instead of relying on SUSY as a way to suppress radiative corrections,
we can try to arrange for special interactions of the inflaton with
the environment fields such as to be able to control the thermal
corrections to the inflaton potential.  This can also be used to avoid
the large multiplicities that would otherwise be required to produce
sufficiently large dissipation, such that WI can be realized. To avoid
these complications,  we can make use of other simpler models
exploring  well motivated symmetry properties for the
interactions. This is the case for the next three particle physics
model constructions for WI to be discussed below. 

\subsection{The warm little inflaton model}

In the warm little inflaton (WLI) model, first introduced in
Ref.~\cite{Bastero-Gil:2016qru}, the inflaton is assumed to be a
pseudo-Nambu-Goldstone boson (PNGB) of a gauge symmetry that is
collectively broken. The
idea is reminiscent of the  ``Little Higgs" models of electroweak
symmetry
breaking~\cite{ArkaniHamed:2001nc,Kaplan:2003aj,ArkaniHamed:2003mz},
where the Higgs boson is also considered to be a PNGB coming from a collectively
broken global symmetry.  A feature
displayed by the type of construction in these type of models is that the PNGB
has a mass that is naturally protected against large radiative
corrections (for a review, see, e.g., Ref.~\cite{Schmaltz:2005ky}). 

The construction of the WLI model uses a very minimum set of
ingredients.  There are two complex scalar fields, $\phi_1$ and
$\phi_2$, which share the same $U(1)$ symmetry. The scalar potential
for them allow both fields to have a nonzero vacuum expectation value,
$\langle \phi_1\rangle= \equiv M_1/\sqrt{2}$ and $\langle
\phi_2\rangle \equiv M_2/\sqrt{2}$, where $M_1$ can be taken to be
equal to $M_2$ without loss of generality, $M_1=M_2 \equiv M$.  With
both fields sharing the same Abelian charge, after the Higgs
mechanism, only one the phases of the complex scalar fields (the
Nambu-Goldstone boson), or a linear combination of them, is absorbed
as the longitudinal component of the $U(1)$ gauge boson, providing
mass to it.  The other phase (or linear independent combination of the
phases) remains, however, as a physical degree of freedom, becoming,
e.g., a singlet.  The two complex scalar fields, in the broken phase,
can then be expressed in the form
\begin{eqnarray}
\phi_1 = {M\over\sqrt{2}} e^{i\phi/M}~,\qquad \phi_2 =
    {M\over\sqrt{2}} e^{-i\phi/M} .
\label{phi12}
\end{eqnarray}

The radial modes in Eq.~(\ref{phi12}) decouple when $M \gtrsim T$ and
the singlet $\phi$, the PNBG, is taken to be the inflaton.  One notes that
as such, we can assume for $\phi$ an arbitrary scalar potential that
can be sufficiently flat to sustain inflation.  Dissipation in the
model comes from the coupling of $\phi_1$ and $\phi_2$ to other
fields. {}For example, they can be coupled to left-handed fermions
$\psi_{1}$ and $\psi_{2}$ with the same $U(1)$ charge of the scalar
fields. The right-handed components of these fermion fields,
$\psi_{1R}$ and $\psi_{2R}$, can be taken to be gauge singlets, in a
construction similar as in the Glashow-Weinberg-Salam of the standard
model of particle physics. A decay width for these fermions, which
will contribute to the dissipation coefficient, can be generated by
coupling them to additional Yukawa interactions, involving a scalar
singlet $\sigma$ and chiral fermions $\psi_{\sigma R}$, carrying the
same charge of $\psi_{1}$ and $\psi_{2}$, and $\psi_{\sigma L}$, with
zero charge.  {}For all these fields, we impose an interchange
symmetry for the complex scalars $\phi_1$ and $\phi_2$,
$\phi_1\leftrightarrow i\phi_2$, and also for the fermions $\psi_{1}$
and $\psi_{2}$, $\psi_{1L,R}\leftrightarrow \psi_{2L,R}$.  The overall
interaction Lagrangian density is~\cite{Bastero-Gil:2016qru}
\begin{equation}
{\cal L}_{\rm int} = \mathcal{L}_{\phi \psi} +
\mathcal{L}_{\psi\sigma},
\end{equation}
with 
\begin{eqnarray} \label{Yukawa_inflaton}
\mathcal{L}_{\phi \psi} &=& -{g\over
  \sqrt{2}}(\phi_1+\phi_2)\bar\psi_{1L} \psi_{1R} + i{g\over
  \sqrt{2}}(\phi_1-\phi_2)\bar\psi_{2L} \psi_{2R},
\end{eqnarray} 
and
\begin{eqnarray}
\mathcal{L}_{\psi\sigma}= -h\sigma \sum_{i=1,2}\left(
\bar{\psi}_{iL}\psi_{\sigma R}+ \bar{\psi}_{\sigma
  L}\psi_{iR}\right),
\end{eqnarray} 
which respects the symmetries imposed on the model. Due to the
symmetries, and using Eq.~(\ref{phi12}),  the masses for the fermions
$\psi_1$ and $\psi_2$, $m_1$ and $m_2$, respectively, are then given
by
\begin{equation}
m_{1} =  gM\cos(\phi/M), \;\;\;   m_{2} =  gM\sin(\phi/M),
\label{masses}
\end{equation}
and, hence, they remain always bounded, $m_{1,2} \leq g M$, and they
can be arranged to remain light during inflation if $gM\lesssim T
\lesssim M$.  The advantage of the interactions in
Eq.~(\ref{Yukawa_inflaton}) is that they lead to no quadratic
corrections for $\phi$ in the inflaton potential, thus, the inflaton
mass does not receive neither radiative nor thermal corrections. The
inflaton potential is only affected by subleading Coleman-Weinberg
corrections, as it was demonstrated in
Ref.~\cite{Bastero-Gil:2016qru}.

With the above interactions, the dissipation coefficient for the WLI
model becomes~\cite{Bastero-Gil:2016qru}
\begin{eqnarray}
\Upsilon = C_T T~, \qquad C_T\simeq \alpha (h) g^2/h^2~,
\label{UpsilonWLI}
\end{eqnarray}
where $\alpha(h)\simeq 3/[ 1-0.34\ln (h)]$.

The model has also been shown to produce a well-motivated dark matter
candidate~\cite{Rosa:2018iff}.  This can be possible because 
soon after inflation, the fermions
coupled to the inflaton field are no longer light. The dissipation
coefficient Eq.~(\ref{UpsilonWLI}) becomes Boltzmann suppressed. {}For
an inflaton potential augmented by quadratic term with a mass term
$m_\phi$,   then, when $H < m_\phi$, the inflaton  field is
underdamped and it will oscillate around the minimum of its
potential. The equation of state oscillates around $w_\phi = 0$, and
the energy density scales as $\rho_\phi \propto 1/a^3$.  This is the
same behavior as ordinary matter. Due to the symmetries of the model,
the inflaton rarely decays and an abundance of inflaton energy density
can remain in the coherent oscillating state. This makes the inflaton
in the WLI model to be a possible valid dark matter candidate.

\subsection{The warm little inflaton model - scalar version}

A variant~\cite{Bastero-Gil:2019gao} of the WLI model discussed above
is when the complex scalar fields $\phi_1$ and $\phi_2$ are now
coupled directly to other two complex scalar fields $\chi_1$ and
$\chi_2$, instead of fermions.  As in the previous model, it is also
imposed the discrete interchange symmetry $\phi_1\leftrightarrow
i\phi_2$, $\chi_1\leftrightarrow \chi_2$. The complex scalar fields
$\chi_{1,2}$ can also have a Yukawa interaction to fermions,
self-interact and also interact between each other. Then, the
interacting Lagrangian density can now be expressed
as~\cite{Bastero-Gil:2019gao}
\begin{equation}
{\cal L}_{\rm int} = \mathcal{L}_{\phi \chi} + \mathcal{L}_{\chi\psi},
\end{equation}
with 
\begin{equation}
  \mathcal{L}_{\phi\chi}=-{1\over2}g^2|\phi_1+\phi_2|^2|\chi_1|^2-
          {1\over2}g^2|\phi_1-\phi_2|^2|\chi_2|^2,
\label{Lphichi}          
\end{equation}
and
\begin{equation}
\mathcal{L}_{\mathrm{\chi\psi}}\!=\!\!\!\!\!\sum_{\substack{i\neq
    j=1,2}}\!\!\!\left(h\chi_i\bar\psi_L\psi_R\!+\!\mathrm{h.c.}
\!-\!{\lambda\over
  2}|\chi_i|^4\!-\!{\lambda'}|\chi_i|^2|\chi_j|^2\!\right)\!.
\label{Lchipsi}
\end{equation}
One notes that under the parametrization given by Eq.~(\ref{phi12}), the
masses for the scalar fields $\chi_1$ and $\chi_2$ are still of the
same form as in Eq.~(\ref{masses}).  Hence, they are still bounded and
can be light with respect to the temperature just the same way as in
the previous model.  In the high temperature regime, $m_{1,2} \ll T$,
the inflaton potential receives leading order thermal corrections of
the form $m_1^2(\phi) T^2/12 +  m_1^2(\phi) T^2/12 = g^2 M^2 T^2/12$
and, thus, the inflaton mass here also does not receive any thermal
corrections, while also not receiving any important radiative
contributions.

In this scalar field variant of the WLI model, the dissipation
coefficient now becomes~\cite{Bastero-Gil:2019gao}
\begin{equation} \label{dissipscalar}
\Upsilon\simeq {4 g^4\over h^2}{M^2 T^2\over
  \tilde{m}_\chi^3}\left[1+{1\over\sqrt{2\pi}}\left({\tilde{m}_\chi\over
    T}\right)^{3/2}\right]e^{-\tilde{m}_\chi/T},
\end{equation}
where we have taken the average of the oscillatory terms for field
excursions $\Delta\phi\gg M$ and $\tilde{m}_\chi$ is the thermal mass
for the $\chi_{1,2}$ fields, which, under the same average over the
oscillatory terms, is the same for both scalars and given by
$\tilde{m}_\chi^2 \simeq g^2M^2/2+\alpha^2T^2$, where $\alpha^2\simeq
\left[h^2+ 2 \lambda +\lambda' \right]/12$ when taking the
interactions in Eq.~(\ref{Lchipsi}).  It is noticed that when considering the
leading thermal contribution in  $\tilde{m}_\chi$, i.e.,
$\tilde{m}_\chi \sim \alpha T$, the dissipation coefficient
(\ref{dissipscalar}) varies with the temperature as
$\Upsilon(T)\propto T^{-1}$, realizing the case with $c=-1,p=0$ in
Eq.~(\ref{g1}).  On the other hand, as the temperature of the thermal
bath drops and the vacuum term $ gM/\sqrt{2}$ in  $\tilde{m}_\chi $
starts no longer to be negligible, it effectively would correspond to
values of $c > -1$, with a limiting case of $c=2$ when $gM/\sqrt{2}
\gg \alpha T$ and with an exponentially suppressed dissipation.  As
shown in Ref.~\cite{Bastero-Gil:2019gao}, this more complex behavior
for the dissipation  coefficient with the temperature of the
environmental radiation bath makes it possible to produce a large
dissipative regimes for WI, in which $Q \gg 1$ can be achieved in this
model, with perturbations  still consistent with the CMB
measurements. On the other hand, the dissipation coefficient in the
previous model, Eq.~(\ref{UpsilonWLI}), in general can only produce
perturbations consistent with the CMB observations in the weak
dissipative regime of WI, i.e., $Q\ll 1$ (see, e.g.,
Ref.~\cite{Benetti:2016jhf}). As this version of the WLI allows WI to be
realized in the strong dissipative regime, the cosmological phenomenology
that it allows is much richer and has quite appealing features, as we will
see in the next section, Sec.~\ref{section5}.

\subsection{The axion-like warm inflaton model}
\label{axionWI}

A Goldstone boson $\phi$ enjoys a shift symmetry, with only derivatives
of $\phi$ appearing in the action.  This symmetry can still be softly
broken such as to give $\phi$ an ultraviolet (UV) potential and the
Goldstone boson becoming a PNGB, like in axion-like models. But even
with the soft breaking of the shift symmetry, radiative and thermal
corrections to the axion potential are naturally suppressed by the
symmetry properties. This is how axions, for instance, can have very
small masses, but that are still protected from large quantum corrections (for a
general review on axions and their properties, see, e.g.,
Ref.~\cite{Marsh:2015xka}).  Thus, it is natural to try to construct
WI by taking the inflaton as an axion-like field. One such successful
construction was given in Ref.~\cite{Berghaus:2019whh}.  The relevant
interaction of the inflaton $\phi$ in this case is as in an axion
interacting with a Yang-Mills field $A_\mu^a$ and with Lagrangian
density given by
\begin{equation}
{\cal L}_{\rm int} = \frac{\alpha_g}{8 \pi} \frac{\phi}{f}
\tilde{F}^{a\,\mu\nu} F^a_{\mu \nu},
\label{axionL}
\end{equation}
where $\tilde{F}^{a\,\mu\nu}$ is the dual gauge field strength,
$\tilde{F}^{a\,\mu \nu} = \frac{1}{2}\epsilon^{\mu \nu \alpha \beta}
F^a_{\alpha\beta}$, $F_{\mu\nu}^a = \partial_\mu A_\nu^a -
\partial_\nu A_\mu^a  + g C^{abc} A_\mu^b A_\nu^c$, with $g$ the
Yang-Mills coupling and $C^{abc}$ is the structure constant of the
non-Abelian group. The coupling constant $\alpha_g$ is $\alpha_g
\equiv g^2/(4 \pi)$ with $f$ denoting a scale analogous to the axion
decay constant in axion-like models~\cite{Marsh:2015xka}.

The interaction term Eq.~(\ref{axionL}) leads to a dissipation
coefficient that has been shown to be  related to the Chern-Simons
diffusion rate~\cite{Moore:2010jd,Laine:2016hma} and given by
\begin{eqnarray}
\Upsilon= C_\Upsilon \frac{T^3}{f^2},\;\;\;  C_\Upsilon
=\kappa(\alpha_g, N_c, N_f) \alpha_g^5,
\label{upsilonaxion}
\end{eqnarray}
where $N_c$ is the dimension of the gauge group, $N_f$ is the
representation of the fermions if any, and $\kappa$ is a dimensional
quantity depending on $N_c$, $N_f$ and $\alpha_g$. 

Successful WI dynamics have been shown to be possible in this
model~\cite{Berghaus:2019whh,Goswami:2019ehb,Laine:2021ego,DeRocco:2021rzv}.
These studies have also shown the generality of WI in this type of
model. Even by starting with quantum initial conditions for
inflation, e.g., like in CI, the dissipative effects naturally drive
the production of a radiation bath and that can thermalize during
inflation in these axion-type of models. Hence, a WI dynamics
naturally emerges given appropriate parameters in the
model~\cite{Laine:2021ego,DeRocco:2021rzv}.  Besides of these
attractive features, like in the scalar field variant of the WLI
model, the dissipation coefficient Eq.~(\ref{upsilonaxion}) has the
appeal of leading to WI in the strong dissipative regime $Q \gg 1$,
yet, still being fully consistent  with the CMB perturbations for many
different potentials (for a recent exposition on this and a detailed
computation of relevant observable quantities, see, e.g.,
Ref.~\cite{Das:2022ubr}). This will be explicitly shown in the next
section.  Besides, it has a minimum setting of parameters and field
ingredients that are necessary to lead to WI. This makes this
construction specially attractive from the point of view of model
building in WI, as far as its simplicity is concerned.

\section{Cosmological Perturbations in WI}
\label{section4}

As we discussed previously in Sec.\ref{section2}, when $T>H$ thermal
fluctuations of the inflaton field will become dominant. As a
consequence of this, the source of density fluctuations in the WI
scenario is the thermal fluctuations in the radiation field, which are
then transferred to the inflaton field as adiabatic curvature
perturbations. This is much different than in the CI case,
where quantum fluctuations are responsible for generating the seeds
for large scale structure formation. We will discuss here how
dissipative effects change the dynamics of inhomogeneous fluctuations
of the inflaton field. To study the scalar perturbations for WI, we
start with the fully perturbed FLRW metric (here we are including only
scalar perturbations), which is given by
\begin{eqnarray}
ds^2 = -(1+2 \alpha) dt^2 - 2 a \partial_i \beta dx^i dt + a^2 \left[
  \delta_{ij} (1 +2 \varphi) + 2 \partial_i \partial_j \gamma\right]
dx^i dx^j, \label{metric}
\end{eqnarray}
where $\alpha$, $\beta$, $\gamma$, and $\varphi$ are the
spacetime-dependent  perturbed-order variables.  Moreover, one needs
to expand the inflaton field, the radiation energy density and also
the radiation pressure around their background values in the FRLW
metric, such that
\begin{align}
\phi({\bf x}, t) &= \bar\phi(t) + \delta\phi({\bf
  x},t),\\ \rho_{r}({\bf x}, t) &= \bar \rho_{r}(t) + \delta
\rho_{r}({\bf x},t),\\ p_{r}({\bf x}, t) &= \bar p_{r}(t) + \delta
p_{r}({\bf x},t),
\end{align}
where $\bar \phi(t)$, $\bar \rho_{r}(t)$ and $\bar p_{r}(t)$ are the
background values for the inflaton field, the radiation energy density
and the pressure, while $\delta \phi(x, t)$, $\delta\rho_{r}({\bf x},
t)$ and $\delta p_{r}({\bf x}, t)$ are, respectively, their
corresponding perturbations. Since the dissipation coefficient is
generally a function of $\phi$ and $T$, Eq.~(\ref{g1}), it must be
treated similarly, i.e., $\Upsilon({\bf x}, t) = \bar \Upsilon(t) +
\delta \Upsilon({\bf x},t)$. Hence, working in momentum space,
defining the {}Fourier transform with respect to the comoving
coordinates,  the equation of motion for the radiation and momentum
fluctuations with comoving wavenumber $k$ are found to be given
by~\cite{Bastero-Gil:2011rva}
\begin{eqnarray}
 &&\delta \dot{\rho_{r}}+3(1+\omega_{r})H\delta\rho_{r}
  =(1+\omega_r)\rho_{r} \left(\kappa-3H \alpha\right)   +
  \frac{k^{2}}{a^{2}}\Psi_{r}+\delta Q_{r}+Q_{r}\alpha \,, 
\label{energyalpha}
\\  &&\dot{\Psi}_{r}+3H\Psi_{r}= - \omega_r\delta \rho_{r}
-(1+\omega_r)\rho_{r}\alpha+J_{r} ,  
\label{momentumalpha} 
\end{eqnarray}
where $\Psi_{r}$ is the momentum perturbation, $\omega_{r}$ is the
equation of state for the radiation fluid, $ \chi = a ( \beta + a \dot
\gamma) $, $\kappa= 3 (H \alpha - \dot \varphi) + k^2 \chi/a^2 $, and
$J_r = - \Upsilon \dot \phi \delta \phi$ is the momentum
source. {}Furthermore, $Q_{r} = \Upsilon\dot\phi^2$ and its
perturbation $\delta Q_{r}$ is given by 
\begin{eqnarray}
 &&\delta Q_r = \delta \Upsilon \dot \phi^2 + 2 \Upsilon \dot \phi
  \delta \dot \phi - 2 \alpha \Upsilon \dot \phi^2 \,\label{deltaQr} .
\end{eqnarray}

In addition to Eqs.~(\ref{energyalpha}) and (\ref{momentumalpha}), one
also has the evolution equation for the inflaton field fluctuations
$\delta\phi$. Assuming the universe remains near thermal equilibrium
during WI, the fluctuations of the inflaton field obey a
fluctuation-dissipation relation~\cite{Gleiser:1993ea}. Therefore, the
evolution of the inflaton fluctuations is achieved by perturbing the
inflaton field equation, and adding stochastic quantum and thermal
white noise terms, following the fluctuation-dissipation theorem, as follows (for
further details, see also Ref.~\cite{Ramos:2013nsa})
\begin{eqnarray}
&&\delta \ddot \phi + 3 H  \delta \dot \phi + \left(\frac{k^2}{a^2} +
  V_{,\phi\phi}\right) \delta \phi =  \xi_q+\xi_T  - \delta \Upsilon
  \dot \phi   + \dot \phi ( \kappa + \dot \alpha) + (2 \ddot \phi + 3
  H \dot \phi) \alpha  -\Upsilon ( \delta \dot \phi - \alpha \dot
  \phi) \label{field}\,,
\end{eqnarray}
where $\xi_{q, T}$ are stochastic Gaussian sources related to the
quantum and thermal fluctuations with zero mean, $\langle \xi_T\rangle
= \langle \xi_q\rangle =0$, and with appropriate amplitudes, which are
defined by the two-point correlation functions:
\begin{eqnarray}
\!\!\!\!\!\!\!\!\!\!\langle \xi_T({\bf k},t)\xi_T({\bf
  k}',t')\rangle &=& \frac{2 \Upsilon T}{a^3} \delta(t-t') (2 \pi)^3
\delta({\bf k} + {\bf k}'), 
\label{noise1}
\\  \!\!\!\!\!\!\!\!\!\!\langle \xi_q({\bf k},t)\xi_q({\bf
  k}',t')\rangle &=& \frac{H^2(9+12\pi Q)^{1/2}(1+2n_*)}{\pi a^3}
\delta(t-t') (2 \pi)^3 \delta({\bf k} + {\bf
  k}'),
\label{noise2}
\end{eqnarray}
where $n$ denotes the inflaton statistical distribution due to the
 presence of the radiation bath. {}For a thermal equilibrium
 distribution\footnote{As a note concerning the emergence of a thermal equilibrium radiation bath
during WI, this problem has been studied in the context of the solution
of the Boltzmann equation in Ref.~\cite{Bastero-Gil:2017yzb}. Likewise,
the generality of formation of a thermalized radiation bath and that
it can be maintained during inflation, has as been demonstrated recently
in Refs.~\cite{Laine:2021ego,DeRocco:2021rzv} in the context of the model
described in Sec.~\ref{axionWI}.}, it assumes the Bose-Einstein distribution form, i.e.,
 $n = 1/[\exp(H/T)-1]$. 
To find the complete evolution of
 perturbations, one needs to specify the fluctuation of the
 dissipation coefficient. Considering the parametrization defined by
 Eq.~(\ref{g1}), $\delta \Upsilon$ can be written as  
\begin{equation}
\delta \Upsilon =  \Upsilon \left[c \frac{\delta T}{T}+p \frac{\delta
    \phi}{\phi} \right]. \label{dupsilon} 
\end{equation}
Although dissipation implies departures from thermal equilibrium in
the radiation fluid, the system has to be close-to-equilibrium for the
calculation of the dissipative coefficient to hold, therefore, $p_{r}
= \rho_{r}/3$, $\rho_{r} \propto T^4$  and $\delta T/T = \delta
\rho_{r}/(4\rho_{r})$. Hence, the $\delta Q_{r}$ term in
Eq.~(\ref{energyalpha}) can be explicitly written as 
\begin{equation}
 \delta Q_{r}= 3 H Q \dot{\phi}^2 \left(\frac{c \delta \rho_{r}}{4
   \rho_r} +\frac{p \delta \phi}{\phi} \right) + 6 H Q \dot{\phi}
 \delta \dot{\phi} - 6 H Q \dot{\phi}^2 \alpha\, .
\label{deltaQalpha}
\end{equation}
{}From the above relations, the complete system of first-order
perturbation equations for WI are given by
\begin{eqnarray}
\delta \ddot{\phi} &=& - 3H\left(1+Q\right)\delta \dot{\phi} -
\left(\frac{k^{2}}{a^{2}}+V_{,\phi\phi}+\frac{3p
  HQ\dot{\phi}}{\phi}\right)\delta \phi  \nonumber + \xi_q+\xi_T -
\frac{c H}{\dot{\phi}}\delta \rho_{r}+ \dot{\phi}(\kappa+\dot{\alpha})
+ [2\ddot{\phi}+3H(1+Q)\dot{\phi}]\alpha,  \nonumber \\
\label{deltaddotphi} 
\\ \delta \dot{\rho_{r}} &=& -H \left(4 - \frac{3cQ\dot{\phi}^2}{4
  \rho_r}  \right)\delta \rho_{r} + \frac{k^2}{a^2}\Psi_{r} +
6HQ\dot{\phi} \delta \dot{\phi}  \nonumber + \frac{3 p
  HQ\dot{\phi^{2}}}{\phi} \delta \phi + \frac{4  \rho_r}{3}  \kappa -
3 H\left( Q \dot{\phi}^2 + \frac{4\rho_r}{3}  \right) \alpha,
\nonumber \\
\label{deltadotrhor} 
\\ \dot{\Psi}_{r}  &=&- 3H \Psi_{r} - 3 H Q \dot{\phi} \delta \phi -
\frac{1}{3} \delta \rho_r   - 4 \rho_r \frac{\alpha}{3} . 
\label{dotpsir}
\end{eqnarray}
One can immediately realize from Eqs.~(\ref{deltaddotphi}) and
(\ref{deltadotrhor}) that the inflaton fluctuations $\delta\phi$ are
coupled to the radiation fluctuations $\delta \rho_{r}$ when $c\neq 0$. 
As first realized in
Ref.~\cite{Graham:2009bf}, such coupling results in a growing mode, if
$c>0$ or a decreasing mode for $c<0$ in the curvature power spectrum
as the dissipation ratio $Q$ increases. In other words, dissipation
will increase the temperature more in regions where it is already
higher than average making the power spectrum scale dependent and blue
(red) for $c>0$ ($c<0$). The effect of this coupling between inflaton 
and radiation perturbations in WI is typically modeled by a function
$G(Q)$, as we will show below.

The sets of perturbation equations
(\ref{deltaddotphi}), (\ref{deltadotrhor}) and (\ref{dotpsir}) are
gauge ready equations. Hereafter, one can rewrite these equations in
terms of gauge invariant quantities  (see
Ref.~\cite{Bastero-Gil:2011rva}) or use an appropriate gauge 
choice~\cite{Kodama:1984ziu,Hwang:1991aj}. 
Although any appropriate gauge can be chosen, here we
will make use of the zero-shear gauge, i.e., $\chi =0$. This in
particular have been shown to be more advantageous, as far as
numerical stability is concerned when solving the complete system of
equations. Thus, in the zero-shear gauge the relevant metric equations
become
\begin{eqnarray}
&&\kappa= \frac{3}{2 M_{\rm Pl}^2} ( \dot{\phi} \delta \phi - \Psi_r
  )\,,
\label{kappachi=0}\\
&& \alpha = -\varphi \,,
\label{alphachi=0}\\
&& \dot{\varphi} = - H \varphi -\frac{1}{3} \kappa\,.
\label{dotvarphichi=0}
\end{eqnarray}
Once all the relevant perturbation equations have been defined, the
power spectrum is determined from the definition of the  comoving
curvature perturbation~\cite{Baumann:2009ds}
\begin{equation}
\Delta_{\cal R}(k)= \frac{k^3}{2 \pi^2} \langle |{\cal R}|^2
\rangle\,, \label{PR}
\end{equation}
where $\langle \ldots \rangle$ means here the ensemble average over
different realization of the noise terms in Eq.~(\ref{deltaddotphi})
and which satisfy Eqs.~(\ref{noise1}) and (\ref{noise2}).  {}Finally,
the genral expression for $\mathcal{R}$ is composed of contributions
from the inflaton momentum perturbations and from the radiation
momentum perturbations~\cite{Bastero-Gil:2011rva}
\begin{eqnarray}
&&{\cal R}= \sum_{i=\phi,r} \frac{\bar \rho_i + {\bar p}_i}{\bar
    \rho+{\bar p}}{\cal R}_i\;,
\label{R}\\
&&{\cal R}_i = - \varphi - \frac{H}{\bar \rho_i + {\bar p}_i}
\Psi_i\,,
\label{Ri}
\end{eqnarray}
with ${\bar p}=\bar p_\phi + \bar p_r$, $\bar \rho_{\phi}+ \bar
p_{\phi} = \dot\phi^2$ and $\bar \rho_{r}+ \bar p_{r} = 4\bar
\rho_{r}/3 = Q\dot \phi^2$. As we have already mentioned, the inflaton
field fluctuations and the radiation field fluctuations are coupled
together in the case when $c\neq 0$. Therefore, to obtain the power spectrum
for the inflaton fluctuations, one typically needs to numerically
solve Eqs.~(\ref{deltaddotphi}), (\ref{deltadotrhor}) and
(\ref{dotpsir}) along with the appropriate set of metric perturbation
equations. However, an explicit analytic expression for the scalar of
curvature power spectrum can be obtained for dissipation coefficients
which are independent of temperature, i.e., when $c=0$. In this case,
the equations for the inflaton and the radiation fluctuations become
decoupled and one can obtain the curvature power spectrum using Green
function techniques (see Refs.~\cite{Ramos:2013nsa,Graham:2009bf} for
details). This leads to the explicit result for the curvature
perturbation,
\begin{eqnarray}
{\Delta_{\cal R}}\bigr|_{c=0}&=&
\frac{H_{*}^3T_{*}}{4\pi^2 \dot{\phi_{*}}^2}  \left[
  \frac{3Q_{*}}{2\sqrt{\pi}} 2^{2\alpha} \frac{\Gamma\left( \alpha
    \right)^2\Gamma\left( \nu -1\right)
    \Gamma\left(\alpha-\nu+3/2\right)}{\Gamma\left(\nu
    -\frac{1}{2}\right) \Gamma\left(\alpha+\nu-1/2\right)} \right.  +
  \left.  \frac{H_{*}}{T_{*}}\coth\left(\frac{H_{*}}{2T_{*}} \right)
  \right],
\label{Pphi}
\end{eqnarray}
where $\nu = 3(1+Q)/2$, $\alpha = \sqrt{\nu^2 + 3 \beta_V Q/(1+Q) - 3
  \eta_{V}}$, with $\beta_V$ and $\eta_V$ as already defined before,
e.g., below Eq.~(\ref{c4}). In Eq.~(\ref{Pphi}) we also have that
$\Gamma(x)$ is the Gamma-function, while the subindex $*$ means that
all quantities are to be evaluated at the Hubble crossing time, e.g.,
when $k_{*}=a_{*}H_{*}$. This specific point during inflation and
where the relevant modes crosses the Hubble radius can be defined  as
follows.  {}First we note that we can relate the mode with comoving
wavenumber $k_*$ that crossed the Hubble horizon, $a_* H_*=k_*$, with
the one at present time, $a_0 H_0$, as~\cite{Liddle:2003as}
\begin{equation}
\frac{k_*}{a_0 H_0} = \frac{a_*}{a_{\rm end}} \frac{a_{\rm
    end}}{a_{\rm reh}} \frac{a_{\rm reh}}{a_0}  \frac{H_*}{H_0},
\label{scales}
\end{equation}
where $a_*/a_{\rm end}= \exp(-N_*)$ and $N_*$ is the number of e-folds
lasting from the point where the modes left the Hubble radius  until
the end of inflation, where the scale factor is $a_{\rm end}$. Now, as
we have already seen in Sec.~\ref{section2}, WI ends when the
radiation energy density takes over and starts dominating. Since in
WI there is no need for a reheating phase after the end of inflation,
this removes a significant source of uncertainty that is present in CI
models, which is related the specific duration of the reheating phase and
which affects the predictions that (cold) inflation can make regarding
observables,  like, for instance the tensor-to-scalar ratio and the
spectral tilt of the spectrum. Hence, in WI we can simply set $a_{\rm
  end}/a_{\rm reh} = 1$ in Eq.~(\ref{scales}). This uniquely specifies
the relevant number of e-folds $N_*$ in WI, which from
Eq.~(\ref{scales}), can be shown to be obtained through the
relation~\cite{Das:2020xmh} 
\begin{equation}
\frac{k_*}{a_0 H_0} = e^{-N_*} \left[ \frac{43}{11 g_s(T_{\rm end})}
  \right]^{1/3} \frac{T_0}{T_{\rm end}} \frac{H_*}{H_0},
\label{N*}
\end{equation}
where $T_{\rm end}$ is the temperature at the end of WI and
$g_s(T_{\rm end})$ is the number of relativistic degrees of freedom at
that temperature.

Thus, by returning again to Eq.~(\ref{Pphi}) and dropping the
slow-roll coefficients as a first order approximation,  we have that
$\alpha \simeq \nu$ and Eq. (\ref{Pphi}) can be very well approximated
by the result\footnote{One should note that there is another
  implementation for calculating the curvature power spectrum first
  developed in Ref.~\cite{DeOliveira:2002wk} where the velocity field
  was used instead of momentum perturbation. In
  Ref.~\cite{Bastero-Gil:2019rsp}, it was shown that the power
  spectrum obtained in the Ref.~\cite{DeOliveira:2002wk}  differs by a
  factor $Q/4$ from the result mentioned here. However, in the
  Ref.~\cite{Bastero-Gil:2019rsp} it was discussed that such
  discrepancy comes from the fact that Ref.~\cite{DeOliveira:2002wk}
  did not consider the variation of momentum perturbation with
  expansion. Once this is done, the discrepancy disappears and both
  approaches are consistent.}
\begin{align}
\Delta_{\mathcal{R}}\bigr|_{c=0}\simeq \left(\frac{H_{*}}{\dot\phi_{*}}\right)^2
\left(\frac{H_{*}}{2\pi}\right)^2  \left(1+2n_{*} + \frac{2\sqrt{3}\pi
  Q_{*}}{\sqrt{3+4\pi Q_*}}{T_{*}\over H_{*}}\right) .
\label{powers}
\end{align}
In the case of dissipation coefficients that have an explicit
temperature dependence, one needs to solve Eqs.~(\ref{deltaddotphi}),
(\ref{deltadotrhor}) and (\ref{dotpsir}) to fully determine what is
the effect of the coupling to radiation on the inflaton power
spectrum.  This then leads to a change of the result given by
Eq.~(\ref{powers}), which is usually parametrized by a multiplicative
function of the dissipation coefficient, $G(Q)$, in
Eq.~(\ref{powers}), such that
\begin{equation}
\Delta_{\mathcal{R}}\bigr|_{c\neq 0}= \Delta_{\mathcal{R}}\bigr|_{c=0} G(Q_*).
\label{powersG}
\end{equation}
Neglecting both the metric perturbations and the slow-roll parameters
 and also the field dependence in Eq. (\ref{dupsilon}),  in
 Ref.~\cite{Graham:2009bf} it was found that $G(Q)$ has an asymptotic behavior
for $Q\gg 1$ given by $G(Q) \sim
 (Q/Q_{c})^{3c}$, where $Q_{c}$ is a constant depending on the value
 of $c$. Later, in Ref.~\cite{Bastero-Gil:2011rva}, it was shown that
 accounting for the neglected slow-roll first order quantities as
 considered in Ref.~\cite{Graham:2009bf} could actually overestimate
 the growing function $G(Q) $ by many orders of magnitude at large $Q$
 values\footnote{We note here that taking into account viscosity
   effects  that can be present in the radiation fluid, the growing
   mode can suffer considerable
   damping~\cite{Bastero-Gil:2011rva,Bastero-Gil:2014jsa}. Similar
   damping effects were also reported in
   Refs.~\cite{Motaharfar:2021egj,Motaharfar:2018mni} when
   perturbations propagate with small sound speed, which is typical
   for non-canonical kinetic terms.}. {}For now, we only know how to
 obtain  the growing function $G(Q) $ by numerically solving the full
 set of coupled  background and perturbation equations and then
 numerically fitting the spectrum for a given form of the dissipation
 coefficient. {}For some specific representative cases of dissipation
 coefficients, for instance, for those with powers in the temperature
 $c=3$ (cubic), $c=1$ (linear) and $c=-1$ (inverse), the function
 $G(Q) $ for not too large values of $Q$, $Q_* \lesssim 200$, can be
 well approximated by\footnote{See also
   Refs.~\cite{Kamali:2019xnt,Das:2020xmh} for more specific forms of
   $G(Q) $ valid for higher dissipation values and which can be used for
more precise estimation of cosmological parameters.}
\begin{align}
G_{\rm cubic}(Q_{*}) &\simeq 1+ 4.981 Q_{*}^{1.946} + 0.127
Q_{*}^{4.330},\label{c}\\ G_{\rm linear }(Q_{*}) &\simeq 1+ 0.335
Q_{*}^{1.364} + 0.0185 Q_{*}^{2.315},\label{l}\\ G_{\rm inverse
}(Q_{*}) & \simeq \frac{1+ 0.4 Q_{*}^{0.77}}{(1+ 0.15
  Q_{*}^{1.09})^2}\label{i}.
\end{align}

Once the growing function $G(Q)$ is specified, then from the expression
for the power spectrum in WI, we can for instance obtain the scalar
spectral index $n_s$, which is defined in general as
\begin{align}\label{spectral}
n_{s}-1 \equiv \lim_{k\rightarrow k_{*}} \frac{d \ln
  \Delta_{\mathcal{R}}}{d \ln k} \simeq  \lim_{k\rightarrow k_{*}}
\frac{d \ln \Delta_{\mathcal{R}}}{d N}.
\end{align}
Using that 
\begin{equation}
\frac{d\ln k}{dN}\approx  1-\epsilon_V/(1+Q),
\end{equation}
we obtain for instance that~\cite{Das:2022ubr}
\begin{eqnarray}
n_s&=& 1+\frac{(1+Q)}{1+Q-\epsilon_V} \left[\frac{d \ln
    \left(\frac{T}{H}\right)}{dN} \right.  \nonumber\\ &+&
  \left. \frac{\frac{d \ln Q}{dN}  \left(-3+Q \left\{3+2 \pi
    \left[-1+Q \left(3+\sqrt{9+12 Q \pi
      }\right)\right]\right\}\right)}{(1+Q) \left[3+Q \pi
      \left(4+\sqrt{9+12 Q \pi }\right)\right]} \right.  \nonumber
  \\ &+& \left.\frac{d \ln Q}{dN} {\cal A}(Q)+\frac{-6 \epsilon_V+2
    \eta_V}{1+Q}\right],
\label{ns}
\end{eqnarray}
where the function ${\cal A}(Q)$ is given by
\begin{eqnarray}
\mathcal{A}(Q)=\frac{3+2\pi Q}{3+4\pi Q}+Q\frac{d\ln G(Q)}{dQ},
\label{AQ}
\end{eqnarray}
and the derivatives with respect to the number of e-folds appearing in Eq.~(\ref{ns})
are given by Eqs.~(\ref{b3}) and (\ref{b4}). One can easily observe from Eq.~(\ref{ns}) 
that it reduces to the
spectral index for CI, i.e., $n_{s}-1 = - 6 {\epsilon_{V}} + 2
{\eta_{V}} $, when $Q\rightarrow 0$ and $T\to 0$.

Precise expressions for the spectral tilt and also for the running,
$\alpha_s=dn_s(k)/d\ln(k)$, and for the running of the running,
$\beta_s=d\alpha_s(k)/d\ln(k)$, have been given in
Ref.~\cite{Das:2022ubr} and which are specially useful to analyze WI
in the strong dissipative regime.

While the curvature
power spectrum is significantly modified by dissipative effects, the
tensor power spectrum remains unaffected since the gravitational
interaction are weak\footnote{See, however, Ref.~\cite{Qiu:2021ytc},
  where it was discussed that the radiation thermal bath can also
  produce gravitational waves and this production would enhance the
  tensor power spectrum. See also
  Ref.~\cite{Li:2018wno}, in which  thermal corrections to tensor
  power spectrum were computed and it was found that these corrections
  are, however, small for $Q<100$.}. Therefore, we can write the tensor power
spectrum as in CI,
\begin{align}\label{tensor}
\Delta_{T} = \frac{8}{M^2_{\rm
    Pl}}\left(\frac{H_{*}}{2\pi}\right)^2.
\end{align}
{}From Eq.~(\ref{tensor}), one can define the tensor spectral index
similar to the scalar spectral one, 
\begin{align}
n_{t}\equiv \lim_{k\rightarrow k_{*}} \frac{d \ln \Delta_{T}}{d\ln
  k} \simeq -2 \epsilon_V.
\end{align}
{}Finally, the tensor-to-scalar ratio $r$ is given by
\begin{align}
r = \frac{\Delta_{T}}{\Delta_{\mathcal{R}}} = \frac{16
  \epsilon_{V}}{1+Q} \mathcal{F}^{-1},
\label{tensor-to-scalar}
\end{align}
which one can see that the CI result gets suppressed in WI by the
factors $1+Q$ and $\mathcal{F}$, where
\begin{equation}
\mathcal{F} = \left(1+2n_{*} + \frac{2\sqrt{3}\pi Q_{*}}{\sqrt{3+4\pi
    Q_*}}{T_{*}\over H_{*}}\right) G(Q_*).
\end{equation}
In particular, as a consequence of this, WI can produce a much reduced
tensor-to-scalar ratio as compared to CI for the same type of
primordial inflaton potentials. In particular, well-motivated
potentials that became excluded in CI, like the simple quadratic and
quartic power law inflaton potentials, can be made perfectly consistent
with the CMB data, e.g., from Planck (see, for instance
Refs.~\cite{Benetti:2016jhf,Motaharfar:2018zyb}). The
Eq.~(\ref{tensor-to-scalar}) also implies that the Lyth
bound~\cite{Baumann:2009ds} found in the context of CI and which
relates the inflaton field excursion to the tensor-to-scalar ratio, $\Delta
\phi/M_{\rm Pl} \sim \sqrt{r}$, will be violated in the WI
scenario. Similarly, the dissipative effects will violate the
consistency relation of CI where $r = -8 n_{t}$. The violation of this
relation happens in WI even in the weak dissipative regime, $Q<1$,
since the fluctuations are in a thermally excited state for $T>H$,
i.e.,   $n_{*}\neq 0$. We will see below, in Sec.~\ref{section5}, that
such smoking gun features, together with the much richer dynamics allowed by WI,
can make WI to satisfy the swampland conjectures inspired from string
theory.

It is useful to compare the typical energy scales from WI with those from 
CI\footnote{Similar to the discussions, e.g., in Ref.~\cite{Arraut:2013lca}, 
which compares the scales of standard gravity with some alternatives.}.
We note that the amplitude of gravitational wave perturbations, as well as the potential 
energy scale of CI, is constrained by the CMB. In particular, in CI we can relate the energy scale
of the inflaton potential $V_*$ directly in terms of the tensor-to-scalar ratio as
\begin{equation}
V_*^{\rm CI} = \frac{3 \pi^2 \Delta_{\mathcal{R}} }{2} r M_{\rm Pl}^4 < \left( 1.4 \times 10^{16}{\rm GeV}\right)^4,
\label{V*CI}
\end{equation}
where we have used Eq.~(\ref{tensor}), with the value for the amplitude of the scalar power
spectrum~\cite{Planck:2018vyg},  $\ln\left(10^{10} \Delta_{{\cal R}} \right) \simeq
3.047$ (TT,TE,EE-lowE+lensing+BAO 68$\%$ limits), i.e., $\Delta_{{\cal R}} \simeq 2.1\times 10^{-9}$,
and that $r < 0.036 $, from the combined BICEP/Keck and Planck results~\cite{BICEP:2021xfz}.
In WI, from Eq.~(\ref{tensor-to-scalar}) we have that
\begin{eqnarray}
	\Delta_T=\frac{16\epsilon_V^{wi}}{1+Q}\mathcal{F}^{-1}\Delta_R.
\end{eqnarray}
Using once again Eq.~(\ref{tensor}) and the first {}Friedmann equation in the slow-roll limit, 
we can find the energy scale in the case of WI (at pivot scale $k_{*}$) as
\begin{eqnarray}
	V_{*}^{\rm WI}=\frac{3\pi^2}{2}M_{\rm Pl}^4\frac{16\epsilon^{ci}}{1+Q_{*}}\mathcal{F}^{-1}\Delta_R.
	\label{Warm energy scale}
\end{eqnarray}
This can be compared with the equivalent result in CI\footnote{Note, as already emphasized in 
Section~\ref{section2}, $\epsilon^{wi}_V$ is related to $\epsilon^{ci}$ by $\epsilon^{ci} \simeq \epsilon^{wi}_V/(1+Q)$.}, 
\begin{eqnarray}
	V_{*}^{\rm CI}=\frac{3\pi^2}{2}M_{\rm Pl}^4 16 \epsilon^{ci}_V \Delta_R.
	\label{Cold Energy scale}
\end{eqnarray}
Thus,

\begin{eqnarray}
	\frac{V_{*}^{\rm WI}}{V_{*}^{\rm CI}}=\mathcal{F}^{-1},
\end{eqnarray}
which means that the energy scale for the inflaton in WI is always smaller than the one
expected from CI, Eq.~(\ref{V*CI}).
Likewise, we have for the ratio between Hubble scales in WI and in CI is given by
\begin{eqnarray}
	\frac{H^{\rm WI}_{*}}{H_{*}^{\rm CI}}=\frac{1}{\sqrt{\mathcal{F}}}.
\end{eqnarray}
As WI tends in general to predict a smaller tensor-to-scalar ratio than that of CI,
this then translates in smaller energy scales in WI than in CI.

To conclude this section, let us briefly discuss about the amount of
non-Gaussianity that is produced by dissipative effects during WI. The
first rigorous attempt to estimate the non-linearity effects by WI was
done in Ref.~\cite{Moss:2007cv}. In that reference, by considering the
dissipation coefficient just as a function of the inflaton field
($c=0$), it was found that the non-linearity parameter could be
approximated as $f_{NL}^{\rm warmS} = - 15 \ln \left(1+ Q/14\right) -
5/2$ in the strong dissipative regime. Then, the results were later
generalized in  Ref.~\cite{Moss:2011qc} for the case of temperature
dependent dissipation coefficients ($c\neq 0$), while still in the
strong dissipative regime. It was also realized that the coupling of
inflaton and the radiation field fluctuations due to the temperature
dependent dissipation coefficient would potentially make the
non-linearity effects stronger. In fact, it was found that the
non-linearity parameter will be larger for larger values for the
exponent $c$, the power of the temperature in the dissipation
coefficient. However, the most concrete results were obtained in
Ref.~\cite{Bastero-Gil:2014raa}. Through a numerical method developed
in Ref.~\cite{Bastero-Gil:2014raa}, the non-linearity parameter was
calculated for the general form of the dissipation coefficient and in
both weak and strong dissipative regimes. The results obtained for the
non-linearity parameter turned out to be smaller than what was found
originally in Ref.~\cite{Moss:2007cv} for the strong dissipative
regime. According to the results reported in
Ref.~\cite{Bastero-Gil:2014raa}, the non-Gaussianity parameter can
significantly depend on the values of $T/H$ and also $Q$ in the weak
dissipation regime, i.e., for $Q<0.1$ and it reaches its maximum at $Q
\sim10^{-3}$, when the thermal fluctuations start to become
dominant. However, when the dissipation ratio is large, $Q>1$, the
non-linearity parameter is independent of $T/H$ and it only mildly
depends on $Q$. This happens because of the existence of the growing
mode in the power spectrum, which enhances the amplitude of the power
spectrum by a factor $Q^{\alpha}$ and  modifies the bispectrum by a
factor $Q^{2\alpha}$, while the effects partially cancel out in the
$f_{NL}$. {}Furthermore, the non-linearity parameter is larger for
large value of $c$ in the weak dissipation regime. However, as the
dissipation ratio increases, it results in comparable non-linearity
parameters in the  strong dissipative regime. Apart from the magnitude
of the non-linearity parameter, it was also found in
Ref.~\cite{Bastero-Gil:2014raa} that WI predicts two distinct shapes
for the non-linearity, depending whether WI happens in the weak or in
the strong dissipative regime. This in particular makes WI
distinguishable from  CI if non-Gaussianities are detected in near
future~\footnote{Recently, the non-Gaussianity has also been
  investigated in Ref.~\cite{Mirbabayi:2022cbt} in an axion-type of
  model and it was pointed out some distinct features in the squeezed
  and folded limits.}.

\section{Swampland Criteria, observational constraints and other applications}
\label{section5}

In this section we briefly review the implications of the swampland
conjectures that have been discussed recently and put forward when applied to inflation models.
We discuss how the swampland
constraints are indicative of ruling out most, if not all, single
field CI models, while WI is able to be in accordance with these swampland 
constraints and at the same time also satisfy all current observational
constraints. {}Finally, we conclude the section by discussing about
some of the recent applications concerning WI and the efforts to using
it to elucidate some of the outstanding pre/post-inflationary phenomena.

\subsection{WI and swampland conjectures}

It is always assumed that inflation can be described using low energy
effective field theory (EFT), since the energy scale of inflation is
below the Planck scale when the currently observable scale exits the
Hubble horizon. However, this does not mean that any inflationary
model can be ultraviolate complete. Therefore, it is important to
try to distinguish those EFTs that can be consistently embedded into a
quantum theory of gravity from those that cannot. This was the starting point of the
swampland program and aiming at distinguishing those EFTs,  which
belongs to the landscape of string theory, from those that inhibit in
the swampland. By having the correct criteria to identify the boundary
between the landscape and the swampland, this resulted in a series of
conjectures known as the {\it swampland conjectures} (see, e.g.,
Ref.~\cite{Palti:2019pca} for a review of these ideas). In fact, these
conjectures, although speculative, are the very first theoretical
constraints originating from string theory and that can have direct
implication on the inflationary cosmology. 

As discussed in Ref.~\cite{Ooguri:2006in}, low-energy EFTs 
can become inapplicable during compactification in string theory. This is because the mass of 
quantum gravity states decreases exponentially rapid as the field excursion in 
the moduli space increases, i.e., a tower of massive states become exponentially light as  
$\exp(-\alpha \Delta \phi/M_{\rm Pl})$, with $\alpha \sim {\cal O}(1)$.
As a result, if the scalar field has excursions beyond the Planck scale, 
$\Delta \phi > M_{\rm Pl}$, a large number of new light states must be considered.
Thus, it has then been speculated that  the change of any scalar field arising
in the EFT in its field space is confined by a positive constant
number of order one,
\begin{eqnarray}
	\frac{\Delta\phi}{M_{\rm Pl}} < c_1  .
	\label{distance}
\end{eqnarray} 
The above constraint (\ref{distance}) is called the {\it swampland
  distance conjecture (SDC)}. This bound arises from the fact that if
the scalar field moves by a range that is larger than the upper bound given by
Eq.~(\ref{distance}), thus making a super-Planckian field excursion,
then a tower of new string states becomes light, as said above,  and 
they must be included
in the low energy EFT. The SDC has an immediate implication for inflation
models. In fact, SDC indicates that large field inflationary models,
which generically are known to lead to super-Planckian field
excursions in the context of CI (we will see below how WI can evade
the constraint given by Eq.~(\ref{distance})), tend to be ruled out as
a consistent EFT model. Since the field excursion is related to the
tensor-to-scalar ratio through the Lyth bound, hence one may infer
that SDC is indicating that the Lyth bound should be violated in a
non-trivial way such as to make inflation consistent with SDC and the
observational data. As we have already discussed in the previous
section~\ref{section4}, WI naturally violates the Lyth
bound. {}Furthermore, in WI the inflaton field can be slowed down not
only because of the Hubble friction but also from the explicit
dissipation term. If the dissipation term is large enough, in
particular in the strong dissipative regime $Q\gg 1$, the inflaton
excursions can be made sufficiently short such as to satisfy
Eq.~(\ref{distance}). This feature of WI in the strong dissipative regime
has been observed by many recent realizations of this 
regime (see, for instance, Refs.~\cite{Kamali:2019xnt,Das:2020xmh,Motaharfar:2021egj}).
 
 Several years after introducing the SDC, it was discussed that
 constructing metastable de Sitter vacuum is notoriously difficult in
 string theory. This has lead to the conjecture that metastable de
 Sitter vacua should belong to the swampland rather than to the
 landscape of string theory. As a consequence of this difficulty in
 building appropriate de Sitter vacua, it can then be translated in
 conditions  that potentials for scalar fields should satisfy and
 which are known as the  swampland {\it  de Sitter conjecture} (SdSC).
 These conditions imply in bounds on the slope of the scalar
 potentials in an EFT and which can be expressed in
 the form~\cite{Garg:2018reu, Ooguri:2018wrx},
\begin{eqnarray}
	\frac{\mid\nabla V\mid}{V} \, \geq \, \frac{c_2}{M_{\rm Pl}},
	\label{de Sitter1}
\end{eqnarray}
or
\begin{eqnarray}
	\frac{min(\nabla_i\nabla_j V)}{V} \, \leq \,
        -\frac{c_3}{M_{\rm Pl}^2} ,
	\label{de Sitter2}
\end{eqnarray}
where $\nabla$ is the gradient in field space, $c_2$ and $c_3$ are
universal and positive constants of order one and
$min(\nabla_i\nabla_j V)$ is the minimum eigenvalue of the Hessian
$\nabla_i\nabla_{j} V$ in an orthonormal frame.  The conditions given
by Eqs.~(\ref{de Sitter1}) and (\ref{de Sitter2})  mean that either a
useful potential has to be sufficiently steep, or else sufficiently
tachyonic near its maximum, if it has one. It was shown in
Ref.~\cite{Ooguri:2018wrx} that the first de Sitter condition, given by
Eq.~(\ref{de Sitter1}), is related to the distance condition,
Eq.~(\ref{distance}), in the weak coupling regime and which can be
demonstrated by using Bousso’s covariant entropy
bound~\cite{Bousso:1999xy}.  Looking at the conditions given by
Eqs.~(\ref{de Sitter1}) and (\ref{de Sitter2}), one can easily realize
that these conditions can be translated into conditions on the
slow-roll coefficients $\epsilon_V$ and $\eta_V$, such that they must
satisfy~\cite{Garg:2018reu},
\begin{eqnarray}
\epsilon_V\equiv \frac{M_{\rm
    Pl}^2}{2}\left(\frac{V,_\phi}{V}\right)^2\geq
\frac{c_2^2}{2},\quad {\rm or}\quad \eta_V\equiv M_{\rm
  Pl}^2\frac{V,_{\phi\phi}}{V}\leq-c_3,
\end{eqnarray}
thus, requiring  the potential slow-roll parameters to be of order
unity.  This is obviously in contrast to the slow-roll conditions
imposed in CI models, where $\epsilon_V \ll 1$ and $\eta_V \ll
1$. Hence, the first condition rules out all single field CI models, while
the second condition is still consistent with those inflationary
potential with tachyonic instability, such as hiltop
potentials~\cite{Kinney:2018kew}. Recalling that $\epsilon_H = -\dot H/H^2$, 
and that in the context of CI
$\epsilon_{H} \simeq \epsilon_{V}$, hence, in the CI scenario one
cannot achieve slow-roll inflationary phase for steep potentials.
Moreover, the SdSC also excludes any extrema of scalar field potentials in
field space, i.e., $|\nabla_{\phi} V|/V \rightarrow 0$. This is in
clear contrast with the reheating phase after the end of inflation,
which requires the inflaton potential to have a minimum. Therefore,
the SdSC also indicates that inflation should terminate using a new
mechanism rather than the so-called reheating phase in CI scenario.  Both
of these issues are again possible to be overcame in the context of
WI. Recalling again that in WI both slow-roll parameters $\epsilon_V$
and $\eta_V$ are replaced by $\epsilon_H\simeq
\epsilon_{wi}=\epsilon_V/(1+Q)$ (see, e.g., Eq.~(\ref{c2})) and
$\eta_{wi}=\eta_V/(1+Q)$. Hence, in WI there is no problem of having
$\epsilon_V$ and $\eta_V$ larger than 1, provided that $Q \gg 1$. WI
in the strong dissipative regime comes again to the rescue, making
inflation perfectly consistent with the SdSC as shown explicitly in
the models worked out in
Refs.~\cite{Kamali:2019xnt,Das:2020xmh,Motaharfar:2021egj} for
example. 

It is also important to note that
the so-called $\eta$-problem of cold inflation~\cite{Baumann:2014nda},
is reminiscent of the SdSC. The $\eta$-problem in related to the
fact that to have inflation, one requires $V_{\phi\phi}\equiv m_\phi^2 \ll H^2$,
such that Hubble friction dominates during inflation. 
However, the inflaton potential is prone to receive large corrections and which can drive
the inflaton mass above the Hubble scale $H$.  This is the "eta-problem", which appears in,
 e.g., {}F-term supergravity and string theory. If the inflaton mass is driven to 
super-Hubble values, we have a tension with the slow-roll conditions. 
As argued a long time ago already~\cite{Berera:1999ws,Berera:2004vm,BasteroGil:2009ec},
WI in the strong dissipative regime provides a natural solution for the
$\eta$-problem, as we have seen also here when the SdSC constraint is satisfied by WI.

As WI intrinsically leads to entropy
production as a consequence of particle production, it might change
the above swampland conditions. This is specially for the case of the
de Sitter conjecture, which, as mentioned above, can be related to
entropic phenomena.  It has been shown in
Ref.~\cite{Brandenberger:2020oav} that extra entropy produced as a
result of particle production in the context of WI has a negligible
effect and WI remains fully consistent with the SDC and SdSC.

In addtion to the above swampland conjectures, very recently, another
conjecture called the {\it Trans-Planckian censorship Conjecture}
(TCC) was also been proposed~\cite{Bedroya:2019snp,
  Bedroya:2019tba}. The TCC  proposal is based on the trans-Planckian
problem of inflationary models,  which severely constraint
inflationary models. The TCC requests that the Hubble horizon must
hide sub-Planckian modes during the early stages of  accelerated
expansion,
\begin{eqnarray}
\left(\frac{a_f}{a_i}\right)\ell_{\rm Pl}<\frac{1}{H_f},
\label{TCC}
\end{eqnarray}
where $a_i$ and $a_f$ are, respectively, the scale factors at the
beginning and at the end of the evolution, $H_f$ is the Hubble
parameter at the end of that evolution and $\ell_{\rm Pl}$ is a length
scale of the order of the Planck scale\footnote{One should also note that there
  are modified versions of the TCC, as, e.g., in
  Ref.~\cite{Aalsma:2020aib}, and which can allow for larger values of
  $H_f$ than the suggested from Eq.~(\ref{TCC}), which alleviates
  appreciably the TCC bound. Additionally, there are other recent
  discussions in the literature concerning the TCC bound, e.g., in
  Refs.~\cite{Mizuno:2019bxy,Kamali:2019gzr,Berera:2019zdd,Berera:2020dvn}
  on how it also be relaxed.}.  The TCC bound can be translated into
an upper bound on the duration of inflation. If we assume a constant
Hubble horizon, or a period of quasi-de Sitter inflation and instant
reheating phase at the end of inflation, the TCC implies in an upper
limit for the energy scale of inflation~\cite{Bedroya:2019tba} 
\begin{eqnarray}\label{energy scale}
	V^{\frac{1}{4}}<6\times 10^{8} \rm{GeV} \sim 3 \times 10^{-10}
        M_{\rm Pl},
\end{eqnarray}
which, in turn, it can be translated to an upper bound on the
tensor-to-scalar ratio parameter,
\begin{eqnarray}\label{TCC1}
	r<10^{-31}.
	\end{eqnarray}
The result given by Eq.~(\ref{TCC1})  tightly constrains the slow-roll
epoch needed to resolve the shortcomings of the standard big bang cosmology
according to the standard CI picture. In fact, to make CI models
compatible with the TCC, one needs to modify the dynamics of the CI scenario
in such a way that the tensor-to-scalar ratio can be suppressed
significantly. By accounting for a non-standard evolution after the end
of inflation, it has been shown~\cite{Mizuno:2019bxy} that it can lead to a more accessible
result for $r$, $r<10^{-8}$, which is weaker than the upper bound
given by Eq.~(\ref{TCC1}). But the result is still very constraining on the energy
scale of inflation.  Turning now to WI, one can see how it can evade the
TCC bound discussed here. As explained in Sec.~\ref{section4}, the
dissipative effects in WI produce a modified primordial scalar of
curvature power spectrum, Eq.~(\ref{powers}), which is enhanced by the
dissipation and thermal effects. Again, for $Q\gg 1$, this results in
a highly suppressed tensor-to-scalar ratio, Eq.~(\ref{tensor-to-scalar}),
which can make WI also
consistent with the
TCC~\cite{Kamali:2019xnt,Das:2020xmh,Motaharfar:2021egj}. All of this
is achieved with WI also leading to consistent results for the
spectral tilt $n_s$ and also for its runnings~\cite{Das:2022ubr},
thus consistent with the current results from
Planck~\cite{Planck:2018jri}.

There have been also discussions concerning eternal inflation and the
swampland and whether eternal inflation would be in the landscape or in the
swamp of EFTs in the context of quantum gravity theories (see, e.g., 
Refs.~\cite{Kinney:2018kew,Dimopoulos:2018upl,Brahma:2019iyy}). Interestingly,
WI again here seems to play a significant role. The dissipative and thermal
characteristics of WI have been shown to provide a way of suppressing the
emergence of eternal inflation and, in the strong dissipative regime, even 
possibly avoiding it to happen~\cite{Vicente:2015hga}.

Although all these swampland conjectures explained above are speculative and there 
are not strong evidences to strengthen them, it is worth to see that WI is able to
satisfy and to be consistent with all of the constraints that these
conditions bring and also in a natural way.

\subsection{Other applications and recent results in WI}

As we have discussed in Sections~\ref{section2} and \ref{section4}, WI
enlarges the class of potentials that can be made observationally
consistent with the observational data in comparison to CI
scenario. {}For instance, simple potentials for the inflaton, like the
quadratic and quartic monomials potentials, the Higgs-like symmetry
breaking potential and the Coleman-Weinberg type of potentials, which are all
well motivated in the context of particle physics, have been shown to
the excluded by the Planck data in the context
of CI~\cite{Planck:2018jri}.  However, in WI, because of the different dynamics resulted
from the dissipative effects, they can all be made fully consistent
with the observations and this can happen for a large range of space of parameters, as
discussed, e.g., in Ref.~\cite{Benetti:2016jhf}. 

Considering also the natural inflation potential~\cite{Freese:1990rb}, 
i.e., $V(\phi) = \Lambda^4[1 +
  \cos\left(\phi/f\right)]$ in the WI scenario,  it was found that
although in CI the model is now borderline consistent with the observations for
super-Planckian values for the decay constant $f$, it can become consistent
with the observations even in the case of sub-Planckian values for $f$ in
the context of the WI picture, as shown in case of a simplified constant
dissipation coefficient form~\cite{Visinelli:2011jy,Mishra:2011vh,Kamali:2019ppi}. A
sub-Planckian value for $f$   is favorable from a model building and
EFT perspective. However, very recently~\cite{Montefalcone:2022jfw},
this model was again reconsidered in WI in the cases of linear and cubic
temperature dependent dissipation coefficients and it was found that
due to the growing mode function $G(Q)$, the model is not consistent
with the observations in the case of sub-Planckian values for $f$. 

The results and studies in WI can also be extended to the case of noncanonical type of
models.  {}For instance, it has been shown that for the quartic
potential, $V=\lambda \phi^4$, the combination of dissipative effects
with G-inflation make this potential consistent with observation even
for large values for the self-coupling $\lambda$~\cite{Motaharfar:2018mni}.

Moreover, there are several papers in the literature which studied WI
using some quasi-de Sitter form of scale factor, e.g., intermediate
and logamediate inflation, and other noncanonical versions of field
theory, e.g., tachyonic fields, in the high dissipative
regime~\cite{Herrera:2006ck,delCampo:2009xi,Herrera:2011zz,Setare:2012fg,Setare:2013ula,Setare:2014oka,Kamali:2016frd,Motaharfar:2016dqt,Mohammadi:2020vgs}, while there have been also many
studies involving different forms of inflaton potentials and in other
interesting
contexts~\cite{Dymnikova:2001ga,Dymnikova:2001jy,Vicente:2015hga,Gashti:2022pvu,Santos:2022exm,Payaka:2022jtb,Bouabdallaoui:2022wyp,Reyimuaji:2020bkm,Graef:2018ulg,Berera:2018tfc,Herrera:2018cgi,Bastero-Gil:2017yzb,Wang:2019ozs,Harko:2021gnz,Bose:2022wla,Harko:2020cev,Sheikhahmadi:2019gzs,Bertolami:2022knd,Motaharfar:2021gwi,AlHallak:2022haa,Montefalcone:2022owy}. 

The analysis of the dynamics from the point of view of a dynamical system realization is important
for many reasons. It not only can bring general information about the dynamics that the system can present but also
regarding its stability. In this context, several studies have analyzed  WI 
from a dynamical system perspective, as, for example in the 
Refs.~\cite{deOliveira:1997jt,Moss:2008yb,delCampo:2010by,Bastero-Gil:2012vuu,Li:2018sfs,Alho:2022qri}.
WI has also been contrasted with the observational data through explicit
statistical analysis, as in
Refs.~\cite{Benetti:2016jhf,Arya:2017zlb,Bastero-Gil:2017wwl,Arya:2018sgw,Benetti:2019kgw}.
The inflaton itself in WI can be a source and responsible for cosmic
magnetic field generation~\cite{Berera:1998hv}, and in combination
with the intrinsic dissipative effects lead to a successful
baryogenesis
scenario~\cite{Brandenberger:2003kc,Bastero-Gil:2011clw,Bastero-Gil:2014oga}.
Moreover, the gravitino problem was also considered in the context of
WI~\cite{BuenoSanchez:2010ygd,Bartrum:2012tg}.  More recently, a focus
has been given to WI in topics like dark
energy~\cite{Dimopoulos:2019gpz,Rosa:2019jci,Lima:2019yyv,Gangopadhyay:2020bxn,Basak:2021cgk,Saleem:2021ytj},
dark
matter~\cite{Rosa:2018iff,Levy:2020zfo,Zhang:2021zol,Sa:2020fvn,DAgostino:2021vvv}
and in problems related to primoridial black hole formation and in providing additional
sources of gravitational
waves~\cite{Arya:2019wck,Bastero-Gil:2021fac,Correa:2022ngq,Arya:2022xzc,Ballesteros:2022hjk},
just to cite a few of the many recent developments in the context of WI.

\section{Conclusions}
\label{section6}

Warm inflation provides a framework for understanding the dynamics of
the early universe from a perspective that differs from that of the
standard cold inflation.  Warm inflation accounts for the explicit
dissipative effects that are expected to be present in any dynamical
physical theory with interacting particles and fields. When these
dissipative effects are strong enough to overcome the dilution of the
radiation during the rapid expansion in the primordial inflation epoch,
a rich dynamics and new phenomena emerge that are not present during
the usual cold inflation picture. 

In warm inflation the generation of density perturbations can be
entirely classical, formed by thermal fluctuations instead of quantum
fluctuations in the case of cold inflation. This overcomes several
issues related to classicalization of perturbations in the case of
cold inflation. {}Furthermore, warm inflation predicts a very small,
if not substantially small, tensor-to-scalar ratio due to the thermal
and dissipative effects.  As a result, simple inflaton potential
models that are well motivated in the context of renormalizable
quantum field theory and particle physics can be made fully consistent
with the cosmological observations from Planck. On the other hand, in
cold inflation all these simple potentials and models are ruled out by
the current cosmological data. 

With the advent of the use of the swampland conjectures, which are
derived from string theory and give indications of what a consistent
effective field theory should satisfy to be  able to descent from a
quantum gravity theory, have posed strong constraints on cold
inflation models.  Warm inflation is also here shown to solve these
issues. Warm inflation, in particular in the regime of strong
dissipation, provides a way to overcome all the constraints brought by
the swampland conjectures.  Thus, warm inflation provides a viable
solution for resolving the tension between quantum field theory and
quantum gravity.  {}From the swampland perspective, warm inflation is
then more natural and less prone to falling into the swampland. 

At its conception about 28 years ago, warm inflation appeared to be an
amusement, not expected to have any significant protagonism in cosmology 
when compared to the  mainstream cold inflation picture. Since then, 
the warm inflation idea has evolved and
moved to a mature and relevant subject, providing a picture towards
which an understanding of some of the fundamental  problems in
cosmology might have a better chance to be understood than in terms of 
the cold inflation scenario.  Ideas in the context of warm inflation 
has been developing rapidly in the recent years, as this review tried to
demonstrate. These have been happening both from a more fundamental quantum field theory
perspective, but also from the wealthy of applications that have now
been considered. Warm inflation has now become a promising and growing
area of research, with the potential to provide new insights into the
physics of the early universe.

\section*{acknowledgements}

V.K. would like to acknowledge the McGill University Physics Department 
for hospitality and partial financial support.
M. M. work is supported by the NSF grant PHY-211020.
R.O.R. acknowledges financial support of the Coordena\c{c}\~ao de
Aperfei\c{c}oamento de Pessoal de N\'{\i}vel Superior (CAPES) -
Finance Code 001 and by research grants from Conselho Nacional de
Desenvolvimento Cient\'{\i}fico e Tecnol\'ogico (CNPq), Grant
No. 307286/2021-5, and from Funda\c{c}\~ao Carlos Chagas Filho de
Amparo \`a Pesquisa do Estado do Rio de Janeiro (FAPERJ), Grant
No. E-26/201.150/2021. 



\begin{thebibliography}{999} 

\bibitem{Berera:1995wh}
A.~Berera and L.~Z.~Fang,
Thermally induced density perturbations in the inflation era,
Phys. Rev. Lett. \textbf{74} (1995), 1912-1915
doi:10.1103/PhysRevLett.74.1912
[arXiv:astro-ph/9501024 [astro-ph]].

\bibitem{Berera:1995ie}
A.~Berera,
Warm inflation,
Phys. Rev. Lett. \textbf{75} (1995), 3218-3221
doi:10.1103/PhysRevLett.75.3218
[arXiv:astro-ph/9509049 [astro-ph]].

\bibitem{Berera:1998px}
A.~Berera, M.~Gleiser and R.~O.~Ramos,
A First principles warm inflation model that solves the cosmological horizon / flatness problems,
Phys. Rev. Lett. \textbf{83} (1999), 264-267
doi:10.1103/PhysRevLett.83.264
[arXiv:hep-ph/9809583 [hep-ph]].

\bibitem{Berera:2008ar}
A.~Berera, I.~G.~Moss and R.~O.~Ramos,
Warm Inflation and its Microphysical Basis,
Rept. Prog. Phys. \textbf{72} (2009), 026901
doi:10.1088/0034-4885/72/2/026901
[arXiv:0808.1855 [hep-ph]].

\bibitem{BasteroGil:2009ec}
M.~Bastero-Gil and A.~Berera,
Warm inflation model building,
Int. J. Mod. Phys. A \textbf{24} (2009), 2207-2240
doi:10.1142/S0217751X09044206
[arXiv:0902.0521 [hep-ph]].

\bibitem{Guth:1980zm}
A.~H.~Guth,
The Inflationary Universe: A Possible Solution to the Horizon and Flatness Problems,
Phys. Rev. D \textbf{23} (1981), 347-356
doi:10.1103/PhysRevD.23.347

\bibitem{Sato:1981ds}
K.~Sato,
Cosmological Baryon Number Domain Structure and the First Order Phase Transition of a Vacuum,
Phys. Lett. B \textbf{99} (1981), 66-70
doi:10.1016/0370-2693(81)90805-4

\bibitem{Albrecht:1982wi}
A.~Albrecht and P.~J.~Steinhardt,
Cosmology for Grand Unified Theories with Radiatively Induced Symmetry Breaking,
Phys. Rev. Lett. \textbf{48} (1982), 1220-1223
doi:10.1103/PhysRevLett.48.1220

\bibitem{Linde:1981mu}
A.~D.~Linde,
A New Inflationary Universe Scenario: A Possible Solution of the Horizon, Flatness, Homogeneity, Isotropy and Primordial Monopole Problems,
Phys. Lett. B \textbf{108} (1982), 389-393
doi:10.1016/0370-2693(82)91219-9

\bibitem{Linde:1983gd}
A.~D.~Linde,
Chaotic Inflation,
Phys. Lett. B \textbf{129} (1983), 177-181
doi:10.1016/0370-2693(83)90837-7

\bibitem{Bartrum:2013fia}
S.~Bartrum, M.~Bastero-Gil, A.~Berera, R.~Cerezo, R.~O.~Ramos and J.~G.~Rosa,
The importance of being warm (during inflation),
Phys. Lett. B \textbf{732} (2014), 116-121
doi:10.1016/j.physletb.2014.03.029
[arXiv:1307.5868 [hep-ph]].

\bibitem{Bastero-Gil:2012akf}
M.~Bastero-Gil, A.~Berera, R.~O.~Ramos and J.~G.~Rosa,
General dissipation coefficient in low-temperature warm inflation,
JCAP \textbf{01} (2013), 016
doi:10.1088/1475-7516/2013/01/016
[arXiv:1207.0445 [hep-ph]].

\bibitem{Bastero-Gil:2011zxb}
M.~Bastero-Gil, A.~Berera and J.~G.~Rosa,
Warming up brane-antibrane inflation,
Phys. Rev. D \textbf{84} (2011), 103503
doi:10.1103/PhysRevD.84.103503
[arXiv:1103.5623 [hep-th]].

\bibitem{Matsuda:2012kc}
T.~Matsuda,
Particle production and dissipation caused by the Kaluza-Klein tower,
Phys. Rev. D \textbf{87} (2013) no.2, 026001
doi:10.1103/PhysRevD.87.026001
[arXiv:1212.3030 [hep-th]].

\bibitem{Berera:2002sp}
A.~Berera and R.~O.~Ramos,
Construction of a robust warm inflation mechanism,
Phys. Lett. B \textbf{567} (2003), 294-304
doi:10.1016/j.physletb.2003.06.028
[arXiv:hep-ph/0210301 [hep-ph]].

\bibitem{Bastero-Gil:2016qru}
M.~Bastero-Gil, A.~Berera, R.~O.~Ramos and J.~G.~Rosa,
Warm Little Inflaton,
Phys. Rev. Lett. \textbf{117} (2016) no.15, 151301
doi:10.1103/PhysRevLett.117.151301
[arXiv:1604.08838 [hep-ph]].

\bibitem{Bastero-Gil:2019gao}
M.~Bastero-Gil, A.~Berera, R.~O.~Ramos and J.~G.~Rosa,
Towards a reliable effective field theory of inflation,
Phys. Lett. B \textbf{813} (2021), 136055
doi:10.1016/j.physletb.2020.136055
[arXiv:1907.13410 [hep-ph]].

\bibitem{Berghaus:2019whh}
K.~V.~Berghaus, P.~W.~Graham and D.~E.~Kaplan,
Minimal Warm Inflation,
JCAP \textbf{03} (2020), 034
doi:10.1088/1475-7516/2020/03/034
[arXiv:1910.07525 [hep-ph]].

\bibitem{Berera:1999ws}
A.~Berera,
Warm inflation at arbitrary adiabaticity: A Model, an existence proof for inflationary dynamics in quantum field theory,
Nucl. Phys. B \textbf{585} (2000), 666-714
doi:10.1016/S0550-3213(00)00411-9
[arXiv:hep-ph/9904409 [hep-ph]].

\bibitem{Berera:2004vm}
A.~Berera,
Warm inflation solution to the eta problem,
PoS \textbf{AHEP2003} (2003), 069
doi:10.22323/1.010.0069
[arXiv:hep-ph/0401139 [hep-ph]].

\bibitem{Das:2018hqy}
S.~Das,
Note on single-field inflation and the swampland criteria,
Phys. Rev. D \textbf{99} (2019) no.8, 083510
doi:10.1103/PhysRevD.99.083510
[arXiv:1809.03962 [hep-th]].

\bibitem{Motaharfar:2018zyb}
M.~Motaharfar, V.~Kamali and R.~O.~Ramos,
Warm inflation as a way out of the swampland,
Phys. Rev. D \textbf{99} (2019) no.6, 063513
doi:10.1103/PhysRevD.99.063513
[arXiv:1810.02816 [astro-ph.CO]].

\bibitem{Das:2018rpg}
S.~Das,
Warm Inflation in the light of Swampland Criteria,
Phys. Rev. D \textbf{99} (2019) no.6, 063514
doi:10.1103/PhysRevD.99.063514
[arXiv:1810.05038 [hep-th]].

\bibitem{Das:2019hto}
S.~Das,
Distance, de Sitter and Trans-Planckian Censorship conjectures: the status quo of Warm Inflation,
Phys. Dark Univ. \textbf{27} (2020), 100432
doi:10.1016/j.dark.2019.100432
[arXiv:1910.02147 [hep-th]].

\bibitem{Kamali:2019xnt}
V.~Kamali, M.~Motaharfar and R.~O.~Ramos,
Warm brane inflation with an exponential potential: a consistent realization away from the swampland,
Phys. Rev. D \textbf{101} (2020) no.2, 023535
doi:10.1103/PhysRevD.101.023535
[arXiv:1910.06796 [gr-qc]].

\bibitem{Brandenberger:2020oav}
R.~Brandenberger, V.~Kamali and R.~O.~Ramos,
Strengthening the de Sitter swampland conjecture in warm inflation,
JHEP \textbf{08} (2020), 127
doi:10.1007/JHEP08(2020)127
[arXiv:2002.04925 [hep-th]].

\bibitem{Das:2020xmh}
S.~Das and R.~O.~Ramos,
Runaway potentials in warm inflation satisfying the swampland conjectures,
Phys. Rev. D \textbf{102} (2020) no.10, 103522
doi:10.1103/PhysRevD.102.103522
[arXiv:2007.15268 [hep-th]].

\bibitem{Berera:2019zdd}
A.~Berera and J.~R.~Calder\'on,
Trans-Planckian censorship and other swampland bothers addressed in warm inflation,
Phys. Rev. D \textbf{100} (2019) no.12, 123530
doi:10.1103/PhysRevD.100.123530
[arXiv:1910.10516 [hep-ph]].

\bibitem{Berera:2020dvn}
A.~Berera, S.~Brahma and J.~R.~Calder\'on,
Role of trans-Planckian modes in cosmology,
JHEP \textbf{08} (2020), 071
doi:10.1007/JHEP08(2020)071
[arXiv:2003.07184 [hep-th]].

\bibitem{Kamali:2019hgv}
V.~Kamali,
Reheating After Swampland Conjecture,
JHEP \textbf{01} (2020), 092
doi:10.1007/JHEP01(2020)092
[arXiv:1902.00701 [gr-qc]].

\bibitem{Kamali:2019wdh}
V.~Kamali, M.~Artymowski and M.~R.~Setare,
Constant roll warm inflation in high dissipative regime,
JCAP \textbf{07} (2020) no.07, 002
doi:10.1088/1475-7516/2020/07/002
[arXiv:1905.04814 [gr-qc]].

\bibitem{Berera:2020iyn}
A.~Berera, R.~Brandenberger, V.~Kamali and R.~Ramos,
Thermal, trapped and chromo-natural inflation in light of the swampland criteria and the trans-Planckian censorship conjecture,
Eur. Phys. J. C \textbf{81} (2021) no.5, 452
doi:10.1140/epjc/s10052-021-09240-3
[arXiv:2006.01902 [hep-th]].

\bibitem{Kamali:2018ylz}
V.~Kamali,
Non-minimal Higgs inflation in the context of warm scenario in the light of Planck data,
Eur. Phys. J. C \textbf{78} (2018) no.11, 975
doi:10.1140/epjc/s10052-018-6449-x
[arXiv:1811.10905 [gr-qc]].

\bibitem{Kamali:2021gkl}
V.~Kamali, S.~Ebrahimi and A.~Alaei,
Intermediate class of warm pseudoscalar inflation,
Eur. Phys. J. C \textbf{81} (2021) no.6, 562
doi:10.1140/epjc/s10052-021-09367-3
[arXiv:2111.10540 [gr-qc]].

\bibitem{Benetti:2016jhf}
M.~Benetti and R.~O.~Ramos,
Warm inflation dissipative effects: predictions and constraints from the Planck data,
Phys. Rev. D \textbf{95} (2017) no.2, 023517
doi:10.1103/PhysRevD.95.023517
[arXiv:1610.08758 [astro-ph.CO]].

\bibitem{Obied:2018sgi}
G.~Obied, H.~Ooguri, L.~Spodyneiko and C.~Vafa,
De Sitter Space and the Swampland,
[arXiv:1806.08362 [hep-th]].

\bibitem{Ooguri:2018wrx}
H.~Ooguri, E.~Palti, G.~Shiu and C.~Vafa,
Distance and de Sitter Conjectures on the Swampland,
Phys. Lett. B \textbf{788} (2019), 180-184
doi:10.1016/j.physletb.2018.11.018
[arXiv:1810.05506 [hep-th]].

\bibitem{Kinney:2018nny}
W.~H.~Kinney, S.~Vagnozzi and L.~Visinelli,
The zoo plot meets the swampland: mutual (in)consistency of single-field inflation, string conjectures, and cosmological data,
Class. Quant. Grav. \textbf{36} (2019) no.11, 117001
doi:10.1088/1361-6382/ab1d87
[arXiv:1808.06424 [astro-ph.CO]].

\bibitem{Bedroya:2019snp}
A.~Bedroya and C.~Vafa,
Trans-Planckian Censorship and the Swampland,
JHEP \textbf{09} (2020), 123
doi:10.1007/JHEP09(2020)123
[arXiv:1909.11063 [hep-th]].

\bibitem{Agrawal:2018own}
P.~Agrawal, G.~Obied, P.~J.~Steinhardt and C.~Vafa,
On the Cosmological Implications of the String Swampland,
Phys. Lett. B \textbf{784} (2018), 271-276
doi:10.1016/j.physletb.2018.07.040
[arXiv:1806.09718 [hep-th]].

\bibitem{Bedroya:2019tba}
A.~Bedroya, R.~Brandenberger, M.~Loverde and C.~Vafa,
Trans-Planckian Censorship and Inflationary Cosmology,
Phys. Rev. D \textbf{101} (2020) no.10, 103502
doi:10.1103/PhysRevD.101.103502
[arXiv:1909.11106 [hep-th]].

\bibitem{Bastero-Gil:2011rva}
M.~Bastero-Gil, A.~Berera and R.~O.~Ramos,
Shear viscous effects on the primordial power spectrum from warm inflation,
JCAP \textbf{07} (2011), 030
doi:10.1088/1475-7516/2011/07/030
[arXiv:1106.0701 [astro-ph.CO]].


\bibitem{Moss:2008yb}
I.~G.~Moss and C.~Xiong,
On the consistency of warm inflation,
JCAP \textbf{11} (2008), 023
doi:10.1088/1475-7516/2008/11/023
[arXiv:0808.0261 [astro-ph]].

\bibitem{Bastero-Gil:2012vuu}
M.~Bastero-Gil, A.~Berera, R.~Cerezo, R.~O.~Ramos and G.~S.~Vicente,
Stability analysis for the background equations for inflation with dissipation and in a viscous radiation bath,
JCAP \textbf{11} (2012), 042
doi:10.1088/1475-7516/2012/11/042
[arXiv:1209.0712 [astro-ph.CO]].

\bibitem{delCampo:2010by}
S.~del Campo, R.~Herrera, D.~Pav\'on and J.~R.~Villanueva,
On the consistency of warm inflation in the presence of viscosity,
JCAP \textbf{08} (2010), 002
doi:10.1088/1475-7516/2010/08/002
[arXiv:1007.0103 [astro-ph.CO]].

\bibitem{Zhang:2013waa}
X.~M.~Zhang and J.~Y.~Zhu,
Consistency of the tachyon warm inflationary universe models,
JCAP \textbf{02} (2014), 005
doi:10.1088/1475-7516/2014/02/005
[arXiv:1311.5327 [gr-qc]].

\bibitem{Zhang:2014dja}
X.~M.~Zhang and j.~Y.~Zhu,
Extension of warm inflation to noncanonical scalar fields,
Phys. Rev. D \textbf{90} (2014) no.12, 123519
doi:10.1103/PhysRevD.90.123519
[arXiv:1402.0205 [gr-qc]].

\bibitem{Motaharfar:2017dxh}
M.~Motaharfar, E.~Massaeli and H.~R.~Sepangi,
Power spectra in warm G-inflation and its consistency: stochastic approach,
Phys. Rev. D \textbf{96} (2017) no.10, 103541
doi:10.1103/PhysRevD.96.103541
[arXiv:1705.04049 [gr-qc]].

\bibitem{Cid:2015ota}
A.~Cid,
On the consistency of tachyon warm inflation with viscous pressure,
Phys. Lett. B \textbf{743} (2015), 127-133
doi:10.1016/j.physletb.2015.02.025
[arXiv:1503.00714 [gr-qc]].

\bibitem{Peng:2016yvb}
Z.~P.~Peng, J.~N.~Yu, X.~M.~Zhang and J.~Y.~Zhu,
Consistency of warm $k$-inflation,
Phys. Rev. D \textbf{94} (2016) no.10, 103531
doi:10.1103/PhysRevD.94.103531
[arXiv:1611.02789 [gr-qc]].

\bibitem{Das:2020lut}
S.~Das and R.~O.~Ramos,
Graceful exit problem in warm inflation,
Phys. Rev. D \textbf{103} (2021) no.12, 12
doi:10.1103/PhysRevD.103.123520
[arXiv:2005.01122 [gr-qc]].

\bibitem{Gleiser:1993ea}
M.~Gleiser and R.~O.~Ramos,
Microphysical approach to nonequilibrium dynamics of quantum fields,
Phys. Rev. D \textbf{50} (1994), 2441-2455
doi:10.1103/PhysRevD.50.2441
[arXiv:hep-ph/9311278 [hep-ph]].

\bibitem{Berera:1998gx}
A.~Berera, M.~Gleiser and R.~O.~Ramos,
Strong dissipative behavior in quantum field theory,
Phys. Rev. D \textbf{58} (1998), 123508
doi:10.1103/PhysRevD.58.123508
[arXiv:hep-ph/9803394 [hep-ph]].

\bibitem{Copeland:1997et}
E.~J.~Copeland, A.~R.~Liddle and D.~Wands,
Exponential potentials and cosmological scaling solutions,
Phys. Rev. D \textbf{57} (1998), 4686-4690
doi:10.1103/PhysRevD.57.4686
[arXiv:gr-qc/9711068 [gr-qc]].

\bibitem{Caldeira:1982iu}
A.~O.~Caldeira and A.~J.~Leggett,
Path integral approach to quantum Brownian motion,
Physica A \textbf{121} (1983), 587-616
doi:10.1016/0378-4371(83)90013-4

\bibitem{Berera:2001gs}
A.~Berera and R.~O.~Ramos,
The Affinity for scalar fields to dissipate,
Phys. Rev. D \textbf{63} (2001), 103509
doi:10.1103/PhysRevD.63.103509
[arXiv:hep-ph/0101049 [hep-ph]].

\bibitem{Berera:2004kc}
A.~Berera and R.~O.~Ramos,
Dynamics of interacting scalar fields in expanding space-time,
Phys. Rev. D \textbf{71} (2005), 023513
doi:10.1103/PhysRevD.71.023513
[arXiv:hep-ph/0406339 [hep-ph]].

\bibitem{Bellac:2011kqa}
M.~L.~Bellac,
Thermal Field Theory,
Cambridge University Press, 2011,
ISBN 978-0-511-88506-8, 978-0-521-65477-7
doi:10.1017/CBO9780511721700

\bibitem{Calzetta:2008iqa}
E.~A.~Calzetta and B.~L.~B.~Hu,
Nonequilibrium Quantum Field Theory,
Cambridge University Press, 2022,
ISBN 978-1-00-929003-6, 978-1-00-928998-6, 978-1-00-929002-9, 978-0-511-42147-1, 978-0-521-64168-5
doi:10.1017/9781009290036

\bibitem{Berera:2007qm}
A.~Berera, I.~G.~Moss and R.~O.~Ramos,
Local Approximations for Effective Scalar Field Equations of Motion,
Phys. Rev. D \textbf{76} (2007), 083520
doi:10.1103/PhysRevD.76.083520
[arXiv:0706.2793 [hep-ph]].

\bibitem{Bastero-Gil:2010dgy}
M.~Bastero-Gil, A.~Berera and R.~O.~Ramos,
Dissipation coefficients from scalar and fermion quantum field interactions,
JCAP \textbf{09} (2011), 033
doi:10.1088/1475-7516/2011/09/033
[arXiv:1008.1929 [hep-ph]].

\bibitem{Berera:1999wt}
A.~Berera and T.~W.~Kephart,
The Ubiquitous Inflaton in String-Inspired Models,
Phys. Rev. Lett. \textbf{83} (1999), 1084-1087
doi:10.1103/PhysRevLett.83.1084
[arXiv:hep-ph/9904410 [hep-ph]].

\bibitem{Bastero-Gil:2018yen}
M.~Bastero-Gil, A.~Berera, R.~Hern\'andez-Jim\'enez and J.~G.~Rosa,
Warm inflation within a supersymmetric distributed mass model,
Phys. Rev. D \textbf{99} (2019) no.10, 103520
doi:10.1103/PhysRevD.99.103520
[arXiv:1812.07296 [hep-ph]].

\bibitem{Hall:2004zr}
L.~M.~H.~Hall and I.~G.~Moss,
Thermal effects on pure and hybrid inflation,
Phys. Rev. D \textbf{71} (2005), 023514
doi:10.1103/PhysRevD.71.023514
[arXiv:hep-ph/0408323 [hep-ph]].

\bibitem{Yokoyama:1998ju}
J.~Yokoyama and A.~D.~Linde,
Is warm inflation possible?,
Phys. Rev. D \textbf{60} (1999), 083509
doi:10.1103/PhysRevD.60.083509
[arXiv:hep-ph/9809409 [hep-ph]].




\bibitem{ArkaniHamed:2001nc}
N.~Arkani-Hamed, A.~G.~Cohen and H.~Georgi,
Electroweak symmetry breaking from dimensional deconstruction,
Phys. Lett. B \textbf{513} (2001), 232-240
doi:10.1016/S0370-2693(01)00741-9
[arXiv:hep-ph/0105239 [hep-ph]].

\bibitem{Kaplan:2003aj}
D.~E.~Kaplan and N.~J.~Weiner,
Little inflatons and gauge inflation,
JCAP \textbf{02} (2004), 005
doi:10.1088/1475-7516/2004/02/005
[arXiv:hep-ph/0302014 [hep-ph]].

\bibitem{ArkaniHamed:2003mz}
N.~Arkani-Hamed, H.~C.~Cheng, P.~Creminelli and L.~Randall,
Pseudonatural inflation,
JCAP \textbf{07} (2003), 003
doi:10.1088/1475-7516/2003/07/003
[arXiv:hep-th/0302034 [hep-th]].

\bibitem{Schmaltz:2005ky}
M.~Schmaltz and D.~Tucker-Smith,
Little Higgs review,
Ann. Rev. Nucl. Part. Sci. \textbf{55} (2005), 229-270
doi:10.1146/annurev.nucl.55.090704.151502
[arXiv:hep-ph/0502182 [hep-ph]].

\bibitem{Rosa:2018iff}
J.~G.~Rosa and L.~B.~Ventura,
Warm Little Inflaton becomes Cold Dark Matter,
Phys. Rev. Lett. \textbf{122} (2019) no.16, 161301
doi:10.1103/PhysRevLett.122.161301
[arXiv:1811.05493 [hep-ph]].

\bibitem{Marsh:2015xka}
D.~J.~E.~Marsh,
Axion Cosmology,
Phys. Rept. \textbf{643} (2016), 1-79
doi:10.1016/j.physrep.2016.06.005
[arXiv:1510.07633 [astro-ph.CO]].

\bibitem{Moore:2010jd}
G.~D.~Moore and M.~Tassler,
The Sphaleron Rate in SU(N) Gauge Theory,
JHEP \textbf{02} (2011), 105
doi:10.1007/JHEP02(2011)105
[arXiv:1011.1167 [hep-ph]].

\bibitem{Laine:2016hma}
M.~Laine and A.~Vuorinen,
Basics of Thermal Field Theory,
Lect. Notes Phys. \textbf{925} (2016), pp.1-281,
doi:10.1007/978-3-319-31933-9
[arXiv:1701.01554 [hep-ph]].

\bibitem{Goswami:2019ehb}
S.~Das, G.~Goswami and C.~Krishnan,
Swampland, axions, and minimal warm inflation,
Phys. Rev. D \textbf{101} (2020) no.10, 103529
doi:10.1103/PhysRevD.101.103529
[arXiv:1911.00323 [hep-th]].

\bibitem{Laine:2021ego}
M.~Laine and S.~Procacci,
Minimal warm inflation with complete medium response,
JCAP \textbf{06} (2021), 031
doi:10.1088/1475-7516/2021/06/031
[arXiv:2102.09913 [hep-ph]].

\bibitem{DeRocco:2021rzv}
W.~DeRocco, P.~W.~Graham and S.~Kalia,
Warming up cold inflation,
JCAP \textbf{11} (2021), 011
doi:10.1088/1475-7516/2021/11/011
[arXiv:2107.07517 [hep-ph]].

\bibitem{Das:2022ubr}
S.~Das and R.~O.~Ramos,
Running and running of the running of the scalar spectral index in warm inflation,
Universe \textbf{9} (2023), 76
doi:10.3390/universe9020076
[arXiv:2212.13914 [astro-ph.CO]].

\bibitem{Ramos:2013nsa}
R.~O.~Ramos and L.~A.~da Silva,
Power spectrum for inflation models with quantum and thermal noises,
JCAP \textbf{03} (2013), 032
doi:10.1088/1475-7516/2013/03/032
[arXiv:1302.3544 [astro-ph.CO]].

\bibitem{Bastero-Gil:2017yzb}
M.~Bastero-Gil, A.~Berera, R.~O.~Ramos and J.~G.~Rosa,
Adiabatic out-of-equilibrium solutions to the Boltzmann equation in warm inflation,
JHEP \textbf{02} (2018), 063
doi:10.1007/JHEP02(2018)063
[arXiv:1711.09023 [hep-ph]].

\bibitem{Graham:2009bf}
C.~Graham and I.~G.~Moss,
Density fluctuations from warm inflation,
JCAP \textbf{07} (2009), 013
doi:10.1088/1475-7516/2009/07/013
[arXiv:0905.3500 [astro-ph.CO]].

\bibitem{Kodama:1984ziu}
H.~Kodama and M.~Sasaki,
Cosmological Perturbation Theory,
Prog. Theor. Phys. Suppl. \textbf{78} (1984), 1-166
doi:10.1143/PTPS.78.1

\bibitem{Hwang:1991aj}
J.~c.~Hwang,
Perturbations of the Robertson-Walker space - Multicomponent sources and generalized gravity,
Astrophys. J. \textbf{375} (1991), 443-462
doi:10.1086/170206

\bibitem{Baumann:2009ds}
D.~Baumann,
Inflation,
doi:10.1142/9789814327183\_0010
[arXiv:0907.5424 [hep-th]].

\bibitem{Liddle:2003as}
A.~R.~Liddle and S.~M.~Leach,
How long before the end of inflation were observable perturbations produced?,
Phys. Rev. D \textbf{68} (2003), 103503
doi:10.1103/PhysRevD.68.103503
[arXiv:astro-ph/0305263 [astro-ph]].

\bibitem{DeOliveira:2002wk}
H.~P.~De Oliveira,
Density perturbations in warm inflation and COBE normalization,
Phys. Lett. B \textbf{526} (2002), 1
doi:10.1016/S0370-2693(01)01496-4
[arXiv:gr-qc/0202045 [gr-qc]].

\bibitem{Bastero-Gil:2019rsp}
M.~Bastero-Gil, A.~Berera and J.~R.~Calder\'on,
Reexamination of the warm inflation curvature perturbations spectrum,
JCAP \textbf{07} (2019), 019
doi:10.1088/1475-7516/2019/07/019
[arXiv:1904.04086 [astro-ph.CO]].

\bibitem{Bastero-Gil:2014jsa}
M.~Bastero-Gil, A.~Berera, I.~G.~Moss and R.~O.~Ramos,
Cosmological fluctuations of a random field and radiation fluid,
JCAP \textbf{05} (2014), 004
doi:10.1088/1475-7516/2014/05/004
[arXiv:1401.1149 [astro-ph.CO]].

\bibitem{Motaharfar:2021egj}
M.~Motaharfar and R.~O.~Ramos,
Dirac-Born-Infeld warm inflation realization in the strong dissipation regime,
Phys. Rev. D \textbf{104} (2021) no.4, 043522
doi:10.1103/PhysRevD.104.043522
[arXiv:2105.01131 [hep-th]].

\bibitem{Motaharfar:2018mni}
M.~Motaharfar, E.~Massaeli and H.~R.~Sepangi,
Warm Higgs G-inflation: predictions and constraints from Planck 2015 likelihood,
JCAP \textbf{10} (2018), 002
doi:10.1088/1475-7516/2018/10/002
[arXiv:1807.09548 [gr-qc]].

\bibitem{Qiu:2021ytc}
Y.~Qiu and L.~Sorbo,
Spectrum of tensor perturbations in warm inflation,
Phys. Rev. D \textbf{104} (2021) no.8, 083542
doi:10.1103/PhysRevD.104.083542
[arXiv:2107.09754 [astro-ph.CO]].

\bibitem{Li:2018wno}
X.~B.~Li, H.~Wang and J.~Y.~Zhu,
Gravitational waves from warm inflation,
Phys. Rev. D \textbf{97} (2018) no.6, 063516
doi:10.1103/PhysRevD.97.063516
[arXiv:1803.10074 [gr-qc]].

\bibitem{Arraut:2013lca}
I.~Arraut,
The Astrophysical Scales Set by the Cosmological Constant, Black-Hole Thermodynamics and Non-Linear Massive Gravity,
Universe \textbf{3}, no.2, 45 (2017)
doi:10.3390/universe3020045
[arXiv:1305.0475 [gr-qc]].

\bibitem{Planck:2018vyg}
N.~Aghanim \textit{et al.} [Planck],
Planck 2018 results. VI. Cosmological parameters,
Astron. Astrophys. \textbf{641} (2020), A6
[erratum: Astron. Astrophys. \textbf{652} (2021), C4]
doi:10.1051/0004-6361/201833910
[arXiv:1807.06209 [astro-ph.CO]].


\bibitem{BICEP:2021xfz}
P.~A.~R.~Ade \textit{et al.} [BICEP and Keck],
Phys. Rev. Lett. \textbf{127} (2021) no.15, 151301
doi:10.1103/PhysRevLett.127.151301
[arXiv:2110.00483 [astro-ph.CO]].

\bibitem{Moss:2007cv}
I.~G.~Moss and C.~Xiong,
Non-Gaussianity in fluctuations from warm inflation,
JCAP \textbf{04} (2007), 007
doi:10.1088/1475-7516/2007/04/007
[arXiv:astro-ph/0701302 [astro-ph]].

\bibitem{Moss:2011qc}
I.~G.~Moss and T.~Yeomans,
Non-gaussianity in the strong regime of warm inflation,
JCAP \textbf{08} (2011), 009
doi:10.1088/1475-7516/2011/08/009
[arXiv:1102.2833 [astro-ph.CO]].

\bibitem{Bastero-Gil:2014raa}
M.~Bastero-Gil, A.~Berera, I.~G.~Moss and R.~O.~Ramos,
Theory of non-Gaussianity in warm inflation,
JCAP \textbf{12} (2014), 008
doi:10.1088/1475-7516/2014/12/008
[arXiv:1408.4391 [astro-ph.CO]].

\bibitem{Mirbabayi:2022cbt}
M.~Mirbabayi and A.~Gruzinov,
Shapes of non-Gaussianity in warm inflation,
[arXiv:2205.13227 [astro-ph.CO]].

\bibitem{Palti:2019pca}
E.~Palti,
The Swampland: Introduction and Review,
Fortsch. Phys. \textbf{67} (2019) no.6, 1900037
doi:10.1002/prop.201900037
[arXiv:1903.06239 [hep-th]].

\bibitem{Ooguri:2006in}
H.~Ooguri and C.~Vafa,
On the Geometry of the String Landscape and the Swampland,
Nucl. Phys. B \textbf{766} (2007), 21-33
doi:10.1016/j.nuclphysb.2006.10.033
[arXiv:hep-th/0605264 [hep-th]].

\bibitem{Garg:2018reu}
S.~K.~Garg and C.~Krishnan,
Bounds on Slow Roll and the de Sitter Swampland,
JHEP \textbf{11} (2019), 075
doi:10.1007/JHEP11(2019)075
[arXiv:1807.05193 [hep-th]].

\bibitem{Bousso:1999xy}
R.~Bousso,
A Covariant entropy conjecture,
JHEP \textbf{07} (1999), 004
doi:10.1088/1126-6708/1999/07/004
[arXiv:hep-th/9905177 [hep-th]].

\bibitem{Kinney:2018kew}
W.~H.~Kinney,
Eternal Inflation and the Refined Swampland Conjecture,
Phys. Rev. Lett. \textbf{122} (2019) no.8, 081302
doi:10.1103/PhysRevLett.122.081302
[arXiv:1811.11698 [astro-ph.CO]].

\bibitem{Baumann:2014nda}
D.~Baumann and L.~McAllister,
Inflation and String Theory,
Cambridge University Press, 2015,
ISBN 978-1-107-08969-3, 978-1-316-23718-2
doi:10.1017/CBO9781316105733
[arXiv:1404.2601 [hep-th]].

\bibitem{Aalsma:2020aib}
L.~Aalsma and G.~Shiu,
Chaos and complementarity in de Sitter space,
JHEP \textbf{05} (2020), 152
doi:10.1007/JHEP05(2020)152
[arXiv:2002.01326 [hep-th]].

\bibitem{Mizuno:2019bxy}
S.~Mizuno, S.~Mukohyama, S.~Pi and Y.~L.~Zhang,
Universal Upper Bound on the Inflationary Energy Scale from the Trans-Planckian Censorship Conjecture,
Phys. Rev. D \textbf{102} (2020) no.2, 021301
doi:10.1103/PhysRevD.102.021301
[arXiv:1910.02979 [astro-ph.CO]].

\bibitem{Kamali:2019gzr}
V.~Kamali and R.~Brandenberger,
Relaxing the TCC Bound on Inflationary Cosmology?,
Eur. Phys. J. C \textbf{80} (2020) no.4, 339
doi:10.1140/epjc/s10052-020-7908-8
[arXiv:2001.00040 [hep-th]].

\bibitem{Planck:2018jri}
Y.~Akrami \textit{et al.} [Planck],
Planck 2018 results. X. Constraints on inflation,
Astron. Astrophys. \textbf{641} (2020), A10
doi:10.1051/0004-6361/201833887
[arXiv:1807.06211 [astro-ph.CO]].

\bibitem{Dimopoulos:2018upl}
K.~Dimopoulos,
Phys. Rev. D \textbf{98} (2018) no.12, 123516
doi:10.1103/PhysRevD.98.123516
[arXiv:1810.03438 [gr-qc]].

\bibitem{Brahma:2019iyy}
S.~Brahma and S.~Shandera,
JHEP \textbf{11} (2019), 016
doi:10.1007/JHEP11(2019)016
[arXiv:1904.10979 [hep-th]].

\bibitem{Vicente:2015hga}
G.~S.~Vicente, L.~A.~da Silva and R.~O.~Ramos,
Eternal inflation in a dissipative and radiation environment: Heated demise of eternity,
Phys. Rev. D \textbf{93} (2016) no.6, 063509
doi:10.1103/PhysRevD.93.063509
[arXiv:1509.08983 [astro-ph.CO]].

\bibitem{Freese:1990rb}
K.~Freese, J.~A.~Frieman and A.~V.~Olinto,
Natural inflation with pseudo - Nambu-Goldstone bosons,
Phys. Rev. Lett. \textbf{65} (1990), 3233-3236
doi:10.1103/PhysRevLett.65.3233

\bibitem{Visinelli:2011jy}
L.~Visinelli,
Natural Warm Inflation,
JCAP \textbf{09} (2011), 013
doi:10.1088/1475-7516/2011/09/013
[arXiv:1107.3523 [astro-ph.CO]].

\bibitem{Mishra:2011vh}
H.~Mishra, S.~Mohanty and A.~Nautiyal,
Warm natural inflation,
Phys. Lett. B \textbf{710} (2012), 245-250
doi:10.1016/j.physletb.2012.02.005
[arXiv:1106.3039 [hep-ph]].

\bibitem{Kamali:2019ppi}
V.~Kamali,
Warm pseudoscalar inflation,
Phys. Rev. D \textbf{100} (2019) no.4, 043520
doi:10.1103/PhysRevD.100.043520
[arXiv:1901.01897 [gr-qc]].

\bibitem{Montefalcone:2022jfw}
G.~Montefalcone, V.~Aragam, L.~Visinelli and K.~Freese,
Observational Constraints on Warm Natural Inflation,
[arXiv:2212.04482 [gr-qc]].

\bibitem{Herrera:2006ck}
R.~Herrera, S.~del Campo and C.~Campuzano,
Tachyon warm inflationary universe models,
JCAP \textbf{10} (2006), 009
doi:10.1088/1475-7516/2006/10/009
[arXiv:astro-ph/0610339 [astro-ph]].

\bibitem{delCampo:2009xi}
S.~del Campo and R.~Herrera,
Warm-Intermediate inflationary universe model,
JCAP \textbf{04} (2009), 005
doi:10.1088/1475-7516/2009/04/005
[arXiv:0903.4214 [astro-ph.CO]].

\bibitem{Herrera:2011zz}
R.~Herrera and E.~San Martin,
Warm-intermediate inflationary universe model in braneworld cosmologies,
Eur. Phys. J. C \textbf{71} (2011), 1701
doi:10.1140/epjc/s10052-011-1701-7
[arXiv:1108.1371 [gr-qc]].

\bibitem{Setare:2012fg}
M.~R.~Setare and V.~Kamali,
Tachyon Warm-Intermediate Inflationary Universe Model in High Dissipative Regime,
JCAP \textbf{08} (2012), 034
doi:10.1088/1475-7516/2012/08/034
[arXiv:1210.0742 [hep-th]].

\bibitem{Setare:2013ula}
M.~R.~Setare and V.~Kamali,
Tachyon Warm-Logamediate Inflationary Universe Model in High Dissipative Regime,
Phys. Rev. D \textbf{87} (2013), 083524
doi:10.1103/PhysRevD.87.083524
[arXiv:1305.0740 [hep-th]].

\bibitem{Setare:2014oka}
M.~R.~Setare and V.~Kamali,
Warm-Intermediate Inflationary Universe Model with Viscous Pressure in High Dissipative Regime,
Gen. Rel. Grav. \textbf{46} (2014), 1698
doi:10.1007/s10714-014-1698-y
[arXiv:1403.0186 [gr-qc]].

\bibitem{Kamali:2016frd}
V.~Kamali, S.~Basilakos and A.~Mehrabi,
Tachyon warm-intermediate inflation in the light of Planck data,
Eur. Phys. J. C \textbf{76} (2016) no.10, 525
doi:10.1140/epjc/s10052-016-4380-6
[arXiv:1604.05434 [gr-qc]].

\bibitem{Motaharfar:2016dqt}
M.~Motaharfar and H.~R.~Sepangi,
Warm-tachyon Gauss\textendash{}Bonnet inflation in the light of Planck 2015 data,
Eur. Phys. J. C \textbf{76} (2016) no.11, 646
doi:10.1140/epjc/s10052-016-4474-1
[arXiv:1604.00453 [gr-qc]].

\bibitem{Mohammadi:2020vgs}
A.~Mohammadi, T.~Golanbari, H.~Sheikhahmadi, K.~Sayar, L.~Akhtari, M.~A.~Rasheed and K.~Saaidi,
Warm tachyon inflation and swampland criteria,
Chin. Phys. C \textbf{44} (2020) no.9, 095101
doi:10.1088/1674-1137/44/9/095101
[arXiv:2001.10042 [gr-qc]].

\bibitem{Dymnikova:2001ga}
I.~Dymnikova and M.~Khlopov,
Decay of cosmological constant as Bose condensate evaporation,
Mod. Phys. Lett. A \textbf{15} (2000), 2305-2314
doi:10.1142/S0217732300002966
[arXiv:astro-ph/0102094 [astro-ph]].

\bibitem{Dymnikova:2001jy}
I.~Dymnikova and M.~Khlopov,
Decay of cosmological constant in selfconsistent inflation,
Eur. Phys. J. C \textbf{20} (2001), 139-146
doi:10.1007/s100520100625

\bibitem{Gashti:2022pvu}
S.~N.~Gashti and J.~Sadeghi,
Refined swampland conjecture in warm vector hybrid inflationary scenario,
Eur. Phys. J. Plus \textbf{137} (2022) no.6, 731
doi:10.1140/epjp/s13360-022-02961-8

\bibitem{Santos:2022exm}
F.~B.~M.~d.~Santos, R.~Silva, S.~S.~da Costa, M.~Benetti and J.~S.~Alcaniz,
Warm $\beta$-exponential inflation and the Swampland Conjectures,
[arXiv:2209.06153 [astro-ph.CO]].

\bibitem{Payaka:2022jtb}
A.~Payaka, W.~Amaek and P.~Channuie,
Warm deformed R2 inflation,
Nucl. Phys. B \textbf{986} (2023), 116052
doi:10.1016/j.nuclphysb.2022.116052
[arXiv:2203.11041 [gr-qc]].

\bibitem{Bouabdallaoui:2022wyp}
Z.~Bouabdallaoui, A.~Errahmani, M.~Bouhmadi-L\'opez and T.~Ouali,
Scalar warm inflation in holographic cosmology,
Phys. Rev. D \textbf{105} (2022) no.4, 043513
doi:10.1103/PhysRevD.105.043513
[arXiv:2203.05047 [hep-th]].

\bibitem{Reyimuaji:2020bkm}
Y.~Reyimuaji and X.~Zhang,
Warm-assisted natural inflation,
JCAP \textbf{04} (2021), 077
doi:10.1088/1475-7516/2021/04/077
[arXiv:2012.07329 [astro-ph.CO]].

\bibitem{Graef:2018ulg}
L.~L.~Graef and R.~O.~Ramos,
Probability of Warm Inflation in Loop Quantum Cosmology,
Phys. Rev. D \textbf{98} (2018) no.2, 023531
doi:10.1103/PhysRevD.98.023531
[arXiv:1805.05985 [gr-qc]].

\bibitem{Berera:2018tfc}
A.~Berera, J.~Mabillard, M.~Pieroni and R.~O.~Ramos,
Identifying Universality in Warm Inflation,
JCAP \textbf{07} (2018), 021
doi:10.1088/1475-7516/2018/07/021
[arXiv:1803.04982 [astro-ph.CO]].

\bibitem{Herrera:2018cgi}
R.~Herrera,
Reconstructing warm inflation,
Eur. Phys. J. C \textbf{78} (2018) no.3, 245
doi:10.1140/epjc/s10052-018-5741-0
[arXiv:1801.05138 [gr-qc]].

\bibitem{Wang:2019ozs}
Y.~y.~Wang, J.~Y.~Zhu and X.~M.~Zhang,
Observational Constraints on Two-field Warm Inflation,
Phys. Rev. D \textbf{99} (2019) no.10, 103529
doi:10.1103/PhysRevD.99.103529
[arXiv:1905.02414 [gr-qc]].

\bibitem{Harko:2021gnz}
T.~Harko and H.~Sheikhahmadi,
Warm inflation with non-comoving scalar field and radiation fluid,
Eur. Phys. J. C \textbf{81} (2021) no.2, 165
doi:10.1140/epjc/s10052-021-08964-6
[arXiv:2102.04728 [gr-qc]].

\bibitem{Bose:2022wla}
A.~Bose and S.~Chakraborty,
Does fractal universe favour warm inflation: Observational support?,
Nucl. Phys. B \textbf{978} (2022), 115767
doi:10.1016/j.nuclphysb.2022.115767
[arXiv:2204.05712 [gr-qc]].

\bibitem{Harko:2020cev}
T.~Harko and H.~Sheikhahmadi,
Irreversible thermodynamical description of warm inflationary cosmological models,
Phys. Dark Univ. \textbf{28} (2020), 100521
doi:10.1016/j.dark.2020.100521
[arXiv:2003.02257 [gr-qc]].

\bibitem{Sheikhahmadi:2019gzs}
H.~Sheikhahmadi, A.~Mohammadi, A.~Aghamohammadi, T.~Harko, R.~Herrera, C.~Corda, A.~Abebe and K.~Saaidi,
Constraining chameleon field driven warm inflation with Planck 2018 data,
Eur. Phys. J. C \textbf{79} (2019) no.12, 1038
doi:10.1140/epjc/s10052-019-7571-0
[arXiv:1907.10966 [gr-qc]].

\bibitem{Bertolami:2022knd}
O.~Bertolami and P.~M.~S\'a,
Multi-field cold and warm inflation and the de Sitter swampland conjectures,
JCAP \textbf{09} (2022), 001
doi:10.1088/1475-7516/2022/09/001
[arXiv:2204.13794 [gr-qc]].

\bibitem{Motaharfar:2021gwi}
M.~Motaharfar and P.~Singh,
Role of dissipative effects in the quantum gravitational onset of warm Starobinsky inflation in a closed universe,
Phys. Rev. D \textbf{104} (2021) no.10, 106006
doi:10.1103/PhysRevD.104.106006
[arXiv:2102.09578 [gr-qc]].

\bibitem{AlHallak:2022haa}
M.~AlHallak, K.~K.~A.~Said, N.~Chamoun and M.~S.~El-Daher,
On Warm Natural Inflation and Planck 2018 constraints,
[arXiv:2211.07775 [gr-qc]].

\bibitem{Montefalcone:2022owy}
G.~Montefalcone, V.~Aragam, L.~Visinelli and K.~Freese,
Constraints on the scalar-field potential in warm inflation,
[arXiv:2209.14908 [gr-qc]].

\bibitem{deOliveira:1997jt}
H.~P.~de Oliveira and R.~O.~Ramos,
Dynamical system analysis for inflation with dissipation,
Phys. Rev. D \textbf{57} (1998), 741-749
doi:10.1103/PhysRevD.57.741
[arXiv:gr-qc/9710093 [gr-qc]].

\bibitem{Li:2018sfs}
X.~B.~Li, Y.~Y.~Wang, H.~Wang and J.~Y.~Zhu,
Dynamic analysis of noncanonical warm inflation,
Phys. Rev. D \textbf{98} (2018) no.4, 043510
doi:10.1103/PhysRevD.98.043510
[arXiv:1804.05360 [gr-qc]].

\bibitem{Alho:2022qri}
A.~Alho, V.~Bessa and F.~C.~Mena,
Dynamics of interacting monomial scalar field potentials and perfect fluids,
[arXiv:2212.02942 [gr-qc]].

\bibitem{Arya:2017zlb}
R.~Arya, A.~Dasgupta, G.~Goswami, J.~Prasad and R.~Rangarajan,
Revisiting CMB constraints on warm inflation,
JCAP \textbf{02} (2018), 043
doi:10.1088/1475-7516/2018/02/043
[arXiv:1710.11109 [astro-ph.CO]].

\bibitem{Bastero-Gil:2017wwl}
M.~Bastero-Gil, S.~Bhattacharya, K.~Dutta and M.~R.~Gangopadhyay,
Constraining Warm Inflation with CMB data,
JCAP \textbf{02} (2018), 054
doi:10.1088/1475-7516/2018/02/054
[arXiv:1710.10008 [astro-ph.CO]].

\bibitem{Arya:2018sgw}
R.~Arya and R.~Rangarajan,
Study of warm inflationary models and their parameter estimation from CMB,
Int. J. Mod. Phys. D \textbf{29} (2020) no.08, 2050055
doi:10.1142/S0218271820500558
[arXiv:1812.03107 [astro-ph.CO]].

\bibitem{Benetti:2019kgw}
M.~Benetti, L.~Graef and R.~O.~Ramos,
Observational Constraints on Warm Inflation in Loop Quantum Cosmology,
JCAP \textbf{10} (2019), 066
doi:10.1088/1475-7516/2019/10/066
[arXiv:1907.03633 [astro-ph.CO]].

\bibitem{Berera:1998hv}
A.~Berera, T.~W.~Kephart and S.~D.~Wick,
GUT cosmic magnetic fields in a warm inflationary universe,
Phys. Rev. D \textbf{59} (1999), 043510
doi:10.1103/PhysRevD.59.043510
[arXiv:hep-ph/9809404 [hep-ph]].

\bibitem{Brandenberger:2003kc}
R.~H.~Brandenberger and M.~Yamaguchi,
Spontaneous baryogenesis in warm inflation,
Phys. Rev. D \textbf{68} (2003), 023505
doi:10.1103/PhysRevD.68.023505
[arXiv:hep-ph/0301270 [hep-ph]].

\bibitem{Bastero-Gil:2011clw}
M.~Bastero-Gil, A.~Berera, R.~O.~Ramos and J.~G.~Rosa,
Warm baryogenesis,
Phys. Lett. B \textbf{712} (2012), 425-429
doi:10.1016/j.physletb.2012.05.032
[arXiv:1110.3971 [hep-ph]].

\bibitem{Bastero-Gil:2014oga}
M.~Bastero-Gil, A.~Berera, R.~O.~Ramos and J.~G.~Rosa,
Observational implications of mattergenesis during inflation,
JCAP \textbf{10} (2014), 053
doi:10.1088/1475-7516/2014/10/053
[arXiv:1404.4976 [astro-ph.CO]].

\bibitem{BuenoSanchez:2010ygd}
J.~C.~Bueno Sanchez, M.~Bastero-Gil, A.~Berera, K.~Dimopoulos and K.~Kohri,
The gravitino problem in supersymmetric warm inflation,
JCAP \textbf{03} (2011), 020
doi:10.1088/1475-7516/2011/03/020
[arXiv:1011.2398 [hep-ph]].

\bibitem{Bartrum:2012tg}
S.~Bartrum, A.~Berera and J.~G.~Rosa,
Gravitino cosmology in supersymmetric warm inflation,
Phys. Rev. D \textbf{86} (2012), 123525
doi:10.1103/PhysRevD.86.123525
[arXiv:1208.4276 [hep-ph]].

\bibitem{Dimopoulos:2019gpz}
K.~Dimopoulos and L.~Donaldson-Wood,
Warm quintessential inflation,
Phys. Lett. B \textbf{796} (2019), 26-31
doi:10.1016/j.physletb.2019.07.017
[arXiv:1906.09648 [gr-qc]].

\bibitem{Rosa:2019jci}
J.~G.~Rosa and L.~B.~Ventura,
Warm Little Inflaton becomes Dark Energy,
Phys. Lett. B \textbf{798} (2019), 134984
doi:10.1016/j.physletb.2019.134984
[arXiv:1906.11835 [hep-ph]].

\bibitem{Lima:2019yyv}
G.~B.~F.~Lima and R.~O.~Ramos,
Unified early and late Universe cosmology through dissipative effects in steep quintessential inflation potential models,
Phys. Rev. D \textbf{100} (2019) no.12, 123529
doi:10.1103/PhysRevD.100.123529
[arXiv:1910.05185 [astro-ph.CO]].

\bibitem{Gangopadhyay:2020bxn}
M.~R.~Gangopadhyay, S.~Myrzakul, M.~Sami and M.~K.~Sharma,
Paradigm of warm quintessential inflation and production of relic gravity waves,
Phys. Rev. D \textbf{103} (2021) no.4, 043505
doi:10.1103/PhysRevD.103.043505
[arXiv:2011.09155 [astro-ph.CO]].

\bibitem{Basak:2021cgk}
S.~Basak, S.~Bhattacharya, M.~R.~Gangopadhyay, N.~Jaman, R.~Rangarajan and M.~Sami,
The paradigm of warm quintessential inflation and spontaneous baryogenesis,
JCAP \textbf{03} (2022) no.03, 063
doi:10.1088/1475-7516/2022/03/063
[arXiv:2110.00607 [astro-ph.CO]].

\bibitem{Saleem:2021ytj}
R.~Saleem and F.~Mehmood,
A study of warm inflation model inspired by some inhomogeneous dark energy fluids,
Eur. Phys. J. Plus \textbf{136} (2021) no.5, 570
doi:10.1140/epjp/s13360-021-01535-4

\bibitem{Levy:2020zfo}
M.~Levy, J.~G.~Rosa and L.~B.~Ventura,
Warm inflation, neutrinos and dark matter: a minimal extension of the Standard Model,
JHEP \textbf{12} (2021), 176
doi:10.1007/JHEP12(2021)176
[arXiv:2012.03988 [hep-ph]].

\bibitem{Zhang:2021zol}
X.~M.~Zhang, K.~Li, Y.~F.~Guo, P.~C.~Chu, H.~Liu and J.~Y.~Zhu,
Two models unifying warm inflation with dark matter and dark energy,
Phys. Rev. D \textbf{104} (2021) no.10, 103513
doi:10.1103/PhysRevD.104.103513
[arXiv:2111.14138 [gr-qc]].

\bibitem{Sa:2020fvn}
P.~M.~S\'a,
Triple unification of inflation, dark energy, and dark matter in two-scalar-field cosmology,
Phys. Rev. D \textbf{102} (2020) no.10, 103519
doi:10.1103/PhysRevD.102.103519
[arXiv:2007.07109 [gr-qc]].

\bibitem{DAgostino:2021vvv}
R.~D'Agostino and O.~Luongo,
Cosmological viability of a double field unified model from warm inflation,
Phys. Lett. B \textbf{829} (2022), 137070
doi:10.1016/j.physletb.2022.137070
[arXiv:2112.12816 [astro-ph.CO]].

\bibitem{Arya:2019wck}
R.~Arya,
Formation of Primordial Black Holes from Warm Inflation,
JCAP \textbf{09} (2020), 042
doi:10.1088/1475-7516/2020/09/042
[arXiv:1910.05238 [astro-ph.CO]].

\bibitem{Bastero-Gil:2021fac}
M.~Bastero-Gil and M.~S.~D\'\i{}az-Blanco,
Gravity waves and primordial black holes in scalar warm little inflation,
JCAP \textbf{12} (2021) no.12, 052
doi:10.1088/1475-7516/2021/12/052
[arXiv:2105.08045 [hep-ph]].

\bibitem{Correa:2022ngq}
M.~Correa, M.~R.~Gangopadhyay, N.~Jaman and G.~J.~Mathews,
Primordial black-hole dark matter via warm natural inflation,
Phys. Lett. B \textbf{835} (2022), 137510
doi:10.1016/j.physletb.2022.137510
[arXiv:2207.10394 [gr-qc]].

\bibitem{Arya:2022xzc}
R.~Arya and A.~K.~Mishra,
Scalar induced gravitational waves from warm inflation,
Phys. Dark Univ. \textbf{37} (2022), 101116
doi:10.1016/j.dark.2022.101116
[arXiv:2204.02896 [astro-ph.CO]].

\bibitem{Ballesteros:2022hjk}
G.~Ballesteros, M.~A.~G.~Garc\'\i{}a, A.~P.~Rodr\'\i{}guez, M.~Pierre and J.~Rey,
Primordial black holes and gravitational waves from dissipation during inflation,
JCAP \textbf{12} (2022), 006
doi:10.1088/1475-7516/2022/12/006
[arXiv:2208.14978 [astro-ph.CO]].

\end{thebibliography}
\end{document}